\documentclass[11pt]{article}

\usepackage{amsfonts,amssymb,chet,dsfont,graphicx,mathrsfs,pgfplots,tikz}
\allowdisplaybreaks

\newcommand{\1}{\mathds{1}}

\newcommand{\Op}[2]{\mathcal{O}_{#1}(\eta_{#2})}

\newcommand{\ee}[3]{(\eta_{#1}\cdot\eta_{#2})^{#3}}
\newcommand{\e}[3]{\eta_{#1}^{#2_{#3}}}
\newcommand{\D}{\mathcal{D}}
\newcommand{\A}{\mathcal{A}}
\newcommand{\cOPE}[4]{{}_{#1}c_{#2#3}^{\phantom{#2#3}#4}}

\newcommand{\tOPE}[6]{{}_{#1}t_{#2#3}^{#5#6#4}}
\newcommand{\cCF}[4]{{}_{#1}c_{#2#3#4}}
\newcommand{\tCF}[6]{{}_{#1}t_{#2#3#4}^{#5#6}}

\newcommand{\Vev}[1]{\left\langle{#1}\right\rangle}

\title{Conformal Four-Point Correlation Functions\\from the Operator Product Expansion}

\author{Jean-Fran\c{c}ois Fortin$^{\ast,}$\email{jean-francois.fortin@phy.ulaval.ca}, Valentina Prilepina$^{\ast,}$\email{valentina.prilepina.1@ulaval.ca} and Witold Skiba$^{\dagger,}$\email{witold.skiba@yale.edu}}

\affiliation{
$^\ast$D\'epartement de Physique, de G\'enie Physique et d'Optique\\Universit\'e Laval, Qu\'ebec, QC G1V 0A6, Canada\\
$^\dagger$Department of Physics, Yale University, New Haven, CT 06520, USA
}

\abstract{We show how to compute conformal blocks of operators in arbitrary Lorentz representations using the formalism described in \cite{Fortin:2019fvx,Fortin:2019dnq} and present several explicit examples of blocks derived via this method.  The procedure for obtaining the blocks has been reduced to (1) determining the relevant group theoretic structures and (2) applying appropriate predetermined substitution rules.  The most transparent expressions for the blocks we find are expressed in terms of specific substitutions on the Gegenbauer polynomials.  In our examples, we study operators which transform as scalars, symmetric tensors, two-index antisymmetric tensors, as well as mixed representations of the Lorentz group.}

\date{July 2019} 

\begin{document}

\maketitle



\section{Introduction}\label{SecIntro}

Applications of conformal field theory (CFT) in high-energy and condensed matter physics are well-known, as is the connection between gravity and CFTs.  Motivation for renewed interest in CFTs includes the plethora of fruitful advances of the bootstrap program \cite{Ferrara:1973yt,Polyakov:1974gs} in more than two dimensions.  (The modern bootstrap literature is vast.  It spans many interesting numerical results \cite{Rattazzi:2008pe,Rychkov:2009ij,Caracciolo:2009bx,Poland:2010wg,Rattazzi:2010gj,Poland:2011ey,Rychkov:2011et,ElShowk:2012ht,Liendo:2012hy,ElShowk:2012hu,Gliozzi:2013ysa,Alday:2013opa,Gaiotto:2013nva,El-Showk:2014dwa,Chester:2014fya,Kos:2014bka,Caracciolo:2014cxa,Paulos:2014vya,Beem:2014zpa,Simmons-Duffin:2015qma,Bobev:2015jxa,Beem:2015aoa,Iliesiu:2015qra,Poland:2015mta,Lemos:2015awa,Lin:2015wcg,Chester:2016wrc,Behan:2016dtz,El-Showk:2016mxr,Lin:2016gcl,Lemos:2016xke,Beem:2016wfs,Li:2017ddj,Collier:2017shs,Cornagliotto:2017dup,Rychkov:2017tpc,Nakayama:2017vdd,Chang:2017xmr,Dymarsky:2017yzx,Karateev:2019pvw}, a variety of impressive analytic results \cite{Cornalba:2007fs,Cornalba:2009ax,Pappadopulo:2012jk,Costa:2012cb,Hogervorst:2013sma,Hartman:2015lfa,Kim:2015oca,Li:2015itl,Hartman:2016dxc,Hofman:2016awc,Hartman:2016lgu,Afkhami-Jeddi:2016ntf,Gadde:2017sjg,Hogervorst:2017sfd,Caron-Huot:2017vep,Hogervorst:2017kbj,Kulaxizi:2017ixa,Li:2017lmh,Cuomo:2017wme,Dey:2017oim,Simmons-Duffin:2017nub,Elkhidir:2017iov,Kravchuk:2018htv,Karateev:2018oml,Liendo:2019jpu,Albayrak:2019gnz,Li:2019twz}, work involving global symmetries \cite{Rattazzi:2010yc,Vichi:2011ux,Kos:2013tga,Berkooz:2014yda,Nakayama:2014lva,Nakayama:2014yia,Nakayama:2014sba,Bae:2014hia,Chester:2014gqa,Kos:2015mba,Chester:2015qca,Chester:2015lej,Dey:2016zbg,Nakayama:2016knq,Li:2016wdp,Pang:2016xno,Dymarsky:2017xzb,Stergiou:2018gjj,Kousvos:2018rhl,Stergiou:2019dcv} and higher-spin fields \cite{Alday:2015eya,Alday:2015ota,Alday:2015ewa,Alday:2016mxe,Alday:2016njk,Alday:2016jfr}, as well as lectures and reviews \cite{Rychkov:2016iqz,Simmons-Duffin:2016gjk,Poland:2018epd,Chester:2019wfx}.)  The starting point for the bootstrap are the conformal blocks, which are the building blocks of the four-point correlation functions.  Calculating conformal blocks beyond two dimensions has proved daunting, and only a few cases were successfully worked out almost twenty years ago~\cite{Dolan:2000ut,Dolan:2003hv} (see also \cite{Ferrara:1973vz,Ferrara:1974nf,Dobrev:1977qv,Exton_1995} for earlier work).  With the revival of interest in the conformal bootstrap, several new results for conformal blocks were developed more recently \cite{Giombi:2011rz,Costa:2011mg,Dolan:2011dv,Costa:2011dw,SimmonsDuffin:2012uy,Costa:2014rya,Elkhidir:2014woa,Echeverri:2015rwa,Hijano:2015zsa,Rejon-Barrera:2015bpa,Penedones:2015aga,Iliesiu:2015akf,Echeverri:2016dun,Isachenkov:2016gim,Costa:2016hju,Costa:2016xah,Chen:2016bxc,Nishida:2016vds,Cordova:2016emh,Schomerus:2016epl,Kravchuk:2016qvl,Gliozzi:2017hni,Castro:2017hpx,Dyer:2017zef,Sleight:2017fpc,Chen:2017yia,Pasterski:2017kqt,Cardoso:2017qmj,Karateev:2017jgd,Kravchuk:2017dzd,Dey:2017fab,Hollands:2017chb,Schomerus:2017eny,Isachenkov:2017qgn,Faller:2017hyt,Rong:2017cow,Chen:2017xdz,Sleight:2018epi,Costa:2018mcg,Kobayashi:2018okw,Bhatta:2018gjb,Lauria:2018klo,Liu:2018jhs,Gromov:2018hut,Rosenhaus:2018zqn,Zhou:2018sfz,Kazakov:2018gcy,Li:2019dix,Goncalves:2019znr,Jepsen:2019svc} using a variety of different methods.

A different approach for the computation of conformal blocks was recently proposed in \cite{Fortin:2019fvx,Fortin:2019dnq}.  It relies on using the operator product expansion (OPE) in the embedding space~\cite{Dirac:1936fq,Mack:1969rr,Weinberg:2010fx,Weinberg:2012mz}.  The framework for embedding space OPE was introduced in \cite{Ferrara:1971vh,Ferrara:1971zy,Ferrara:1972cq,Ferrara:1973eg,Dobrev:1975ru,Mack:1976pa}, with further developments presented in \cite{Fortin:2016lmf,Fortin:2016dlj,Comeau:2019xco}.  This approach can be applied to yield any conformal block in general spacetime dimensions.  In this formalism, operators in arbitrary Lorentz representations are uplifted to the embedding space in a uniform manner using products of spinor representations alone.  Derivatives naturally occur in the OPE, and hence it is of interest to fully determine their action in order to directly obtain the blocks.  These were evaluated explicitly in \cite{Fortin:2019fvx,Fortin:2019dnq} for any expression that may potentially arise in any $M$-point function.  With the action of derivatives already in hand, computing conformal blocks just requires finding the projection operators for irreducible Lorentz representations and then performing appropriate replacements of terms with the corresponding expressions obtained from derivatives in the OPE.  

In this work, we derive several four-point conformal blocks using the approach developed in \cite{Fortin:2019fvx,Fortin:2019dnq}.  We have two main goals here.  One is to illustrate how the formalism performs in practice.  Another is to validate the approach by comparing the results with the existing ones in the literature whenever available.  Some of the ingredients needed here, in particular, the projection operators and three-point tensor structures, were studied in detail in \cite{Fortin:2019xyr,Fortin:2019pep}; we rely on those results in this paper.

An interesting aspect of the present approach is that all conformal blocks computed here can be expressed in terms of the Gegenbauer polynomials onto which particular substitution rules are then applied.  The Gegenbauer polynomials are functions of a variable $X$, and a set of substitution rules transforms $X$ into the final answer.

This paper is organized as follows: Section \ref{SecFour} expresses all four-point correlation functions in terms of the conformal blocks.  The conformal blocks themselves are obtained by contracting two tensor structures, each originating from the OPE, with the so-called ``pre-conformal blocks''.  These pre-conformal blocks depend primarily on the Lorentz quantum numbers of the exchanged quasi-primary operator.  They are computed in two steps using the corresponding hatted projection operators.  In the first step, the projection operators are transformed using the three-point tensorial function.  In the second step, the result is transformed further by a four-point conformal substitution rule yielding the proper conformal quantity.  The resulting pre-conformal blocks are linear combinations of tensorial objects, which involve the generalized Exton $G$-functions of the conformal cross-ratios.  The contractions of the pre-conformal blocks with the two tensor structures can be facilitated with the help of several contiguous relations, leading to the standard conformal blocks.  In this work, all pre-conformal blocks and conformal blocks are computed in the $s$-channel.  Section \ref{SecEx} illustrates how the formalism can be applied to derive pre-conformal blocks and conformal blocks in a series of examples.  The conformal blocks are all written in terms of appropriate conformal substitutions on the Gegenbauer polynomials.  As such, the conformal blocks presented here are the final answers that do not contain any derivatives.  Comparison with the existing literature demonstrates the validity of the approach.  Finally, Section \ref{SecConc} concludes, pointing out the importance of hatted projection operators and tensor structures in the computation of pre-conformal blocks and conformal blocks, respectively.  The reader interested in the general method based on the OPE is referred to \cite{Fortin:2019fvx,Fortin:2019dnq} for an extensive exposition of the formalism.

For certain computations, the answers are applicable for $d\geq3$ only, since in that case extra tensor structures appear which must be taken into account appropriately.  Those cases should be clear from the context.  Moreover, although the formalism works for any spacetime signature, the emphasis here is on Lorentz signature.


\section{Four-Point Correlation Functions}\label{SecFour}

In this section, we compute four-point correlation functions in the embedding space with the help of the OPE, as laid out in \cite{Fortin:2019fvx,Fortin:2019dnq}.  The procedure is analogous to the one used to obtain three-point correlation functions from the OPE \cite{Fortin:2019pep}.  The result combines a group-theoretic part, which depends on the Lorentz irreducible representation of the exchanged quasi-primary operator, and a scalar part, which involves simple powers of the conformal cross-ratios.  The latter is fixed by the conformal dimensions of the exchanged and the external quasi-primary operators.  Afterward, some simple substitution rules are introduced to transform these objects into tensorial functions appearing in four-point correlation functions, namely the conformal blocks.


\subsection{OPE and Four-Point Correlation Functions}

Four-point correlation functions can be computed from the OPE \cite{Fortin:2019fvx,Fortin:2019dnq}
\eqn{
\begin{gathered}
\Op{i}{1}\Op{j}{2}=(\mathcal{T}_{12}^{\boldsymbol{N}_i}\Gamma)(\mathcal{T}_{21}^{\boldsymbol{N}_j}\Gamma)\cdot\sum_k\sum_{a=1}^{N_{ijk}}\frac{\cOPE{a}{i}{j}{k}\tOPE{a}{i}{j}{k}{1}{2}}{\ee{1}{2}{p_{ijk}}}\cdot\D_{12}^{(d,h_{ijk}-n_a/2,n_a)}(\mathcal{T}_{12\boldsymbol{N}_k}\Gamma)*\Op{k}{2},\\
p_{ijk}=\frac{1}{2}(\tau_i+\tau_j-\tau_k),\qquad h_{ijk}=-\frac{1}{2}(\chi_i-\chi_j+\chi_k),\\
\tau_\mathcal{O}=\Delta_\mathcal{O}-S_\mathcal{O},\qquad\chi_\mathcal{O}=\Delta_\mathcal{O}-\xi_\mathcal{O},\qquad\xi_\mathcal{O}=S_\mathcal{O}-\lfloor S_\mathcal{O}\rfloor,
\end{gathered}
}[EqOPE]
in terms of three-point correlation functions as
\eqna{
\Vev{\Op{i}{1}\Op{j}{2}\Op{k}{3}\Op{\ell}{4}}&=(\mathcal{T}_{12}^{\boldsymbol{N}_i}\Gamma)(\mathcal{T}_{21}^{\boldsymbol{N}_j}\Gamma)\cdot\sum_m\sum_{a=1}^{N_{ijm}}(-1)^{2\xi_m}\frac{\cOPE{a}{i}{j}{m}\tOPE{a}{i}{j}{m}{1}{2}}{\ee{1}{2}{p_{ijm}}}\cdot\D_{12}^{(d,h_{ijm}-n_a/2,n_a)}\\
&\phantom{=}\qquad\cdot(\mathcal{T}_{12\boldsymbol{N}_m}\Gamma)*\Vev{\Op{k}{3}\Op{\ell}{4}\Op{m}{2}}.
}[Eq4from3]
Three-point correlation functions can also be obtained from the OPE \eqref{EqOPE}, see \cite{Fortin:2019pep}, as can be the two-point correlation functions \cite{Fortin:2019xyr}.

Upon inserting the result of \cite{Fortin:2019pep} in \eqref{Eq4from3}, the four-point correlation functions assume the form
\begingroup\makeatletter\def\f@size{10}\check@mathfonts\def\maketag@@@#1{\hbox{\m@th\large\normalfont#1}}%
\eqna{
&\Vev{\Op{i}{1}\Op{j}{2}\Op{k}{3}\Op{l}{4}}\\
&\qquad=\frac{(\mathcal{T}_{12}^{\boldsymbol{N}_i}\Gamma)^{\{Aa\}}(\mathcal{T}_{21}^{\boldsymbol{N}_j}\Gamma)^{\{Bb\}}(\mathcal{T}_{34}^{\boldsymbol{N}_k}\Gamma)^{\{Cc\}}(\mathcal{T}_{43}^{\boldsymbol{N}_l}\Gamma)^{\{Dd\}}}{\ee{1}{2}{\frac{1}{2}(\tau_i-\chi_i+\tau_j+\chi_j)}\ee{1}{3}{\frac{1}{2}(\chi_i-\chi_j+\chi_k-\chi_l)}\ee{1}{4}{\frac{1}{2}(\chi_i-\chi_j-\chi_k+\chi_l)}\ee{3}{4}{\frac{1}{2}(-\chi_i+\chi_j+\tau_k+\tau_l)}}\\
&\qquad\phantom{=}\qquad\times\sum_m\sum_{a=1}^{N_{ijm}}\sum_{b=1}^{N_{klm}}(-1)^{2\xi_m}\lambda_{\boldsymbol{N}_m}\cOPE{a}{i}{j}{m}\cCF{b}{k}{l}{m}(\tOPE{a}{i}{j}{m}{1}{2})_{\{aA\}\{bB\}}^{\phantom{\{aA\}\{bB\}}\{Ee\}\{F\}}(\tCF{b}{k}{l}{m}{3}{4})_{\{cC\}\{dD\}\{e'E'\}\{F'\}}\\
&\qquad\phantom{=}\qquad\times\left[\frac{\ee{1}{2}{}\ee{3}{4}{}}{\ee{1}{3}{}\ee{1}{4}{}}\right]^{h_{ijm}}\D_{12\{F\}}^{(d,h_{ijm}-n_a/2,n_a)}\left[\frac{\ee{1}{2}{}\ee{3}{4}{}}{\ee{1}{4}{}\ee{2}{3}{}}\right]^{-h_{klm}}\left[\frac{\ee{1}{2}{}\ee{3}{4}{}}{\ee{1}{3}{}\ee{2}{4}{}}\right]^{-h_{lkm}}\\
&\qquad\phantom{=}\qquad\times\left(\frac{\eta_2\cdot\Gamma\,\hat{\mathcal{P}}_{21}^{\boldsymbol{N}_m}\cdot\hat{\mathcal{P}}_{23}^{\boldsymbol{N}_m}\,\eta_3\cdot\Gamma}{\ee{2}{3}{}}\right)_{\{eE\}}^{\phantom{\{eE\}}\{E''e''\}}(\bar{\bar{J}}_{34;2}^{(d,h_{klm},n_b,\Delta_m,\boldsymbol{N}_m)})_{\{e''E''\}}^{\phantom{\{e''E''\}}\{E'e'\}\{F'\}},
}[Eq4with3]
\endgroup
where the three-point correlation function quantities are
\eqn{\cCF{a}{i}{j}{k}=\sum_l\cOPE{a}{i}{j}{l}\cOPE{}{l}{k}{\1},\qquad\tCF{a}{i}{j}{k}{1}{2}=\tOPE{a}{i}{j}{k^C}{1}{2}[(C_\Gamma^{-1})]^{2\xi_k}(g)^{n_v^k}(g)^{n_a},}[EqCoeff]
and $\lambda_{\boldsymbol{N}_k}$ is a normalization constant orthonormalizing the two-point tensor structures \cite{Fortin:2019xyr}.

Before discussing the conformal substitution rule, it is necessary to explicitly exhibit the three-point tensorial function.


\subsection{Three-Point Tensorial Function}

In \eqref{Eq4with3}, the three-point tensorial quantity $\bar{\bar{J}}_{34;2}^{(d,h,n,\Delta,\boldsymbol{N})}$ is known from the three-point correlation functions \cite{Fortin:2019pep} and is obtained by a simple conformal substitution, namely\footnote{Departing from the notation used in \cite{Fortin:2019pep}, homogeneized quantities for three-point correlation functions are denoted by double bars to avoid confusion with homogeneized quantities for four-point correlation functions, denoted by single bars.}
\eqna{
\bar{\bar{J}}_{34;2}^{(d,h,n,\Delta,\boldsymbol{N})}&=(\bar{\bar{\eta}}_2\cdot\Gamma\,\hat{\mathcal{P}}_{24}^{\boldsymbol{N}}\cdot\hat{\mathcal{P}}_{34}^{\boldsymbol{N}}\,\bar{\bar{\eta}}_4\cdot\Gamma)_{cs_3}\\
&\equiv\left.\bar{\bar{\eta}}_2\cdot\Gamma\,\hat{\mathcal{P}}_{24}^{\boldsymbol{N}}\cdot\hat{\mathcal{P}}_{34}^{\boldsymbol{N}}\,\bar{\bar{\eta}}_4\cdot\Gamma\right|_{\substack{(g)^{s_0}(\bar{\bar{\eta}}_2)^{s_2}(\bar{\bar{\eta}}_3)^{s_3}(\bar{\bar{\eta}}_4)^{s_4}\to(g)^{s_0}(\bar{\bar{\eta}}_2)^{s_2}(\bar{\bar{\eta}}_3)^{s_3}\\\times\bar{I}_{34}^{(d,h-n/2-s_4,n+s_4;\chi+s_2/2-s_3/2+s_4/2)}}},
}[EqJbSub3]
where the three-point tensorial function is \cite{Fortin:2019fvx,Fortin:2019dnq}
\eqn{\bar{I}_{34}^{(d,h,n;p)}=\sum_{\substack{q_0,q_2,q_3,q_4\geq0\\\bar{q}=2q_0+q_2+q_3+q_4=n}}S_{(q_0,q_2,q_3,q_4)}\rho^{(d,h;p)}K^{(d,h;p;q_0,q_3,q_4,q_2)}.}[EqIb3]
The totally symmetric tensor, the prefactor and the $K$-function appearing in \eqref{EqIb3} are
\eqna{
S_{(q_0,q_2,q_3,q_4)}^{A_1\cdots A_{\bar{q}}}&=g^{(A_1A_2}\cdots g^{A_{2q_0-1}A_{2q_0}}\bar{\bar{\eta}}_2^{A_{2q_0+1}}\cdots\bar{\bar{\eta}}_2^{A_{2q_0+q_2}}\\
&\phantom{=}\qquad\times\bar{\bar{\eta}}_3^{A_{2q_0+q_2+1}}\cdots\bar{\bar{\eta}}_3^{A_{2q_0+q_2+q_3}}\bar{\bar{\eta}}_4^{A_{2q_0+q_2+q_3+1}}\cdots\bar{\bar{\eta}}_4^{A_{\bar{q}})},\\
\rho^{(d,h;p)}&=(-2)^h(p)_h(p+1-d/2)_h,\\
K^{(d,h;p;q_0,q_3,q_4,q_2)}&=\frac{(-1)^{\bar{q}-q_0-q_3-q_4}(-2)^{\bar{q}-q_0}\bar{q}!}{q_0!q_2!q_3!q_4!}\frac{(-h-\bar{q})_{\bar{q}-q_0-q_4}(p+h)_{\bar{q}-q_0-q_3}}{(p+1-d/2)_{-q_0-q_3-q_4}},
}[EqK3]
with $\bar{q}=2q_0+q_2+q_3+q_4$.  In the totally symmetric tensor, the homogeneized embedding space coordinates are defined as
\eqn{\bar{\bar{\eta}}_i^A=\frac{\ee{j}{k}{\frac{1}{2}}}{\ee{i}{j}{\frac{1}{2}}\ee{i}{k}{\frac{1}{2}}}\e{i}{A}{},}[Eqetab3]
with $(i,j,k)$ a cyclic permutation of $(2,3,4)$.  Clearly, the three-point tensorial function is totally symmetric and traceless with respect to the embedding space metric.  As such, it satisfies the following contiguous relations \cite{Fortin:2019fvx,Fortin:2019dnq}:
\eqna{
g\cdot\bar{I}_{34}^{(d,h,n;p)}&=0,\\
\bar{\bar{\eta}}_3\cdot\bar{I}_{34}^{(d,h,n;p)}&=\bar{I}_{34}^{(d,h+1,n-1;p)},\\
\bar{\bar{\eta}}_4\cdot\bar{I}_{34}^{(d,h,n;p)}&=\rho^{(d,1;-h-n)}\bar{I}_{34}^{(d,h,n-1;p)},\\
\bar{\bar{\eta}}_2\cdot\bar{I}_{34}^{(d,h,n;p)}&=\bar{I}_{34}^{(d,h+1,n-1;p-1)}.
}[EqCont3]
Since $\bar{\bar{J}}_{34;2}^{(d,h,n,\Delta,\boldsymbol{N})}$ is contracted with the tensor structure $\tCF{b}{k}{\ell}{m}{3}{4}$ in \eqref{Eq4with3} and the latter commutes through the differential operator $\D_{12}^{(d,h-n/2,n)}$, the contiguous relations \eqref{EqCont3} can be very handy in simplifying the product $\bar{\bar{J}}_{34;2}^{(d,h_{klm},n_b,\Delta_m,\boldsymbol{N}_m)}\cdot\tCF{b}{k}{l}{m}{3}{4}$ when computing conformal blocks.  One can also express $\bar{\bar{J}}_{34;2}^{(d,h_{klm},n_b,\Delta_m,\boldsymbol{N}_m)}\cdot\tCF{b}{k}{l}{m}{3}{4}$ in a generic basis of tensor structures by constructing it with the help of the quantities $\A_{34}$, $\epsilon_{34}$, $\Gamma_{34}$ and $\A_{34}\cdot\bar{\bar{\eta}}_4$.

For future convenience, we also define $\widetilde{K}^{(d,h;p;q_0,q_3,q_4,q_2)}=\rho^{(d,h;p)}K^{(d,h;p;q_0,q_3,q_4,q_2)}$, which will appear in the construction of the pre-conformal blocks.


\subsection{Rules for Four-Point Correlation Functions}

The last two lines in \eqref{Eq4with3} are homogeneous of degree zero in all four embedding space coordinates.  Following \cite{Fortin:2019fvx,Fortin:2019dnq}, they can be re-expressed in terms of the homogeneized embedding space coordinates
\eqn{
\begin{gathered}
\bar{\eta}_1^A=\frac{\ee{3}{4}{\frac{1}{2}}}{\ee{1}{3}{\frac{1}{2}}\ee{1}{4}{\frac{1}{2}}}\eta_1^A,\qquad\qquad\bar{\eta}_2^A=\frac{\ee{1}{3}{\frac{1}{2}}\ee{1}{4}{\frac{1}{2}}}{\ee{1}{2}{}\ee{3}{4}{\frac{1}{2}}}\eta_2^A,\\
\bar{\eta}_3^A=\frac{\ee{1}{4}{\frac{1}{2}}}{\ee{3}{4}{\frac{1}{2}}\ee{1}{3}{\frac{1}{2}}}\eta_3^A,\qquad\qquad\bar{\eta}_4^A=\frac{\ee{1}{3}{\frac{1}{2}}}{\ee{3}{4}{\frac{1}{2}}\ee{1}{4}{\frac{1}{2}}}\eta_4^A,
\end{gathered}
}[Eqetab4]
and the conformal cross-ratios
\eqn{x_3=\frac{\ee{1}{2}{}\ee{3}{4}{}}{\ee{1}{4}{}\ee{2}{3}{}}=\frac{u}{v},\qquad\qquad x_4=\frac{\ee{1}{2}{}\ee{3}{4}{}}{\ee{1}{3}{}\ee{2}{4}{}}=u.}[EqCR]
Hence, the last two lines of \eqref{Eq4with3} can be represented by the following function:
\eqna{
\bar{J}_{34;21}^{(d,h_1,n_1,h_2,n_2,\Delta,\boldsymbol{N})}&=\left[\frac{\ee{1}{2}{}\ee{3}{4}{}}{\ee{1}{3}{}\ee{1}{4}{}}\right]^{h_1}\D_{12}^{(d,h_1-n_1/2,n_1)}\left[\frac{\ee{1}{2}{}\ee{3}{4}{}}{\ee{1}{4}{}\ee{2}{3}{}}\right]^{-h_2}\\
&\phantom{=}\qquad\times\left[\frac{\ee{1}{2}{}\ee{3}{4}{}}{\ee{1}{3}{}\ee{2}{4}{}}\right]^{\chi+h_2}\left(\frac{\eta_2\cdot\Gamma\,\hat{\mathcal{P}}_{21}^{\boldsymbol{N}}\cdot\hat{\mathcal{P}}_{23}^{\boldsymbol{N}}\,\eta_3\cdot\Gamma}{\ee{2}{3}{}}\right)\cdot\bar{\bar{J}}_{34;2}^{(d,h_2,n_2,\Delta,\boldsymbol{N})}\\
&=\bar{\D}_{12}^{(d,h_1-n_1/2,n_1)}x_3^{-h_2}x_4^{\chi+h_2}\left(\frac{\eta_2\cdot\Gamma\,\hat{\mathcal{P}}_{21}^{\boldsymbol{N}}\cdot\hat{\mathcal{P}}_{23}^{\boldsymbol{N}}\,\eta_3\cdot\Gamma}{\ee{2}{3}{}}\right)\cdot\bar{\bar{J}}_{34;2}^{(d,h_2,n_2,\Delta,\boldsymbol{N})},
}[EqJb4]
which depends primarily on the exchanged quasi-primary operator, most importantly, on its irreducible representation $\boldsymbol{N}$ under the Lorentz group.  Using the definition of the three-point tensorial function \eqref{EqJbSub3} and the general result of \cite{Fortin:2019fvx,Fortin:2019dnq} for the action of the differential operator, we find that there exists a simple conformal substitution rule for \eqref{EqJb4}, analogous to the one in the three-point case \cite{Fortin:2019pep}.  It can be explicitly and concisely given as
\begingroup\makeatletter\def\f@size{10}\check@mathfonts\def\maketag@@@#1{\hbox{\m@th\large\normalfont#1}}%
\eqna{
\bar{J}_{34;21}^{(d,h_1,n_1,h_2,n_2,\Delta,\boldsymbol{N})}&=2^{2\xi}(\bar{\bar{\eta}}_2\cdot\Gamma\,\hat{\mathcal{P}}_{21}^{\boldsymbol{N}}\cdot\hat{\mathcal{P}}_{23}^{\boldsymbol{N}}\cdot\hat{\mathcal{P}}_{24}^{\boldsymbol{N}}\cdot\hat{\mathcal{P}}_{34}^{\boldsymbol{N}}\,\bar{\bar{\eta}}_4\cdot\Gamma)_{cs_3,cs_4}\\
&\equiv2^{2\xi}\left.(\bar{\bar{\eta}}_2\cdot\Gamma\,\hat{\mathcal{P}}_{21}^{\boldsymbol{N}}\cdot\hat{\mathcal{P}}_{23}^{\boldsymbol{N}}\cdot\hat{\mathcal{P}}_{24}^{\boldsymbol{N}}\cdot\hat{\mathcal{P}}_{34}^{\boldsymbol{N}}\,\bar{\bar{\eta}}_4\cdot\Gamma)_{cs_3}\right|_{(\bar{\eta}_2)^{s_2}x_3^{r_3}x_4^{r_4}\to\bar{I}_{12;34}^{(d,h_1-n_1/2-s_2,n_1+s_2;-h_2+r_3,\chi+h_2+r_4)}},
}[EqJbSub4]
\endgroup
where only $\hat{\mathcal{P}}_{24}^{\boldsymbol{N}}$ and $\hat{\mathcal{P}}_{34}^{\boldsymbol{N}}$ are expressed in terms of the homogeneized three-point embedding space coordinates \eqref{Eqetab3} for the three-point conformal substitution \eqref{EqJbSub3}.  After the three-point conformal substitution has been implemented but before the four-point one is performed, all the embedding space coordinates are re-expressed in terms of the homogeneized four-point embedding space coordinates \eqref{Eqetab4} and the conformal cross-ratios \eqref{EqCR}, with the homogeneized three-point quantities \eqref{Eqetab3} given by
\eqn{\bar{\bar{\eta}}_2=\sqrt{x_3x_4}\bar{\eta}_2,\qquad\bar{\bar{\eta}}_3=\sqrt{\frac{x_3}{x_4}}\bar{\eta}_3,\qquad\bar{\bar{\eta}}_4=\sqrt{\frac{x_4}{x_3}}\bar{\eta}_4.}
The four-point tensorial function $\bar{I}_{12;34}^{(d,h,n;p_3,p_4)}$ appearing in the conformal substitution rule \eqref{Eq4with3} is described in more detail below.

After using the conformal substitution rule \eqref{EqJbSub4}, the four-point correlation functions \eqref{Eq4with3} become
\begingroup\makeatletter\def\f@size{10}\check@mathfonts\def\maketag@@@#1{\hbox{\m@th\large\normalfont#1}}%
\eqna{
&\Vev{\Op{i}{1}\Op{j}{2}\Op{k}{3}\Op{l}{4}}\\
&\qquad=\frac{(\mathcal{T}_{12}^{\boldsymbol{N}_i}\Gamma)^{\{Aa\}}(\mathcal{T}_{21}^{\boldsymbol{N}_j}\Gamma)^{\{Bb\}}(\mathcal{T}_{34}^{\boldsymbol{N}_k}\Gamma)^{\{Cc\}}(\mathcal{T}_{43}^{\boldsymbol{N}_l}\Gamma)^{\{Dd\}}}{\ee{1}{2}{\frac{1}{2}(\tau_i-\chi_i+\tau_j+\chi_j)}\ee{1}{3}{\frac{1}{2}(\chi_i-\chi_j+\chi_k-\chi_l)}\ee{1}{4}{\frac{1}{2}(\chi_i-\chi_j-\chi_k+\chi_l)}\ee{3}{4}{\frac{1}{2}(-\chi_i+\chi_j+\tau_k+\tau_l)}}\\
&\qquad\phantom{=}\qquad\times\sum_m\sum_{a=1}^{N_{ijm}}\sum_{b=1}^{N_{klm}}(-1)^{2\xi_m}\lambda_{\boldsymbol{N}_m}\cOPE{a}{i}{j}{m}\cCF{b}{k}{l}{m}(\tOPE{a}{i}{j}{m}{1}{2})_{\{aA\}\{bB\}}^{\phantom{\{aA\}\{bB\}}\{Ee\}\{F\}}(\tCF{b}{k}{l}{m}{3}{4})_{\{cC\}\{dD\}\{e'E'\}\{F'\}}\\
&\qquad\phantom{=}\qquad\times(\bar{J}_{34;21}^{(d,h_{ijm},n_a,h_{klm},n_b,\Delta_m,\boldsymbol{N}_m)})_{\{F\}\{eE\}}^{\phantom{\{F\}\{eE\}}\{E'e'\}\{F'\}},
}
\endgroup
where the four-point $\bar{J}$-functions can be seen as pre-conformal blocks.  They depend primarily on Lorentz group irreducible representation $\boldsymbol{N}$ of the exchanged quasi-primary operator, as well as on three real numbers related to the conformal dimensions of all quasi-primary operators, two integers associated with the two symmetric-traceless irreducible representations appearing in the two tensor structures, and the spacetime dimension.  The equation above is valid for all four-point correlation functions irrespective of the irreducible representations of the quasi-primary operators.  Moreover, the nontrivial part of the computation corresponds to the contraction of the hatted projection operators.  The conformal substitution rule \eqref{EqJbSub4} leading to the pre-conformal blocks is trivial.

Consequently, once the irreducible representation of the exchanged quasi-primary operator is fixed, the pre-conformal blocks (\textit{i.e.} the $\bar{J}$-functions) are completely determined from the corresponding hatted projection operator.\footnote{Hatted projection operators are discussed in \cite{Fortin:2019xyr}.}  The two tensor structures,\footnote{Tensor structures are discussed in \cite{Fortin:2019pep}.} which dictate the two integers mentioned above, are then needed to contract the remaining dummy indices, which leads to 
\eqna{
&\Vev{\Op{i}{1}\Op{j}{2}\Op{k}{3}\Op{l}{4}}\\
&\qquad=\frac{(\mathcal{T}_{12}^{\boldsymbol{N}_i}\Gamma)^{\{Aa\}}(\mathcal{T}_{21}^{\boldsymbol{N}_j}\Gamma)^{\{Bb\}}(\mathcal{T}_{34}^{\boldsymbol{N}_k}\Gamma)^{\{Cc\}}(\mathcal{T}_{43}^{\boldsymbol{N}_l}\Gamma)^{\{Dd\}}}{\ee{1}{2}{\frac{1}{2}(\tau_i-\chi_i+\tau_j+\chi_j)}\ee{1}{3}{\frac{1}{2}(\chi_i-\chi_j+\chi_k-\chi_l)}\ee{1}{4}{\frac{1}{2}(\chi_i-\chi_j-\chi_k+\chi_l)}\ee{3}{4}{\frac{1}{2}(-\chi_i+\chi_j+\tau_k+\tau_l)}}\\
&\qquad\phantom{=}\qquad\times\sum_m\sum_{a=1}^{N_{ijm}}\sum_{b=1}^{N_{klm}}\cOPE{a}{i}{j}{m}\cCF{b}{k}{l}{m}(\mathscr{G}_{(a,b)}^{ij|m|kl})_{\{aA\}\{bB\}\{cC\}\{dD\}},
}[EqCFSub]
with the conformal blocks
\eqn{\mathscr{G}_{(a,b)}^{ij|m|kl}=(-1)^{2\xi_m}\lambda_{\boldsymbol{N}_m}\tOPE{a}{i}{j}{m}{1}{2}\cdot\bar{J}_{34;21}^{(d,h_{ijm},n_a,h_{klm},n_b,\Delta_m,\boldsymbol{N}_m)}\cdot\tCF{b}{k}{l}{m}{3}{4}.}
As mentioned earlier, the contiguous relations \eqref{EqCont3} and \eqref{EqCont4} can be quite helpful in computing the conformal blocks.  Therefore, it might be more efficient to contract the pre-conformal blocks with the appropriate tensor structures before performing all conformal substitutions, which results in the expression
\eqn{\mathscr{G}_{(a,b)}^{ij|m|kl}=\lambda_{\boldsymbol{N}_m}\tOPE{a}{i}{j}{m}{1}{2}\cdot\left((-x_3)^{2\xi_m}\bar{\eta}_2\cdot\Gamma\,\hat{\mathcal{P}}_{21}^{\boldsymbol{N}_m}\cdot\hat{\mathcal{P}}_{23}^{\boldsymbol{N}_m}\,\bar{\eta}_3\cdot\Gamma(\bar{\bar{\eta}}_2\cdot\Gamma\,\hat{\mathcal{P}}_{24}^{\boldsymbol{N}_m}\cdot\hat{\mathcal{P}}_{34}^{\boldsymbol{N}_m}\,\bar{\bar{\eta}}_4\cdot\Gamma)_{cs_3}\cdot\tCF{b}{k}{l}{m}{3}{4}\right)_{cs_4},}[EqCB]
for the conformal blocks, with the conformal substitution rules \eqref{EqJbSub3} and \eqref{EqJbSub4}, respectively.


\subsection{Four-Point Tensorial Function}

From the results of \cite{Fortin:2019fvx,Fortin:2019dnq}, the four-point tensorial function $\bar{I}_{12;34}^{(d,h,n;p_3,p_4)}$ is given by
\eqn{\bar{I}_{12;34}^{(d,h,n;p_3,p_4)}=\sum_{\substack{q_0,q_1,q_2,q_3,q_4\geq0\\\bar{q}=2q_0+q_1+q_2+q_3+q_4=n}}S_{(\boldsymbol{q})}\rho^{(d,h;p_3+p_4)}x_3^{p_3+p_4+h+q_0+q_2+q_3+q_4}K_{12;34;3}^{(d,h;p_3,p_4;q_0,q_1,q_2,q_3,q_4)}(x_3;y_4),}[EqIb4]
with the totally symmetric tensor $S_{(\boldsymbol{q})}$
\eqn{S_{(\boldsymbol{q})}^{A_1\cdots A_{\bar{q}}}=g^{(A_1A_2}\cdots g^{A_{2q_0-1}A_{2q_0}}\bar{\eta}_1^{A_{2q_0+1}}\cdots\bar{\eta}_1^{A_{2q_0+q_1}}\cdots\bar{\eta}_4^{A_{\bar{q}-q_4+1}}\cdots\bar{\eta}_4^{A_{\bar{q}})},}[EqS4]
$\bar{q}=2q_0+q_1+q_2+q_3+q_4$ and $y_4=1-x_3/x_4$.

The $K$-function is simply a shifted version of the Exton $G$-function,
\eqna{
K_{12;34;3}^{(d,h;\boldsymbol{p};\boldsymbol{q})}(x_3;y_4)&=\frac{(-1)^{q_0+q_3+q_4}(-2)^{\bar{q}-q_0}\bar{q}!}{q_0!q_1!q_2!q_3!q_4!}\frac{(-h-\bar{q})_{\bar{q}-q_0-q_2}(p_3)_{q_3}(p_3+p_4+h)_{\bar{q}-q_0-q_1}}{(p_3+p_4)_{q_3+q_4}(p_3+p_4+1-d/2)_{-q_0-q_1-q_2}}(p_4)_{q_4}\\
&\phantom{=}\qquad\times K_{12;34;3}^{(d+2\bar{q}-2q_0,h+q_0+q_2;p_3+q_3,p_4+q_4)}(x_3;y_4),
}[EqK4]
where
\eqna{
K_{12;34;3}^{(d,h;p_3,p_4)}(x_3;y_4)&=\sum_{n_4,n_{34}\geq0}\frac{(-h)_{n_{34}}(p_3)_{n_{34}}(p_3+p_4+h)_{n_4}}{(p_3+p_4)_{n_4+n_{34}}(p_3+p_4+1-d/2)_{n_{34}}}\frac{(p_4)_{n_4}}{n_{34}!(n_4-n_{34})!}y_4^{n_4}\left(\frac{x_3}{y_4}\right)^{n_{34}}\\
&=G(p_4,p_3+p_4+h,p_3+p_4+1-d/2,p_3+p_4;u/v,1-1/v).
}[EqK0]
Here $G(\alpha,\beta,\gamma,\delta;x,y)$ is the usual Exton $G$-function \cite{Exton_1995}.

As was the case for the three-point tensorial function, the four-point tensorial function satisfies contiguous relations that can greatly simplify computations.  They are given by
\eqna{
g\cdot\bar{I}_{12;34}^{(d,h,n;p)}&=0,\\
\bar{\eta}_1\cdot\bar{I}_{12;34}^{(d,h,n;p_3,p_4)}&=\bar{I}_{12;34}^{(d,h+1,n-1;p_3,p_4)},\\
\bar{\eta}_2\cdot\bar{I}_{12;34}^{(d,h,n;p_3,p_4)}&=\rho^{(d,1;-h-n)}\bar{I}_{12;34}^{(d,h,n-1;p_3,p_4)},\\
\bar{\eta}_3\cdot\bar{I}_{12;34}^{(d,h,n;p_3,p_4)}&=\bar{I}_{12;34}^{(d,h+1,n-1;p_3-1,p_4)},\\
\bar{\eta}_4\cdot\bar{I}_{12;34}^{(d,h,n;p_3,p_4)}&=\bar{I}_{12;34}^{(d,h+1,n-1;p_3,p_4-1)}.
}[EqCont4]
Further details are provided in \cite{Fortin:2019fvx,Fortin:2019dnq}.

Finally, for future convenience, we define
\begingroup\makeatletter\def\f@size{10}\check@mathfonts\def\maketag@@@#1{\hbox{\m@th\large\normalfont#1}}%
\eqna{
G_{(n_1,n_2,n_3,n_4,n_5)A_1\cdots A_n}^{ij|m|kl}&=\rho^{(d,(\ell+n_1+s_2-s_3-s_4)/2;-h_{ijm}-(\ell+n_2)/2)}x_3^{-s_3}x_4^{-s_4}\\
&\phantom{\to}\qquad\times\bar{I}_{12;34}^{(d,h_{ijm}-(s_2-s_3-s_4+n_3)/2,n;-h_{klm}+(r_3-r_4+n_4)/2,\chi_m+h_{klm}-(r_3-r_4+n_5)/2)}{}_{A_1\cdots A_n}.
}[EqG]
\endgroup
This quantity will appear frequently in the conformal substitutions for the conformal blocks, where the meaning of $\ell$, $s_i$ and $r_i$ will become clear.


\section{Examples of Four-Point Correlation Functions}\label{SecEx}

In this section, we explicitly demonstrate how to compute the pre-conformal blocks and conformal blocks using the formalism introduced in \cite{Fortin:2019fvx,Fortin:2019dnq}.  Examples illustrating both computational paths explained in the previous section are given: conformal blocks will be computed either directly from pre-conformal blocks or using \eqref{EqCB}.  The advantage of the pre-conformal blocks is that they can be used in any four-point correlation function where one of the exchanged quasi-primary operators is in the appropriate irreducible representation of the Lorentz group.  Moreover, they only require the knowledge of the corresponding hatted projection operator.


\subsection{Pre-Conformal Blocks}

The pre-conformal blocks \eqref{EqJbSub4} are some of the most fundamental objects leading to the conformal blocks.  They are straightforward to compute once the corresponding hatted projection operators are known.  However, due to the proliferation of indices, they are not always expressible in a manner convenient for exposition.  Because the substitution rules \eqref{EqJbSub3} and \eqref{EqJbSub4} are trivial, the pre-conformal blocks can be easily generated with the help of any convenient symbolic computation program.  Hence, in the following, only some simple pre-conformal blocks are shown explicitly.  Once the pre-conformal block for a specific irreducible representation is known, it can subsequently be used to obtain any conformal block with the corresponding exchanged quasi-primary operator.


\subsubsection{Symmetric-Traceless Exchange}

Since the hatted projection operator for quasi-primary operators in the symmetric-traceless irreducible representation $\ell\boldsymbol{e}_1$ is
\eqn{(\hat{\mathcal{P}}^{\ell\boldsymbol{e}_1})_{\mu_\ell\cdots\mu_1}^{\phantom{\mu_\ell\cdots\mu_1}\mu'_1\cdots\mu'_\ell}=\sum_{i=0}^{\lfloor\ell/2\rfloor}\frac{(-\ell)_{2i}}{2^{2i}i!(-\ell+2-d/2)_i}g_{(\mu_1\mu_2}g^{(\mu'_1\mu'_2}\cdots g_{\mu_{2i-1}\mu_{2i}}g^{\mu'_{2i-1}\mu'_{2i}}g_{\mu_{2i+1}}^{\phantom{\mu_{2i+1}}\mu'_{2i+1}}\cdots g_{\mu_\ell)}^{\phantom{\mu_\ell)}\mu'_\ell)},}[EqPle1]
the $\bar{\bar{J}}$-functions \eqref{EqJbSub3} become
\eqna{
(\bar{\bar{J}}_{34;2}^{(d,h_2,n_2,\Delta,\boldsymbol{0})})^{\{F'\}}&=\bar{I}_{34}^{(d,h_2-n_2/2,n_2;\Delta)\{F'\}},\\
(\bar{\bar{J}}_{34;2}^{(d,h_2,n_2,\Delta,\boldsymbol{e}_1)})_{E''}^{\phantom{E''}E'\{F'\}}&=g_{E''}^{\phantom{E''}E'}\bar{I}_{34}^{(d,h_2-n_2/2,n_2;\Delta)\{F'\}}-\bar{\bar{\eta}}_{3E''}\bar{I}_{34}^{(d,h_2-n_2/2-1,n_2+1;\Delta)E'\{F'\}}\\
&\phantom{=}\qquad-\bar{\bar{\eta}}_2^{E'}\bar{I}_{34}^{(d,h_2-n_2/2-1,n_2+1;\Delta+1)}{}_{E''}^{\phantom{E''}\{F'\}}+\bar{I}_{34}^{(d,h_2-n_2/2-2,n_2+2;\Delta+1)}{}_{E''}^{\phantom{E''}E'\{F'\}},
}[EqJb30ande1]
for the irreducible representations $\boldsymbol{0}$ and $\boldsymbol{e}_1$, respectively.  Then, the pre-conformal blocks \eqref{EqJbSub4} are given by
\eqna{
&(\bar{J}_{34;21}^{(d,h_1,n_1,h_2,n_2,\Delta,\boldsymbol{0})})_{\{F\}}^{\phantom{\{F\}}\{F'\}}=(\bar{I}_{34}^{(d,h_2-n_2/2,n_2;\Delta)\{F'\}})_{cs_4}\\
&\qquad=\sum_{\substack{q_0,q_2,q_3,q_4\geq0\\2q_0+q_2+q_3+q_4=n_2}}g^{(F'_1F'_2}\cdots g^{F'_{2q_0-1}F'_{2q_0}}\bar{\eta}_3^{F'_{2q_0+1}}\cdots\bar{\eta}_3^{F'_{2q_0+q_3}}\bar{\eta}_4^{F'_{2q_0+q_3+1}}\cdots\bar{\eta}_4^{F'_{2q_0+q_3+q_4}}\\
&\qquad\phantom{=}\times\widetilde{K}^{(d,h_2-n_2/2;\Delta;q_0,q_3,q_4,q_2)}(x_3^{(q_2+q_3-q_4)/2}x_4^{(q_2-q_3+q_4)/2}\bar{\eta}_2^{F'_{2q_0+q_3+q_4+1}}\cdots\bar{\eta}_2^{F'_{n_2})})_{cs_4}\\
&\qquad=\widetilde{K}^{(d,h_2-n_2/2;\Delta;0,0,0,n_2)}\bar{I}_{12;34}^{(d,h_1-n_1/2-n_2,n_1+n_2;-h_2+n_2/2,\Delta+h_2+n_2/2)}{}_{\{F\}}^{\phantom{\{F\}}\{F'\}},
}[EqJb40]
for scalar exchange and
\eqna{
(\bar{J}_{34;21}^{(d,h_1,n_1,h_2,n_2,\Delta,\boldsymbol{e}_1)})_{\{F\}E}^{\phantom{\{F\}E}E'\{F'\}}&=(\A_{123E}^{\phantom{123E}E'}\bar{I}_{34}^{(d,h_2-n_2/2,n_2;\Delta)\{F'\}})_{cs_4}\\
&\qquad\phantom{=}-(\sqrt{x_3x_4}\bar{\eta}_2^{E'}\A_{123E}^{\phantom{123E}E''}\bar{I}_{34}^{(d,h_2-n_2/2-1,n_2+1;\Delta+1)}{}_{E''}^{\phantom{E''}\{F'\}})_{cs_4}\\
&\qquad\phantom{=}+(\A_{123E}^{\phantom{123E}E''}\bar{I}_{34}^{(d,h_2-n_2/2-2,n_2+2;\Delta+1)}{}_{E''}^{\phantom{E''}E'\{F'\}})_{cs_4},
}[EqJb4e1]
or, more explicitly,
\newpage
\eqna{
&(\bar{J}_{34;21}^{(d,h_1,n_1,h_2,n_2,\Delta,\boldsymbol{e}_1)})_{\{F\}E}^{\phantom{\{F\}E}E'\{F'\}}\\
&\qquad=\left[\widetilde{K}^{(d,h_2-n_2/2;\Delta;0,0,0,n_2)}+\frac{2\widetilde{K}^{(d,h_2-n_2/2-2;\Delta+1;1,0,0,n_2)}}{(n_2+2)(n_2+1)}\right]\\
&\qquad\phantom{=}\times\left[g_E^{\phantom{E}E'}\bar{I}_{12;34}^{(d,h_1-n_1/2-n_2,n_1+n_2;-h_2+n_2/2,\Delta+h_2+n_2/2)}{}_{\{F\}}^{\phantom{\{F\}}\{F'\}}\right.\\
&\qquad\qquad\phantom{=}\left.-\bar{\eta}_1^{E'}\bar{I}_{12;34}^{(d,h_1-n_1/2-n_2-1,n_1+n_2+1;-h_2+n_2/2,\Delta+h_2+n_2/2)}{}_{\{F\}E}^{\phantom{\{F\}E}\{F'\}}\right]\\
&\qquad\phantom{=}-\left[\widetilde{K}^{(d,h_2-n_2/2-1;\Delta;0,0,0,n_2+1)}+\frac{\widetilde{K}^{(d,h_2-n_2/2-1;\Delta+1;0,1,0,n_2)}}{n_2+1}-\frac{\widetilde{K}^{(d,h_2-n_2/2-2;\Delta+1;0,1,0,n_2+1)}}{n_2+2}\right]\\
&\qquad\phantom{=}\times\left[\bar{\eta}_{3E}\bar{I}_{12;34}^{(d,h_1-n_1/2-n_2-1,n_1+n_2+1;-h_2+n_2/2+1,\Delta+h_2+n_2/2)}{}_{\{F\}}^{\phantom{\{F\}}E'\{F'\}}\right.\\
&\qquad\qquad\phantom{=}\left.-\bar{I}_{12;34}^{(d,h_1-n_1/2-n_2-2,n_1+n_2+2;-h_2+n_2/2+1,\Delta+h_2+n_2/2)}{}_{\{F\}E}^{\phantom{\{F\}E}E'\{F'\}}\right]\\
&\qquad\phantom{=}-\left[\frac{\widetilde{K}^{(d,h_2-n_2/2-1;\Delta+1;0,0,1,n_2)}}{n_2+1}-\frac{\widetilde{K}^{(d,h_2-n_2/2-2;\Delta+1;0,0,1,n_2+1)}}{n_2+2}\right]\\
&\qquad\phantom{=}\times\left[\bar{\eta}_{4E}\bar{I}_{12;34}^{(d,h_1-n_1/2-n_2-1,n_1+n_2+1;-h_2+n_2/2,\Delta+h_2+n_2/2+1)}{}_{\{F\}}^{\phantom{\{F\}}E'\{F'\}}\right.\\
&\qquad\qquad\phantom{=}\left.-\bar{I}_{12;34}^{(d,h_1-n_1/2-n_2-2,n_1+n_2+2;-h_2+n_2/2,\Delta+h_2+n_2/2+1)}{}_{\{F\}E}^{\phantom{\{F\}E}E'\{F'\}}\right]\\
&\qquad\phantom{=}-\left[\frac{2\widetilde{K}^{(d,h_2-n_2/2-1;\Delta;1,0,0,n_2-1)}}{n_2+1}-\frac{2\widetilde{K}^{(d,h_2-n_2/2-2;\Delta+1;1,1,0,n_2-1)}}{(n_2+2)(n_2+1)}\right]\\
&\qquad\phantom{=}\times\left[\bar{\eta}_{3E}g^{E'(F'_1}\bar{I}_{12;34}^{(d,h_1-n_1/2-n_2+1,n_1+n_2-1;-h_2+n_2/2,\Delta+h_2+n_2/2-1)}{}_{\{F\}}^{\phantom{\{F\}}F'_2\cdots F'_{n_2})}\right.\\
&\qquad\qquad\phantom{=}\left.-g^{E'(F'_1}\bar{I}_{12;34}^{(d,h_1-n_1/2-n_2,n_1+n_2;-h_2+n_2/2,\Delta+h_2+n_2/2-1)}{}_{\{F\}E}^{\phantom{\{F\}E}F'_2\cdots F'_{n_2})}\right]\\
&\qquad\phantom{=}+\frac{2\widetilde{K}^{(d,h_2-n_2/2-2;\Delta+1;1,0,1,n_2-1)}}{(n_2+2)(n_2+1)}\\
&\qquad\phantom{=}\times\left[\bar{\eta}_{4E}g^{E'(F'_1}\bar{I}_{12;34}^{(d,h_1-n_1/2-n_2+1,n_1+n_2-1;-h_2+n_2/2-1,\Delta+h_2+n_2/2)}{}_{\{F\}}^{\phantom{\{F\}}F'_2\cdots F'_{n_2})}\right.\\
&\qquad\qquad\phantom{=}\left.-g^{E'(F'_1}\bar{I}_{12;34}^{(d,h_1-n_1/2-n_2,n_1+n_2;-h_2+n_2/2-1,\Delta+h_2+n_2/2)}{}_{\{F\}E}^{\phantom{\{F\}E}F'_2\cdots F'_{n_2})}\right]\\
&\qquad\phantom{=}-\left[\frac{2\widetilde{K}^{(d,h_2-n_2/2-1;\Delta+1;1,0,0,n_2-1)}}{n_2+1}-\frac{2n_2\widetilde{K}^{(d,h_2-n_2/2-2;\Delta+1;1,0,0,n_2)}}{(n_2+2)(n_2+1)}\right]\\
&\qquad\phantom{=}\times\left[g_E^{\phantom{E}(F'_1}\bar{I}_{12;34}^{(d,h_1-n_1/2-n_2,n_1+n_2;-h_2+n_2/2,\Delta+h_2+n_2/2)}{}_{\{F\}}^{\phantom{\{F\}}F'_2\cdots F'_{n_2})E'}\right.\\
&\qquad\qquad\phantom{=}\left.-\bar{\eta}_1^{(F'_1}\bar{I}_{12;34}^{(d,h_1-n_1/2-n_2-1,n_1+n_2+1;-h_2+n_2/2,\Delta+h_2+n_2/2)}{}_{\{F\}E}^{\phantom{\{F\}E}F'_2\cdots F'_{n_2})E'}\right]\\
&\qquad\phantom{=}+\frac{8\widetilde{K}^{(d,h_2-n_2/2-2;\Delta+1;2,0,0,n_2-2)}}{(n_2+2)(n_2+1)}\\
&\qquad\phantom{=}\times\left[g^{E'(F'_1}g_E^{\phantom{E}F'_2}\bar{I}_{12;34}^{(d,h_1-n_1/2-n_2+2,n_1+n_2-2;-h_2+n_2/2-1,\Delta+h_2+n_2/2-1)}{}_{\{F\}}^{\phantom{\{F\}}F'_3\cdots F'_{n_2})}\right.\\
&\qquad\qquad\phantom{=}\left.-g^{E'(F'_1}\bar{\eta}_1^{F'_2}\bar{I}_{12;34}^{(d,h_1-n_1/2-n_2+1,n_1+n_2-1;-h_2+n_2/2-1,\Delta+h_2+n_2/2-1)}{}_{\{F\}E}^{\phantom{\{F\}E}F'_3\cdots F'_{n_2})E'}\right],
}
for vector exchange.  Here, we first used the contiguous relations \eqref{EqCont3} and afterwards performed the substitutions to the four-point homogeneized embedding space coordinates \eqref{Eqetab4}.  Finally, we implemented the conformal substitution \eqref{EqJbSub4} to get the pre-conformal blocks, after taking into account the possible simplifications stemming from contraction with the tensor structure $\tCF{b}{k}{l}{m}{3}{4}$, due to its double-transversality and tracelessness.

The corresponding results for $\ell\boldsymbol{e}_1$ with larger $\ell$ are obtained in a similar manner, although they become quite complicated to display due to the proliferation of indices.  The complexity of the pre-conformal blocks stems from their universality: they generate all the corresponding conformal blocks once they are contracted with the appropriate tensor structures.


\subsubsection{\texorpdfstring{$\ell\boldsymbol{e}_1+\boldsymbol{e}_2$}{e1+e2} Exchange}

For the exchange of quasi-primary operators in the $\ell\boldsymbol{e}_1+\boldsymbol{e}_2$ representation, the projection operator is simply \cite{Rejon-Barrera:2015bpa}
\begingroup\makeatletter\def\f@size{10}\check@mathfonts\def\maketag@@@#1{\hbox{\m@th\large\normalfont#1}}%
\eqna{
&(\hat{\mathcal{P}}^{\ell\boldsymbol{e}_1+\boldsymbol{e}_2})_{\nu_2\nu_1\mu_\ell\cdots\mu_1}^{\phantom{\nu_2\nu_1\mu_\ell\cdots\mu_1}\mu'_1\cdots\mu'_\ell\nu_1'\nu_2'}\\
&\qquad=\sum_{i=0}^{\lfloor\ell/2\rfloor}a_ig_{[\nu_1}^{\phantom{\nu_1]}\nu'_1}g_{\nu_2]}^{\phantom{\nu_2]}\nu'_2}g_{(\mu_1\mu_2}g^{(\mu'_1\mu'_2}\cdots g_{\mu_{2i-1}\mu_{2i}}g^{\mu'_{2i-1}\mu'_{2i}}g_{\mu_{2i+1}}^{\phantom{\mu_{2i+1}}\mu'_{2i+1}}\cdots g_{\mu_\ell)}^{\phantom{\mu_\ell)}\mu'_\ell)}\\
&\qquad\phantom{=}\qquad+\sum_{i=0}^{\lfloor(\ell-1)/2\rfloor}b_ig_{[\nu_1}^{\phantom{\nu_1]}[\nu'_1}g_{\nu_2]}^{\phantom{\nu_2]}(\mu'_1}g_{(\mu_1}^{\phantom{\mu_1]}\nu'_2]}g_{\mu_2\mu_3}g^{\mu'_2\mu'_3}\cdots g_{\mu_{2i}\mu_{2i+1}}g^{\mu'_{2i}\mu'_{2i+1}}g_{\mu_{2i+2}}^{\phantom{\mu_{2i+2}}\mu'_{2i+2}}\cdots g_{\mu_\ell)}^{\phantom{\mu_\ell)}\mu'_\ell)}\\
&\qquad\phantom{=}\qquad+\sum_{i=0}^{\lfloor(\ell-1)/2\rfloor}c_ig_{[\nu_1}^{\phantom{\nu_1]}[\nu'_1}g_{\nu_2](\mu_1}g^{\nu'_2](\mu'_1}g_{\mu_2\mu_3}g^{\mu'_2\mu'_3}\cdots g_{\mu_{2i}\mu_{2i+1}}g^{\mu'_{2i}\mu'_{2i+1}}g_{\mu_{2i+2}}^{\phantom{\mu_{2i+2}}\mu'_{2i+2}}\cdots g_{\mu_\ell)}^{\phantom{\mu_\ell)}\mu'_\ell)}\\
&\qquad\phantom{=}\qquad+\sum_{i=0}^{\lfloor(\ell-2)/2\rfloor}d_ig_{[\nu_1(\mu_1}g^{[\nu'_1(\mu'_1}g_{\nu_2]}^{\phantom{\nu_2]}\mu'_2}g_{\mu_2}^{\phantom{\mu_2}\nu'_2]}g_{\mu_3\mu_4}g^{\mu'_3\mu'_4}\cdots g_{\mu_{2i+1}\mu_{2i+2}}g^{\mu'_{2i+1}\mu'_{2i+2}}g_{\mu_{2i+3}}^{\phantom{\mu_{2i+3}}\mu'_{2i+3}}\cdots g_{\mu_\ell)}^{\phantom{\mu_\ell)}\mu'_\ell)}\\
&\qquad\phantom{=}\qquad+\sum_{i=0}^{\lfloor(\ell-2)/2\rfloor}e_i\left(g_{[\nu_1(\mu_1}g_{\nu_2]}^{\phantom{\nu_2]}[\nu'_1}g_{\mu_2}^{\phantom{\mu_2}\nu'_2]}g^{(\mu'_1\mu'_2}+g^{[\nu'_1(\mu'_1}g_{[\nu_1}^{\phantom{[\nu_1}\nu'_2]}g_{\nu_2]}^{\phantom{\nu_2]}\mu'_2}g_{(\mu_1\mu_2}\right)\\
&\qquad\phantom{=}\qquad\times g_{\mu_3\mu_4}g^{\mu'_3\mu'_4}\cdots g_{\mu_{2i+1}\mu_{2i+2}}g^{\mu'_{2i+1}\mu'_{2i+2}}g_{\mu_{2i+3}}^{\phantom{\mu_{2i+3}}\mu'_{2i+3}}\cdots g_{\mu_\ell)}^{\phantom{\mu_\ell)}\mu'_\ell)},
}[EqPle1pe2]
\endgroup
with
\eqn{
\begin{gathered}
a_i=\frac{2}{\ell+2}\frac{(-\ell)_{2i}}{2^{2i}i!(-\ell+2-(d/2+1))_i},\\
c_i=-\frac{(\ell-2i)[(2i+3)d+2(i+2)\ell-4(i+1)]}{(d+\ell-2)(d+2\ell-2i-2)}a_i,\\
b_i=(\ell-2i)a_i,\qquad d_i=\frac{2(i+1)(d+2\ell)}{d+\ell-2}a_{i+1},\qquad e_i=-2(i+1)a_{i+1}.
\end{gathered}
}
It is straightforward to compute the corresponding pre-conformal blocks from the substitution rules \eqref{EqJbSub3} and \eqref{EqJbSub4}.  However, as the number of free indices is already large for $\ell=0$ (four free indices in total), the final result is cumbersome and not necessarily enlightening by itself.  We therefore do not display it directly here, although we did use it to compare with the conformal blocks for $\ell=0$ and $\ell=1$ obtained later.

We would like to note that, apart from the prefactor $2/(\ell+2)$, the coefficients $a_i$ in \eqref{EqPle1pe2} are identical to those appearing in the hatted projection operator for $\ell\boldsymbol{e}_1$ \eqref{EqPle1} with $d\to d+2$.  This observation, which comes about from the equivalent role played by $\ell$ in all towers of irreducible representations $\boldsymbol{N}+\ell\boldsymbol{e}_1$, will have far-reaching consequences later on.


\subsection{Conformal Blocks and Four-Point Correlation Functions}

On the one hand, the conformal blocks can be obtained directly from the pre-conformal blocks.  On the other, they can be computed in two steps, exploiting the contiguous relations to simplify the contraction with the tensor structures after the first conformal substitution.  In both cases, the final result is the same, although it is more efficient to use the contiguous relations to simplify the conformal blocks.  The convenience of the pre-conformal blocks is that they are fully determined as soon as the irreducible representation of the exchanged quasi-primary operator is known.

Here we have computed the conformal blocks for four four-point correlation functions: symmetric-traceless exchange in scalar-scalar-scalar-scalar, symmetric-traceless exchange in scalar-scalar-scalar-$\boldsymbol{e}_2$, symmetric-traceless exchange and $\ell\boldsymbol{e}_1+\boldsymbol{e}_2$ exchange in scalar-vector-scalar-vector, and symmetric-traceless exchange in scalar-scalar-vector-vector.  In all cases, all the possible exchanged quasi-primary operators are considered, and the conformal blocks in all OPE channels are obtained, allowing the implementation of the conformal bootstrap.  The first, third and fourth four-point correlation functions are chosen for comparison with the literature, while the second set of conformal blocks is a proof-of-concept example, which shows that we are able to compute any conformal block, albeit in a simple example with only one tower of exchanged quasi-primary operators with one tensor structure each.\footnote{The number of conformal blocks increases quite quickly for generic four-point correlation functions.  For example, $\boldsymbol{e}_1+\boldsymbol{e}_2$ exchange in spinor-$(\boldsymbol{e}_1+\boldsymbol{e}_r)$-scalar-$\boldsymbol{e}_2$ already has $24$ different blocks.  Such a large number of conformal blocks is not convenient for the format of a typical article.}

The conformal blocks $\mathscr{G}_{(a,b)}^{ij|m|kl}$ \eqref{EqCB} are naturally obtained in the OPE tensor structure basis.  Indeed, the tensor structures used to compute the conformal blocks are the ones appearing in the OPE.  However, since three-point correlation functions appear directly in \eqref{EqCB}, it is possible to obtain the conformal blocks $\mathscr{G}_{[a,b]}^{ij|m|kl}$ in the three-point function tensor structure basis.  Obviously, the conformal blocks obtained from the OPE tensor structures are linear combinations of those obtained from the three-point function tensor structures.  Therefore, the conformal blocks in the latter basis are obtained from the former ones with the help of (invertible) transformation matrices $R_{ijm}$ and $R_{klm}$ as
\eqn{\mathscr{G}_{[a,b]}^{ij|m|kl}=(R_{ijm})_a^{\phantom{a}a'}(R_{klm})_b^{\phantom{b}b'}\mathscr{G}_{(a',b')}^{ij|m|kl}}[EqRR]
The distinction is irrelevant when there is just a single conformal block (the transformation matrices are simply multiplicative factors), but in cases with more than one block, the difference is important.  We will see later that the best way of representing conformal blocks originates from a mixed basis of tensor structures,
\eqn{\mathscr{G}_{(a,b]}^{ij|m|kl}=(R_{klm})_b^{\phantom{b}b'}\mathscr{G}_{(a,b')}^{ij|m|kl},}
where $\tCF{b}{k}{\ell}{m}{3}{4}$ are natural three-point function tensor structures, while $\tOPE{a}{i}{j}{m}{1}{2}$ are natural OPE tensor structures.  The examples below will clarify this distinction.

To simplify the notation, in the following, conformal blocks will be denoted by $\mathscr{G}_{(a,b)}^{\boldsymbol{N}}$, $\mathscr{G}_{[a,b]}^{\boldsymbol{N}}$ or $\mathscr{G}_{(a,b]}^{\boldsymbol{N}}$ for an exchanged quasi-primary operator in the irreducible representation $\boldsymbol{N}$ with the OPE or three-point function tensor structures $a$ and $b$, irrespective of the four-point correlation function under consideration.


\subsubsection{Symmetric-Traceless Exchange in Scalar-Scalar-Scalar-Scalar}

For our first example, we focus on the classic case of symmetric-traceless exchange in the four-point correlation function of four scalars.  It is straightforward to compute the conformal blocks \eqref{EqCB} from the pre-conformal blocks \eqref{EqJb40} and \eqref{EqJb4e1}.  Here we have only one tensor structure of each type; hence, the indices $a$ and $b$ are superfluous.

For scalar exchange, the normalization constant and tensor structures are simply $\lambda_{\boldsymbol{0}}=\tOPE{1}{i}{j}{m}{1}{2}=\tCF{1}{k}{l}{m}{3}{4}=1$.  These result in
\eqna{
\mathscr{G}_{(1,1)}^{\boldsymbol{0}}&=\rho^{(d,h_{klm};\Delta_m)}\bar{I}_{12;34}^{(d,h_{ijm},0;-h_{klm},\Delta_m+h_{klm})}\\
&=\rho^{(d,h_{ijm};\Delta_m)}\rho^{(d,h_{klm};\Delta_m)}x_3^{\Delta_m+h_{ijm}}K_{12;34;3}^{(d,h_{ijm};-h_{klm},\Delta_m+h_{klm})}(x_3;y_4)\\
&=\rho^{(d,h_{ijm};\Delta_m)}\rho^{(d,h_{klm};\Delta_m)}\left(\frac{u}{v}\right)^{\Delta_m+h_{ijm}}\\
&\phantom{=}\qquad\times G(\Delta_m+h_{klm},\Delta_m+h_{ijm},\Delta_m+1-d/2,\Delta_m;u/v,1-1/v),
}[EqCB0in0000]
while for vector exchange, the normalization constant is $\lambda_{\boldsymbol{e}_1}=1/\sqrt{d}$, and the tensor structures are $(\tOPE{1}{i}{j}{m}{1}{2})^{EF}=\A_{12}^{EF}/\sqrt{d}$ and $(\tCF{1}{k}{l}{m}{3}{4})_{E'F'}=\A_{34E'F'}/\sqrt{d}$, giving
\eqna{
\mathscr{G}_{(1,1)}^{\boldsymbol{e}_1}&=\frac{(-2)^{h_{klm}+1/2}(h_{klm}+1/2)(d-1-\Delta_m)}{d^{3/2}}(\Delta_m+1)_{h_{klm}-1/2}(\Delta_m+1-d/2)_{h_{klm}-1/2}\\
&\phantom{=}\qquad\times\left[\frac{1}{x_4}\bar{I}_{12;34}^{(d,h_{ijm}+1/2,0;-h_{klm}-1/2,\Delta_m+h_{klm}+1/2)}\right.\\
&\phantom{=}\qquad+\frac{(2h_{ijm}+\ell)(2h_{ijm}-1+d)}{2}\bar{I}_{12;34}^{(d,h_{ijm}-1/2,0;-h_{klm}-1/2,\Delta_m+h_{klm}+1/2)}\\
&\phantom{=}\qquad-\frac{1}{x_3}\bar{I}_{12;34}^{(d,h_{ijm}+1/2,0;-h_{klm}+1/2,\Delta_m+h_{klm}-1/2)}\\
&\phantom{=}\qquad\left.+\frac{(2h_{ijm}+\ell)(2h_{ijm}-1+d)}{2}\bar{I}_{12;34}^{(d,h_{ijm}-1/2,0;-h_{klm}+1/2,\Delta_m+h_{klm}-1/2)}\right].
}[EqCBe1in0000]
Up to a different normalization, these results match with the usual ones found in the literature \cite{Dolan:2000ut}.

The other conformal blocks for the $\ell\boldsymbol{e}_1$ irreducible representations can be obtained in the same manner, although it is simpler to rely on the contiguous relations after the first conformal substitution.  Indeed, from the three-point correlation functions \cite{Fortin:2019pep}
\eqna{
\lambda_{\ell\boldsymbol{e}_1}(\bar{\bar{J}}_{34;2}^{(d,h_{klm},\ell,\Delta_m,\ell\boldsymbol{e}_1)}\cdot\tCF{1}{k}{l}{m}{3}{4})_{E''_{\ell}\cdots E''_1}&=\frac{(-2)^{h_{klm}-\ell/2}2^{\ell}\ell!(h_{klm}-\ell/2+1)_{\ell}(d-1-\Delta_m)_{\ell}}{(d+2\ell-2)(d-1)_{\ell-1}}\\
&\phantom{=}\qquad\times(\Delta_m+\ell)_{h_{klm}-\ell/2}(\Delta_m+1-d/2)_{h_{klm}-\ell/2}\bar{\bar{\eta}}_{4E''_\ell}\cdots\bar{\bar{\eta}}_{4E''_1},
}
the tensor structure, see \eqref{EqPle1},
\eqna{
(\tOPE{1}{i}{j}{m}{1}{2})^{E_1\cdots E_\ell F_1\cdots F_\ell}&=\lambda_{\ell\boldsymbol{e}_1}\sum_{i=0}^{\lfloor\ell/2\rfloor}\frac{(-\ell)_{2i}}{2^{2i}i!(-\ell+2-d/2)_i}\A_{12}^{(E_1E_2}\A_{12}^{(F_1F_2}\cdots\A_{12}^{E_{2i-1}E_{2i}}\A_{12}^{F_{2i-1}F_{2i}}\\
&\phantom{=}\qquad\times\A_{12}^{E_{2i+1}F_{2i+1}}\cdots\A_{12}^{E_\ell)F_\ell)},}
and the normalization constant $\lambda_{\ell\boldsymbol{e}_1}=\sqrt{\ell!/[(d+2\ell-2)(d-1)_{\ell-1}]}$, the conformal blocks are given by (with $n_a=\ell$)
\eqna{
\mathscr{G}_{(1,1)}^{\ell\boldsymbol{e}_1}&=\frac{(-2)^{h_{klm}-\ell/2}2^{\ell}\ell!(h_{klm}-\ell/2+1)_{\ell}(d-1-\Delta_m)_{\ell}}{(d+2\ell-2)(d-1)_{\ell-1}}(\Delta_m+\ell)_{h_{klm}-\ell/2}(\Delta_m+1-d/2)_{h_{klm}-\ell/2}\\
&\phantom{=}\qquad\times(\tOPE{1}{i}{j}{m}{1}{2})^{E_1\cdots E_\ell F_1\cdots F_\ell}\left(x_3^{-\ell/2}x_4^{\ell/2}(\hat{\mathcal{P}}_{21}^{\ell\boldsymbol{e}_1}\cdot\hat{\mathcal{P}}_{23}^{\ell\boldsymbol{e}_1})_{\{E\}}^{\phantom{\{E\}}\{E''\}}\bar{\eta}_{4E''_{\ell}}\cdots\bar{\eta}_{4E''_1}\right)_{cs_4}\\
&=\frac{(-2)^{h_{klm}-\ell/2}2^{\ell}\ell!(h_{klm}-\ell/2+1)_{\ell}(d-1-\Delta_m)_{\ell}}{(d+2\ell-2)(d-1)_{\ell-1}}(\Delta_m+\ell)_{h_{klm}-\ell/2}(\Delta_m+1-d/2)_{h_{klm}-\ell/2}\\
&\phantom{=}\qquad\times(\tOPE{1}{i}{j}{m}{1}{2})^{E_1\cdots E_\ell F_1\cdots F_\ell}\sum_{i=0}^{\lfloor\ell/2\rfloor}\frac{(-\ell)_{2i}}{2^{2i}i!(-\ell+2-d/2)_i}\left(x_3^{-\ell/2}x_4^{\ell/2}(\bar{\eta}_4\cdot\A_{23}\cdot\bar{\eta}_4)^i\right.\\
&\phantom{=}\qquad\times\left.\A_{12(E_1E_2}\cdots\A_{12E_{2i-1}E_{2i}}(\A_{123}\cdot\bar{\eta}_4)_{E_{2i+1}}\cdots(\A_{123}\cdot\bar{\eta}_4)_{E_\ell)}\right)_{cs_4}.
}[Eqle1in0000p1]
Since the metrics $g_{E_iE_j}$ and the embedding space coordinates $\bar{\eta}_{1E_i}$ commute with the conformal substitution and vanish once contracted with the tensor structure, only the $i=0$ term survives in \eqref{Eqle1in0000p1}.  The expression then simplifies to
\eqna{
\mathscr{G}_{(1,1)}^{\ell\boldsymbol{e}_1}&=\frac{(-2)^{h_{klm}-\ell/2}2^{\ell}\ell!(h_{klm}-\ell/2+1)_{\ell}(d-1-\Delta_m)_{\ell}}{(d+2\ell-2)(d-1)_{\ell-1}}(\Delta_m+\ell)_{h_{klm}-\ell/2}(\Delta_m+1-d/2)_{h_{klm}-\ell/2}\\
&\phantom{=}\qquad\times(\tOPE{1}{i}{j}{m}{1}{2})^{E_1\cdots E_\ell F_1\cdots F_\ell}\left(x_3^{-\ell/2}x_4^{\ell/2}(\A_{123}\cdot\bar{\eta}_4)_{E_1}\cdots(\A_{123}\cdot\bar{\eta}_4)_{E_\ell}\right)_{cs_4}\\
&=\frac{(-2)^{h_{klm}-\ell/2}2^{\ell}\ell!(h_{klm}-\ell/2+1)_{\ell}(d-1-\Delta_m)_{\ell}}{(d+2\ell-2)(d-1)_{\ell-1}}(\Delta_m+\ell)_{h_{klm}-\ell/2}(\Delta_m+1-d/2)_{h_{klm}-\ell/2}\\
&\phantom{=}\qquad\times\lambda_{\ell\boldsymbol{e}_1}\sum_{i=0}^{\lfloor\ell/2\rfloor}\frac{(-2)^i(-\ell)_{2i}}{2^{2i}i!(-\ell+2-d/2)_i}g^{E_1E_2}\bar{\eta}_1^{F_1}\bar{\eta}_2^{F_2}\cdots g^{E_{2i-1}E_{2i}}\bar{\eta}_1^{F_{2i-1}}\bar{\eta}_2^{F_{2i}}\\
&\phantom{=}\qquad\times\A_{12}^{E_{2i+1}F_{2i+1}}\cdots\A_{12}^{E_\ell F_\ell}\left(x_3^{-\ell/2}x_4^{\ell/2}(\A_{123}\cdot\bar{\eta}_4)_{E_1}\cdots(\A_{123}\cdot\bar{\eta}_4)_{E_\ell}\right)_{cs_4}.
}[Eqle1in0000p2]
In the last equality above, we removed the explicit symmetrizations over the sets of $\{F\}$ and $\{E\}$ due to the symmetry properties of the $\bar{I}$-functions and the product of $\A_{123}\cdot\bar{\eta}_4$, respectively.  We also used the fact that only the metrics $g^{E_iE_j}$ in the trace terms do not vanish when contracted.

Moreover, the contiguous relations \eqref{EqCont4} were used to transform $\A_{12}^{F_iF_j}$ into $-2\bar{\eta}_1^{F_i}\bar{\eta}_2^{F_j}$.  Contracting the embedding space metrics and using simple relations for the product of $\A$-metrics, we obtain
\eqna{
\mathscr{G}_{(1,1)}^{\ell\boldsymbol{e}_1}&=\frac{(-2)^{h_{klm}-\ell/2}2^{\ell}\ell!(h_{klm}-\ell/2+1)_{\ell}(d-1-\Delta_m)_{\ell}}{(d+2\ell-2)(d-1)_{\ell-1}}(\Delta_m+\ell)_{h_{klm}-\ell/2}(\Delta_m+1-d/2)_{h_{klm}-\ell/2}\\
&\phantom{=}\qquad\times\lambda_{\ell\boldsymbol{e}_1}\sum_{i=0}^{\lfloor\ell/2\rfloor}\frac{(-2)^i(-\ell)_{2i}}{2^{2i}i!(-\ell+2-d/2)_i}\bar{\eta}_1^{F_1}\bar{\eta}_2^{F_2}\cdots\bar{\eta}_1^{F_{2i-1}}\bar{\eta}_2^{F_{2i}}\A_{12}^{E_{2i+1}F_{2i+1}}\cdots\A_{12}^{E_\ell F_\ell}\\
&\phantom{=}\qquad\times\left(x_3^{-\ell/2}x_4^{\ell/2}(\bar{\eta}_4\cdot\A_{23}\cdot\bar{\eta}_4)^i(\A_{123}\cdot\bar{\eta}_4)_{E_{2i+1}}\cdots(\A_{123}\cdot\bar{\eta}_4)_{E_\ell}\right)_{cs_4}\\
&=\frac{(-2)^{h_{klm}-\ell/2}2^{\ell}\ell!(h_{klm}-\ell/2+1)_{\ell}(d-1-\Delta_m)_{\ell}}{(d+2\ell-2)(d-1)_{\ell-1}}(\Delta_m+\ell)_{h_{klm}-\ell/2}(\Delta_m+1-d/2)_{h_{klm}-\ell/2}\\
&\phantom{=}\qquad\times\lambda_{\ell\boldsymbol{e}_1}\sum_{i=0}^{\lfloor\ell/2\rfloor}\frac{(-\ell)_{2i}}{i!(-\ell+2-d/2)_i}\bar{\eta}_1^{F_1}\bar{\eta}_2^{F_2}\cdots\bar{\eta}_1^{F_{2i-1}}\bar{\eta}_2^{F_{2i}}\A_{12}^{E_{2i+1}F_{2i+1}}\cdots\A_{12}^{E_\ell F_\ell}\\
&\phantom{=}\qquad\times\left(x_3^{-\ell/2+i}x_4^{-\ell/2+i}[x_4(\bar{\eta}_4-\bar{\eta}_2)-x_3(\bar{\eta}_3-\bar{\eta}_2)]_{E_{2i+1}}\cdots[x_4(\bar{\eta}_4-\bar{\eta}_2)-x_3(\bar{\eta}_3-\bar{\eta}_2)]_{E_\ell}\right)_{cs_4}.
}[Eqle1in0000p3]
From the contiguous relations \eqref{EqCont4}, it is clear that the metrics $g^{E_iF_i}$ lead to vanishing contributions.  Indeed, if the conformal substitution is performed on terms containing $\bar{\eta}_{2E_j}$, they lead to traces which vanish identically.  Moreover, if the conformal substitution is done on terms with $(x_4\bar{\eta}_{4}-x_3\bar{\eta}_{3})_{E_j}$, the two contributions cancel due to the contiguous relations \eqref{EqCont4}.  Thus, in \eqref{Eqle1in0000p3} one can replace $\A_{12}^{E_iF_i}$ by $-\bar{\eta}_1^{E_i}\bar{\eta}_2^{F_i}-\bar{\eta}_2^{E_i}\bar{\eta}_1^{F_i}$.  However, the contractions with $-\bar{\eta}_1^{E_i}\bar{\eta}_2^{F_i}$ vanish identically, leading to
\eqna{
\mathscr{G}_{(1,1)}^{\ell\boldsymbol{e}_1}&=\frac{(-2)^{h_{klm}-\ell/2}2^{\ell}\ell!(h_{klm}-\ell/2+1)_{\ell}(d-1-\Delta_m)_{\ell}}{(d+2\ell-2)(d-1)_{\ell-1}}(\Delta_m+\ell)_{h_{klm}-\ell/2}(\Delta_m+1-d/2)_{h_{klm}-\ell/2}\\
&\phantom{=}\qquad\times\lambda_{\ell\boldsymbol{e}_1}\sum_{i=0}^{\lfloor\ell/2\rfloor}\frac{(-1)^\ell(-\ell)_{2i}}{i!(-\ell+2-d/2)_i}\bar{\eta}_1^{F_1}\bar{\eta}_2^{F_2}\cdots\bar{\eta}_1^{F_{2i-1}}\bar{\eta}_2^{F_{2i}}\bar{\eta}_2^{E_{2i+1}}\bar{\eta}_1^{F_{2i+1}}\cdots\bar{\eta}_2^{E_\ell}\bar{\eta}_1^{F_\ell}\\
&\phantom{=}\qquad\times\left(x_3^{-\ell/2+i}x_4^{-\ell/2+i}[x_4(\bar{\eta}_4-\bar{\eta}_2)-x_3(\bar{\eta}_3-\bar{\eta}_2)]_{E_{2i+1}}\cdots[x_4(\bar{\eta}_4-\bar{\eta}_2)-x_3(\bar{\eta}_3-\bar{\eta}_2)]_{E_\ell}\right)_{cs_4}.
}[Eqle1in0000p4]
At this point, we only need to proceed with the conformal substitution \eqref{EqJbSub4} and the contiguous relations \eqref{EqCont4}.  Moreover, the contractions are straightforward since all the $E$-indices are symmetrized and the $\bar{I}$-functions are totally symmetrized.  Hence, the indices can be forgotten and \eqref{Eqle1in0000p4} can be rewritten efficiently as
\eqna{
\mathscr{G}_{(1,1)}^{\ell\boldsymbol{e}_1}&=\frac{(-2)^{h_{klm}+\ell/2}2^{\ell}\ell!(h_{klm}-\ell/2+1)_{\ell}(d-1-\Delta_m)_{\ell}}{(d+2\ell-2)(d-1)_{\ell-1}}(\Delta_m+\ell)_{h_{klm}-\ell/2}(\Delta_m+1-d/2)_{h_{klm}-\ell/2}\\
&\phantom{=}\qquad\times\lambda_{\ell\boldsymbol{e}_1}\left(\sum_{n=0}^{\lfloor\ell/2\rfloor}\frac{(-\ell)_{2n}}{2^{2n}n!(-\ell+2-d/2)_n}\left[\frac{(\alpha_4-\alpha_2)x_4-(\alpha_3-\alpha_2)x_3}{2}\right]^{\ell-2n}\right)_s\\
&=\omega(h_{klm},\Delta_m,\ell)\left(C_\ell^{(d/2-1)}(X)\right)_s,
}[Eqle1in0000]
where the normalization constant is
\eqna{
\omega(h_{klm},\Delta_m,\ell)&=\frac{(-2)^{h_{klm}+\ell/2}2^{\ell}\ell!(h_{klm}-\ell/2+1)_{\ell}(d-1-\Delta_m)_{\ell}}{(d+2\ell-2)(d-1)_{\ell-1}}\\
&\qquad\times(\Delta_m+\ell)_{h_{klm}-\ell/2}(\Delta_m+1-d/2)_{h_{klm}-\ell/2}\frac{\lambda_{\ell\boldsymbol{e}_1}\ell!}{2^\ell(d/2-1)_\ell},
}
the $C_\ell^{(d/2-1)}(X)$ are the usual Gegenbauer polynomials in terms of the variable
\eqn{X=\frac{(\alpha_4-\alpha_2)x_4-(\alpha_3-\alpha_2)x_3}{2},}[EqX]
and the $s$-substitution is
\eqna{
s:\alpha_2^{s_2}\alpha_3^{s_3}\alpha_4^{s_4}x_3^{r_3}x_4^{r_4}&\to G_{(0,0,0,0,0)}^{ij|m|kl}\\
&=\rho^{(d,(\ell+s_2-s_3-s_4)/2;-h_{ijm}-\ell/2)}x_3^{-s_3}x_4^{-s_4}\\
&\phantom{\to}\qquad\times\bar{I}_{12;34}^{(d,h_{ijm}-(s_2-s_3-s_4)/2,0;-h_{klm}+(r_3-r_4)/2,\Delta_m+h_{klm}-(r_3-r_4)/2)}\\
&=\rho^{(d,(\ell+s_2-s_3-s_4)/2;-h_{ijm}-\ell/2)}\rho^{(d,h_{ijm}-(s_2-s_3-s_4)/2;\Delta_m)}x_3^{\Delta_m+h_{ijm}-(s_2+s_3-s_4)/2}\\
&\phantom{=}\qquad\times x_4^{-s_4}K_{12;34;3}^{(d,h_{ijm}-(s_2-s_3-s_4)/2;-h_{klm}+(r_3-r_4)/2,\Delta_m+h_{klm}-(r_3-r_4)/2)}(x_3;y_4).
}
Here, the $\alpha_i$ are placeholders for the $s$-substitution that enable a very convenient form for the conformal blocks.  Indeed, \eqref{Eqle1in0000} gives all the exchanged conformal blocks once the simple $s$-substitution is performed.  The latter is straightforwardly determined by first contracting the $\bar{\eta}_2^{E_i}$ with the $\bar{\eta}_{3E_i}$ and $\bar{\eta}_{4E_i}$, followed by the usual conformal substitution with the contiguous relations for the $\bar{\eta}_1^{F_i}$ and the remaining $\bar{\eta}_2^{E_i}$ and $\bar{\eta}_2^{F_i}$.  Finally, the explicit dependence on the dummy summation index $n$ is transformed into a dependence on $\ell$ and $s_i$ or $r_i$ so that the final substitution can be pulled outside of the sum.  The presence of $\ell$, $s_i$ and $r_i$ in \eqref{EqG} should now be clear.  The explicit form \eqref{Eqle1in0000} in terms of Gegenbauer polynomials with proper substitutions is natural from the $\ell\boldsymbol{e}_1$ projection operator and it is an interesting feature that generalizes to all conformal blocks.  Moreover, it allows for a very effective way of determining conformal blocks for larger $\ell$.

Although \eqref{Eqle1in0000} is our final result, we can obtain more explicit equations for the conformal blocks that can be compared with the literature.  For example, using the binomial expansion for $X$, the conformal blocks \eqref{Eqle1in0000} can be rewritten as \cite{Rejon-Barrera:2015bpa}
\eqna{
\mathscr{G}_{(1,1)}^{\ell\boldsymbol{e}_1}&=\frac{\omega(h_{klm},\Delta_m,\ell)}{\Gamma(d/2-1)}\sum_{n_1=0}^{\lfloor\ell/2\rfloor}\sum_{n_2=0}^{\ell-2n_1}\sum_{n_3=0}^{\ell-2n_1-n_2}\sum_{n_4=0}^{n_2}\frac{(-1)^{n_1+n_2+n_3+n_4}\Gamma(\ell-n_1+d/2-1)}{n_1!\Gamma(\ell-2n_1+1)}\genfrac{(}{)}{0pt}{0}{\ell-2n_1}{n_2}\\
&\phantom{=}\qquad\times\genfrac{(}{)}{0pt}{0}{\ell-2n_1-n_2}{n_3}\genfrac{(}{)}{0pt}{0}{n_2}{n_4}\rho^{(d,n_1+n_2;-h_{ijm}-\ell/2)}x_3^{-n_3}x_4^{-\ell+2n_1+n_2+n_3}\\
&\phantom{=}\qquad\times\bar{I}_{12;34}^{(d,h_{ijm}+\ell/2-n_1-n_2,0;-h_{klm}-\ell/2+n_1+n_3+n_4,\Delta_m+h_{klm}+\ell/2-n_1-n_3-n_4)}.
}[EqCBScalar]
From the recurrence relation for Gegenbauer polynomials, it is also easy to get the recurrence relation for the conformal blocks \eqref{Eqle1in0000} as \cite{Dolan:2000ut}
\eqna{
\mathscr{G}_{(1,1)}^{\ell\boldsymbol{e}_1}&=\omega(h_{klm},\Delta_m,\ell)\left(\frac{2\ell+d-4}{\ell}XC_{\ell-1}^{(d/2-1)}(X)-\frac{\ell+d-4}{\ell}C_{\ell-2}^{(d/2-1)}(X)\right)_s\\
&=\frac{2\ell+d-4}{2\ell}\left[\frac{\omega(h_{klm},\Delta_m,\ell)}{\omega(h_{klm}+1/2,\Delta_m,\ell-1)}\frac{1}{x_4}\left(\mathscr{G}_{(1,1)}^{(\ell-1)\boldsymbol{e}_1}\right)_{\substack{h_{ijm}\to h_{ijm}+1/2\\h_{klm}\to h_{klm}+1/2}}\right.\\
&\phantom{=}\qquad\left.+\frac{\omega(h_{klm},\Delta_m,\ell)}{\omega(h_{klm}+1/2,\Delta_m,\ell-1)}\frac{(2h_{ijm}+\ell)(2h_{ijm}+\ell-2+d)}{2}\left(\mathscr{G}_{(1,1)}^{(\ell-1)\boldsymbol{e}_1}\right)_{\substack{h_{ijm}\to h_{ijm}-1/2\\h_{klm}\to h_{klm}+1/2}}\right.\\
&\phantom{=}\qquad\left.-\frac{\omega(h_{klm},\Delta_m,\ell)}{\omega(h_{klm}-1/2,\Delta_m,\ell-1)}\frac{1}{x_3}\left(\mathscr{G}_{(1,1)}^{(\ell-1)\boldsymbol{e}_1}\right)_{\substack{h_{ijm}\to h_{ijm}+1/2\\h_{klm}\to h_{klm}-1/2}}\right.\\
&\phantom{=}\qquad\left.+\frac{\omega(h_{klm},\Delta_m,\ell)}{\omega(h_{klm}-1/2,\Delta_m,\ell-1)}\frac{(2h_{ijm}+\ell)(2h_{ijm}+\ell-2+d)}{2}\left(\mathscr{G}_{(1,1)}^{(\ell-1)\boldsymbol{e}_1}\right)_{\substack{h_{ijm}\to h_{ijm}-1/2\\h_{klm}\to h_{klm}-1/2}}\right]\\
&\phantom{=}\qquad+\frac{\ell+d-4}{\ell}\frac{\omega(h_{klm},\Delta_m,\ell)}{\omega(h_{klm},\Delta_m,\ell-2)}\frac{(2h_{ijm}+\ell)(2h_{ijm}+\ell-2+d)}{2}\mathscr{G}_{(1,1)}^{(\ell-2)\boldsymbol{e}_1}.
}[EqRRScalar]
Forgetting about the natural OPE normalization and normalizing as done in the literature, the properly-normalized conformal blocks \eqref{EqCBScalar} and the recurrence relation \eqref{EqRRScalar} agree with \cite{Rejon-Barrera:2015bpa} once the $\bar{I}$-functions are re-expressed in terms of the Exton $G$-function, demonstrating that \eqref{Eqle1in0000} is correct.


\subsubsection{Symmetric-Traceless Exchange in Scalar-Scalar-Scalar-\texorpdfstring{$\boldsymbol{e}_2$}{e2}}

In the previous example, the conformal blocks in the natural OPE basis were computed directly from the pre-conformal blocks for $\ell=0$ and $\ell=1$ and from the general definition for all $\ell$.  Here, we will compute the conformal blocks directly in the mixed basis.

For a symmetric-traceless exchange in the four-point correlation function of three scalars and one $\boldsymbol{e}_2$, there is only a single tensor structure per OPE; thus, there is only one conformal block per exchanged quasi-primary operator.  The tensor structure in the OPE basis is given by
\eqn{(\tOPE{1}{i}{j}{m}{1}{2})^{E_1\cdots E_\ell F_1\cdots F_\ell}=\lambda_{\ell\boldsymbol{e}_1}(g)^\ell\hat{\mathcal{P}}_{12}^{\ell\boldsymbol{e}_1},}
where the indices were suppressed on the right-hand side.  Meanwhile, the natural three-point tensor structure is chosen to be
\eqn{\lambda_{\ell\boldsymbol{e}_1}R_\ell(\bar{\bar{J}}_{34;2}^{(d,h_{klm},\ell,\Delta_m,\ell\boldsymbol{e}_1)}\cdot\tCF{1}{k}{l}{m}{3}{4})_{D_2D_1\{E''\}}=g_{D_1E''_1}\bar{\bar{\eta}}_{2D_2}\bar{\bar{\eta}}_{4E''_2}\cdots\bar{\bar{\eta}}_{4E''_\ell},}
where $R_\ell$ is the appropriate transformation matrix, \textit{i.e.} the multiplicative factor that normalizes the three-point correlation functions as on the right-hand side.

Using \eqref{EqCB} and proceeding as in the previous case, the conformal blocks turn out to be
\eqna{
\mathscr{G}_{(1,1]}^{\ell\boldsymbol{e}_1}&=(\tOPE{a}{i}{j}{m}{1}{2})^{E_1\cdots E_\ell F_1\cdots F_\ell}\left(x_3^{-(\ell-2)/2}x_4^{\ell/2}\bar{\eta}_{2D_2}\A_{123E_1D_1}(\A_{123}\cdot\bar{\eta}_4)_{E_2}\cdots(\A_{123}\cdot\bar{\eta}_4)_{E_\ell}\right)_{cs_4}\\
&=\lambda_{\ell\boldsymbol{e}_1}\sum_{i=0}^{\lfloor\ell/2\rfloor}\frac{(-\ell)_{2i}}{2^{2i}i!(-\ell+2-d/2)_i}\A_{12}^{(E_1E_2}\A_{12}^{F_1F_2}\cdots\A_{12}^{E_{2i-1}E_{2i}}\A_{12}^{F_{2i-1}F_{2i}}\A_{12}^{E_{2i+1}F_{2i+1}}\cdots\A_{12}^{E_\ell)F_\ell}\\
&\phantom{=}\qquad\times\left(x_3^{-(\ell-2)/2}x_4^{\ell/2}\bar{\eta}_{2D_2}\A_{123E_1D_1}(\A_{123}\cdot\bar{\eta}_4)_{E_2}\cdots(\A_{123}\cdot\bar{\eta}_4)_{E_\ell}\right)_{cs_4}.
}[Eqle1in000e2]
However, here it is necessary to separate the $E_1$ index from the symmetrized set of indices $\{E\}$, since only $\{E_2,\ldots,E_\ell\}$ are explicitly symmetrized on the last line of \eqref{Eqle1in000e2}.  Extracting the $E_1$ index leads to two different contributions, which would later give two different Gegenbauer polynomials with appropriate conformal substitutions, if it were not for the antisymmetry properties of $\boldsymbol{e}_2$.  Indeed, one has
\eqna{
\mathscr{G}_{(1,1]}^{\ell\boldsymbol{e}_1}&=\lambda_{\ell\boldsymbol{e}_1}\sum_{i=0}^{\lfloor\ell/2\rfloor}\frac{(-\ell)_{2i}}{2^{2i}i!(-\ell+2-d/2)_i}\\
&\phantom{=}\qquad\times\left[\frac{\ell-2i}{\ell}\A_{12}^{(E_\ell E_2}\A_{12}^{F_1F_2}\cdots\A_{12}^{E_{2i-1}E_{2i}}\A_{12}^{F_{2i-1}F_{2i}}\A_{12}^{E_{2i+1}F_{2i+1}}\cdots\A_{12}^{E_{\ell-1})F_{\ell-1}}\A_{12}^{E_1F_\ell}\right.\\
&\phantom{=}\qquad\left.+\frac{2i}{\ell}\A_{12}^{E_1(E_2}\A_{12}^{F_1F_2}\cdots\A_{12}^{E_{2i-1}E_{2i}}\A_{12}^{F_{2i-1}F_{2i}}\A_{12}^{E_{2i+1}F_{2i+1}}\cdots\A_{12}^{E_\ell)F_\ell}\right]\\
&\phantom{=}\qquad\times\left(x_3^{-(\ell-2)/2}x_4^{\ell/2}\bar{\eta}_{2D_2}\A_{123E_1D_1}(\A_{123}\cdot\bar{\eta}_4)_{E_2}\cdots(\A_{123}\cdot\bar{\eta}_4)_{E_\ell}\right)_{cs_4},
}[Eqle1in000e2p1]
where the remaining symmetrization over the set $\{E_2,\ldots,E_\ell\}$ can now be neglected.  At this point, the computation is completely analogous to the one leading to the conformal blocks for scalar exchange in correlation functions of four scalars, and gives
\begingroup\makeatletter\def\f@size{10}\check@mathfonts\def\maketag@@@#1{\hbox{\m@th\large\normalfont#1}}%
\eqna{
\mathscr{G}_{(1,1]}^{\ell\boldsymbol{e}_1}&=\lambda_{\ell\boldsymbol{e}_1}\sum_{i=0}^{\lfloor\ell/2\rfloor}\frac{(-2)^i(-\ell)_{2i}}{2^{2i}i!(-\ell+2-d/2)_i}\left[\frac{\ell-2i}{\ell}g^{E_\ell E_2}\bar{\eta}_1^{F_1}\bar{\eta}_2^{F_2}\cdots g^{E_{2i-1}E_{2i}}\bar{\eta}_1^{F_{2i-1}}\bar{\eta}_2^{F_{2i}}\A_{12}^{E_{2i+1}F_{2i+1}}\cdots\A_{12}^{E_1F_\ell}\right.\\
&\phantom{=}\qquad\left.+\frac{2i}{\ell}g^{E_1E_2}\bar{\eta}_1^{F_1}\bar{\eta}_2^{F_2}\cdots g^{E_{2i-1}E_{2i}}\bar{\eta}_1^{F_{2i-1}}\bar{\eta}_2^{F_{2i}}\A_{12}^{E_{2i+1}F_{2i+1}}\cdots\A_{12}^{E_\ell F_\ell}\right]\\
&\phantom{=}\qquad\times\left(x_3^{-(\ell-2)/2}x_4^{\ell/2}\bar{\eta}_{2D_2}\A_{123E_1D_1}(\A_{123}\cdot\bar{\eta}_4)_{E_2}\cdots(\A_{123}\cdot\bar{\eta}_4)_{E_\ell}\right)_{cs_4}\\
&=\lambda_{\ell\boldsymbol{e}_1}\sum_{i=0}^{\lfloor\ell/2\rfloor}\frac{(-2)^i(-\ell)_{2i}}{2^{2i}i!(-\ell+2-d/2)_i}\left[(-2)^i\frac{\ell-2i}{\ell}\bar{\eta}_1^{F_1}\bar{\eta}_2^{F_2}\cdots\bar{\eta}_1^{F_{2i-1}}\bar{\eta}_2^{F_{2i}}\A_{12}^{E_{2i+1}F_{2i+1}}\cdots\A_{12}^{E_\ell F_\ell}\right.\\
&\phantom{=}\qquad\times\left(x_3^{-(\ell-2)/2+i}x_4^{\ell/2-i}\bar{\eta}_{2D_2}\A_{123E_\ell D_1}(\A_{123}\cdot\bar{\eta}_4)_{E_{2i+1}}\cdots(\A_{123}\cdot\bar{\eta}_4)_{E_{\ell-1}}\right)_{cs_4}\\
&\phantom{=}\qquad+(-2)^{i-1}\frac{2i}{\ell}\bar{\eta}_1^{F_1}\bar{\eta}_2^{F_2}\cdots\bar{\eta}_1^{F_{2i-1}}\bar{\eta}_2^{F_{2i}}\A_{12}^{E_{2i+1}F_{2i+1}}\cdots\A_{12}^{E_\ell F_\ell}\\
&\phantom{=}\qquad\left.\times\left(x_3^{-(\ell-2)/2+i-1}x_4^{\ell/2-i+1}\bar{\eta}_{2D_2}(\A_{23}\cdot\bar{\eta}_4)_{D_1}(\A_{123}\cdot\bar{\eta}_4)_{E_{2i+1}}\cdots(\A_{123}\cdot\bar{\eta}_4)_{E_\ell}\right)_{cs_4}\right].
}[Eqle1in000e2p2]
\endgroup
Clearly, \eqref{Eqle1in000e2p2} implies two different Gegenbauer polynomials, but the second one has a vanishing coefficient since $(\A_{23}\cdot\bar{\eta}_4)_{D_1}$ can be replaced by $-x_3\bar{\eta}_{2D_1}$ without loss of generality due to its contraction with the half-projector for $\boldsymbol{e}_2$.  The antisymmetry of the same half-projector implies that the second term vanishes, leading to
\eqna{
\mathscr{G}_{(1,1]}^{\ell\boldsymbol{e}_1}&=\lambda_{\ell\boldsymbol{e}_1}\sum_{i=0}^{\lfloor\ell/2\rfloor}\frac{(-1)^\ell(\ell-2i)(-\ell)_{2i}}{i!\ell(-\ell+2-d/2)_i}\bar{\eta}_1^{F_1}\bar{\eta}_2^{F_2}\cdots\bar{\eta}_1^{F_{2i-1}}\bar{\eta}_2^{F_{2i}}\bar{\eta}_2^{E_{2i+1}}\bar{\eta}_1^{F_{2i+1}}\cdots\bar{\eta}_2^{E_\ell}\bar{\eta}_1^{F_\ell}\\
&\phantom{=}\qquad\times\left(x_3^{-(\ell-2)/2+i}x_4^{\ell/2-i}\bar{\eta}_{2D_2}\A_{123E_\ell D_1}(\A_{123}\cdot\bar{\eta}_4)_{E_{2i+1}}\cdots(\A_{123}\cdot\bar{\eta}_4)_{E_{\ell-1}}\right)_{cs_4}\\
&=\lambda_{\ell\boldsymbol{e}_1}\frac{(-1)^\ell(\ell-1)!}{(d/2)_{\ell-1}}\left(C_{\ell-1}^{d/2}(X)\right)_s,
}[Eqle1in000e2p3]
with the conformal substitution
\begingroup\makeatletter\def\f@size{10}\check@mathfonts\def\maketag@@@#1{\hbox{\m@th\large\normalfont#1}}%
\eqna{
&s:\alpha_2^{s_2}\alpha_3^{s_3}\alpha_4^{s_4}x_3^{r_3}x_4^{r_4}\to\bar{\eta}_{2D_1}G_{(-1,0,1,1,-1)D_2}^{ij|m|kl}-\bar{\eta}_{1D_1}G_{(1,0,3,1,-1)D_2}^{ij|m|kl}-x_3^{-1}G_{(-1,0,3,3,-1)D_1D_2}^{ij|m|kl}+G_{(1,0,5,3,-1)D_1D_2}^{ij|m|kl}\\
&\qquad=\rho^{(d,(\ell-1+s_2-s_3-s_4)/2;-h_{ijm}-\ell/2)}x_3^{-s_3}x_4^{-s_4}\\
&\qquad\phantom{=}\qquad\times\bar{\eta}_{2D_1}\bar{I}_{12;34}^{(d,h_{ijm}-(s_2-s_3-s_4+1)/2,1;-h_{klm}+(r_3-r_4+1)/2,\Delta_m+h_{klm}-(r_3-r_4-1)/2)}{}_{D_2}\\
&\qquad\phantom{=}\qquad-\rho^{(d,(\ell+1+s_2-s_3-s_4)/2;-h_{ijm}-\ell/2)}x_3^{-s_3}x_4^{-s_4}\\
&\qquad\phantom{=}\qquad\times\bar{\eta}_{1D_1}\bar{I}_{12;34}^{(d,h_{ijm}-(s_2-s_3-s_4+3)/2,1;-h_{klm}+(r_3-r_4+1)/2,\Delta_m+h_{klm}-(r_3-r_4-1)/2)}{}_{D_2}\\
&\qquad=-\rho^{(d,(\ell-1+s_2-s_3-s_4)/2;-h_{ijm}-\ell/2)}\rho^{(d,h_{ijm}-(s_2-s_3-s_4+1)/2;\Delta_m)}\\
&\qquad\phantom{=}\qquad\times\left\{1+\frac{[\Delta_m+h_{ijm}-(s_2-s_3-s_4+1)/2][-h_{ijm}+(s_2-s_3-s_4-1)/2+1-d/2]}{[\Delta_m+h_{ijm}-(s_2-s_3-s_4+3)/2][\Delta_m+h_{ijm}-(s_2-s_3-s_4+3)/2+1-d/2]}\right\}\\
&\qquad\phantom{=}\qquad\times x_3^{\Delta_m+h_{ijm}-(s_2+s_3-s_4+1)/2}x_4^{-s_4}\bar{\eta}_{1[D_1}\bar{\eta}_{2D_2]}\\
&\qquad\phantom{=}\qquad\times K_{12;34;3}^{(d+2,h_{ijm}-(s_2-s_3-s_4+1)/2;-h_{klm}+(r_3-r_4+1)/2,\Delta_m+h_{klm}-(r_3-r_4-1)/2)}(x_3;y_4),
}
\endgroup
In the first equality of \eqref{Eqle1in000e2p3}, a modified version of the argument based on the contiguous relations presented earlier was used to show that $\A_{12}^{E_\ell F_\ell}$ can nonetheless be replaced by $-\bar{\eta}_2^{E_\ell}\bar{\eta}_1^{F_\ell}$.  Moreover, in the conformal substitution, all terms symmetric under the interchange of $D_1$ and $D_2$ were discarded, and the final result was written explicitly in terms of the $K$-function, which is simply the Exton $G$-function.  As shown in the first line, without this simplification, the conformal substitution would have four different contributions, originating from the four different terms appearing in $\A_{123E_\ell D_1}$.

Finally, it is important to note that the conformal blocks \eqref{Eqle1in000e2p3} exist only for $\ell\geq1$, as predicted by the tensor product decomposition.  Furthermore, as expected from general arguments, the conformal blocks can be expressed with the help of Gegenbauer polynomials written in terms of the variable $X$ \eqref{EqX}, which is a very convenient feature.  Obviously, it is always possible to obtain explicit solutions and recurrence relations for the conformal blocks \eqref{Eqle1in000e2p3}, following \eqref{EqCBScalar} and \eqref{EqRRScalar} respectively, although it is unnecessary.


\subsubsection{Symmetric-Traceless Exchange in Scalar-Vector-Scalar-Vector}

To elaborate on the mixed basis, we now return to the pre-conformal blocks \eqref{EqJb40} and \eqref{EqJb4e1} to compute the conformal blocks for symmetric-traceless exchange in scalar-vector-scalar-vector four-point correlation functions.

For scalar exchange, the normalization constant is $\lambda_{\boldsymbol{0}}=1$ and there is only one tensor structure per OPE, given by $(\tOPE{1}{i}{j}{m}{1}{2})_B^{\phantom{B}F}=\A_{12B}^{\phantom{12B}F}/\sqrt{d}$ and $(\tCF{1}{k}{l}{m}{3}{4})_{DF'}=\A_{34DF'}/\sqrt{d}$ respectively.  From the pre-conformal block \eqref{EqJb40}, we find 
\eqna{
\mathscr{G}_{(1,1)}^{\boldsymbol{0}}&=\frac{1}{d}\widetilde{K}^{(d,h_{klm}-1/2;\Delta_m;0,0,0,1)}\A_{12B}^{\phantom{12B}F}\A_{34DF'}\bar{I}_{12;34}^{(d,h_{ijm}-3/2,2;-h_{klm}+1/2,\Delta_m+h_{klm}+1/2)}{}_F^{\phantom{F}F'}\\
&=\frac{1}{d}\widetilde{K}^{(d,h_{klm}-1/2;\Delta_m;0,0,0,1)}g_B^{\phantom{B}F}g_{DF'}\bar{I}_{12;34}^{(d,h_{ijm}-3/2,2;-h_{klm}+1/2,\Delta_m+h_{klm}+1/2)}{}_F^{\phantom{F}F'}\\
&=\frac{1}{d}\widetilde{K}^{(d,h_{klm}-1/2;\Delta_m;0,0,0,1)}\bar{I}_{12;34}^{(d,h_{ijm}-3/2,2;-h_{klm}+1/2,\Delta_m+h_{klm}+1/2)}{}_{BD}.
}[EqCB0in0e10e1]
In the second equality, the transversality of the half-projectors appearing in the four-point correlation function \eqref{EqCFSub} was used to simplify the tensor structures.

This result can obviously be expanded in terms of the Exton $G$-function as in \eqref{EqIb4}, showing that the conformal block agrees with the one found in the literature \cite{Rejon-Barrera:2015bpa}.  However, since the $\bar{I}$-functions have such nice properties, we do not find it useful to do so.

For vector exchange, there are two tensor structures per OPE, leading to four different conformal blocks.  These are given by
\eqn{
\begin{gathered}
(\tOPE{1}{i}{j}{m}{1}{2})_B^{\phantom{B}EF_1F_2}=\sqrt{\frac{2}{(d-1)(d+2)}}\left[\A_{12B}^{\phantom{12B}(F_1}\A_{12}^{F_2)E}-\frac{1}{d}\A_{12B}^{\phantom{12B}E}\A_{12}^{F_1F_2}\right],\\
(\tOPE{2}{i}{j}{m}{1}{2})_B^{\phantom{B}E}=\frac{1}{\sqrt{d}}\A_{12B}^{\phantom{12B}E},\\
(\tCF{1}{k}{l}{m}{3}{4})_{DE'F'_2F'_1}=\sqrt{\frac{2}{(d-1)(d+2)}}\left[\A_{34D(F'_1}\A_{34F'_2)E'}-\frac{1}{d}\A_{34DE'}\A_{34F'_1F'_2}\right],\\
(\tCF{2}{k}{l}{m}{3}{4})_{DE'}=\frac{1}{\sqrt{d}}\A_{34DE'}.
\end{gathered}
}[EqTSe1OPE]
These tensor structures are the natural OPE tensor structures, \textit{i.e.} they are natural from the point of view of the OPE \eqref{EqOPE}.  However, they are not the natural three-point function tensor structures, since they do not lead to simple three-point correlation functions.  With the normalization constant $\lambda_{\boldsymbol{e}_1}=1/\sqrt{d}$, the latter are computed from
\eqna{
\lambda_{\boldsymbol{e}_1}(R_1)_1^{\phantom{1}b}(\bar{\bar{J}}_{34;2}^{(d,h_{klm},n_b,\Delta_m,\boldsymbol{e}_1)}\cdot\tCF{b}{k}{l}{m}{3}{4})_{DE''}&=\bar{\bar{\eta}}_{2D}\bar{\bar{\eta}}_{4E''},\\
\lambda_{\boldsymbol{e}_1}(R_1)_2^{\phantom{2}b}(\bar{\bar{J}}_{34;2}^{(d,h_{klm},n_b,\Delta_m,\boldsymbol{e}_1)}\cdot\tCF{b}{k}{l}{m}{3}{4})_{DE''}&=g_{DE''},
}[EqTSe1TPF]
where the transformation matrix is
\eqna{
R_1&=-\frac{\sqrt{d(d-1)(d/2+1)}\Delta_m}{(\Delta_m-1)(\Delta_m+1-d)\rho^{(d,h_{klm};\Delta_m)}}\\
&\phantom{=}\qquad\times\left(\begin{array}{cc}\frac{(\Delta_m-1)h_{klm}+\Delta_m(\Delta_m-d/2)}{2(\Delta_m+h_{klm})(h_{klm})_2}&\frac{d^2+2(\Delta_m-1)h_{klm}+2\Delta_m^2-d(2\Delta_m+1)}{\sqrt{d(d-1)(d/2+1)}(\Delta_m+h_{klm})}\\\frac{\Delta_m-d/2}{2(h_{klm})_2}&-\frac{(d-2)(\Delta_m-d)}{\sqrt{d(d-1)(d/2+1)}}\end{array}\right).
}
Clearly, the use of both the natural OPE tensor structures $\tOPE{a}{i}{j}{m}{1}{2}$ \eqref{EqTSe1OPE} and three-point function tensor structures $\tCF{b}{k}{l}{m}{3}{4}$ \eqref{EqTSe1TPF} in \eqref{EqCB} simplifies greatly the computation of conformal blocks.  Indeed, the simplest conformal blocks are obtained in this mixed basis.  With the pre-conformal block \eqref{EqJb4e1}, the conformal blocks are thus
\begingroup\makeatletter\def\f@size{10}\check@mathfonts\def\maketag@@@#1{\hbox{\m@th\large\normalfont#1}}%
\eqna{
\mathscr{G}_{(1,1]}^{\boldsymbol{e}_1}&=\frac{(d-2)(h_{ijm}+1)(2h_{ijm}+d)}{d\sqrt{(d-1)(d/2+1)}}\left[\bar{I}_{12;34}^{(d,h_{ijm}-2,2;-h_{klm}+1,\Delta_m+h_{klm})}{}_{BD}-\bar{I}_{12;34}^{(d,h_{ijm}-2,2;-h_{klm},\Delta_m+h_{klm}+1)}{}_{BD}\right.\\
&\phantom{=}\qquad+\frac{2}{d-2}\left(\bar{\eta}_{3B}\bar{I}_{12;34}^{(d,h_{ijm}-1,1;-h_{klm}+1,\Delta_m+h_{klm})}{}_D-\bar{\eta}_{4B}\bar{I}_{12;34}^{(d,h_{ijm}-1,1;-h_{klm},\Delta_m+h_{klm}+1)}{}_D\right)\\
&\phantom{=}\qquad+\frac{d}{(d-2)(h_{ijm}+1)(2h_{ijm}+d)}\left(\frac{1}{x_3}\bar{I}_{12;34}^{(d,h_{ijm}-1,2;-h_{klm}+1,\Delta_m+h_{klm})}{}_{BD}\right.\\
&\phantom{=}\qquad\left.\left.-\frac{1}{x_4}\bar{I}_{12;34}^{(d,h_{ijm}-1,2;-h_{klm},\Delta_m+h_{klm}+1)}{}_{BD}\right)\right],\\
\mathscr{G}_{(1,2]}^{\boldsymbol{e}_1}&=\frac{(d-2)(h_{ijm}+1)(2h_{ijm}+d)}{d\sqrt{(d-1)(d/2+1)}}\left[\bar{I}_{12;34}^{(d,h_{ijm}-2,2;-h_{klm}+1,\Delta_m+h_{klm})}{}_{BD}-\bar{\eta}_{1D}\bar{I}_{12;34}^{(d,h_{ijm}-1,1;-h_{klm},\Delta_m+h_{klm})}{}_B\right.\\
&\phantom{=}\qquad+\frac{2}{d-2}\left(\bar{\eta}_{3B}\bar{I}_{12;34}^{(d,h_{ijm}-1,1;-h_{klm}+1,\Delta_m+h_{klm})}{}_D-g_{BD}\bar{I}_{12;34}^{(d,h_{ijm},0;-h_{klm},\Delta_m+h_{klm})}\right)\\
&\phantom{=}\qquad+\frac{d}{(d-2)(h_{ijm}+1)(2h_{ijm}+d)}\left(\frac{1}{x_3}\bar{I}_{12;34}^{(d,h_{ijm}-1,2;-h_{klm}+1,\Delta_m+h_{klm})}{}_{BD}\right.\\
&\phantom{=}\qquad\left.\left.-\bar{\eta}_{2D}\bar{I}_{12;34}^{(d,h_{ijm},1;-h_{klm},\Delta_m+h_{klm})}{}_B\right)\right],
}[EqCBe1in0e10e1I]
\endgroup
and
\eqna{
\mathscr{G}_{(2,1]}^{\boldsymbol{e}_1}&=\frac{1}{\sqrt{d}}\left[\bar{I}_{12;34}^{(d,h_{ijm}-2,2;-h_{klm}+1,\Delta_m+h_{klm})}{}_{BD}-\bar{I}_{12;34}^{(d,h_{ijm}-2,2;-h_{klm},\Delta_m+h_{klm}+1)}{}_{BD}\right.\\
&\phantom{=}\qquad\left.-\bar{\eta}_{3B}\bar{I}_{12;34}^{(d,h_{ijm}-1,1;-h_{klm}+1,\Delta_m+h_{klm})}{}_D+\bar{\eta}_{4B}\bar{I}_{12;34}^{(d,h_{ijm}-1,1;-h_{klm},\Delta_m+h_{klm}+1)}{}_D\right],\\
\mathscr{G}_{(2,2]}^{\boldsymbol{e}_1}&=\frac{1}{\sqrt{d}}\left[\bar{I}_{12;34}^{(d,h_{ijm}-2,2;-h_{klm}+1,\Delta_m+h_{klm})}{}_{BD}+g_{BD}\bar{I}_{12;34}^{(d,h_{ijm},0;-h_{klm},\Delta_m+h_{klm})}\right.\\
&\phantom{=}\qquad\left.-\bar{\eta}_{1D}\bar{I}_{12;34}^{(d,h_{ijm}-1,1;-h_{klm},\Delta_m+h_{klm})}{}_B-\bar{\eta}_{3B}\bar{I}_{12;34}^{(d,h_{ijm}-1,1;-h_{klm}+1,\Delta_m+h_{klm})}{}_D\right],
}[EqCBe1in0e10e1II]
once the transformation matrix $R_1$ has been used to rotate to the mixed basis.

The remaining symmetric-traceless exchange can be investigated more straightforwardly from the definition \eqref{EqCB}.  In general, there are two tensor structures per OPE, which are simple generalizations of the above, and are given by
\eqn{
\begin{gathered}
(\tOPE{1}{i}{j}{m}{1}{2})_B^{\phantom{B}E_1\cdots E_\ell F_1\cdots F_{\ell+1}}=\lambda_{(\ell+1)\boldsymbol{e}_1}(g)^\ell\hat{\mathcal{P}}_{12}^{(\ell+1)\boldsymbol{e}_1},\qquad(\tOPE{2}{i}{j}{m}{1}{2})_B^{\phantom{B}E_1\cdots E_\ell F_1\cdots F_{\ell-1}}=\lambda_{\ell\boldsymbol{e}_1}(g)^\ell\hat{\mathcal{P}}_{12}^{\ell\boldsymbol{e}_1}g,\\
(\tCF{1}{k}{l}{m}{3}{4})_{DE'_\ell\cdots E'_1F'_{\ell+1}\cdots F'_1}=\lambda_{(\ell+1)\boldsymbol{e}_1}\hat{\mathcal{P}}_{34}^{(\ell+1)\boldsymbol{e}_1}(g)^{\ell+1},\qquad(\tCF{2}{k}{l}{m}{3}{4})_{DE'_\ell\cdots E'_1F'_{\ell-1}\cdots F'_1}=\lambda_{\ell\boldsymbol{e}_1}\hat{\mathcal{P}}_{34}^{\ell\boldsymbol{e}_1}(g)^\ell,
\end{gathered}
}
where the indices have been suppressed on the right-hand side.  Again, these are the natural OPE tensor structures.  However, as mentioned above, the conformal blocks are easiest to display in the mixed basis.  The relation between the natural three-point function tensor structures and the natural OPE tensor structures is given by
\eqna{
\lambda_{\ell\boldsymbol{e}_1}(R_\ell)_1^{\phantom{1}b}(\bar{\bar{J}}_{34;2}^{(d,h_{klm},n_b,\Delta_m,\ell\boldsymbol{e}_1)}\cdot\tCF{b}{k}{l}{m}{3}{4})_{D\{E''\}}&=\bar{\bar{\eta}}_{2D}\bar{\bar{\eta}}_{4E''_1}\cdots\bar{\bar{\eta}}_{4E''_\ell},\\
\lambda_{\ell\boldsymbol{e}_1}(R_\ell)_2^{\phantom{2}b}(\bar{\bar{J}}_{34;2}^{(d,h_{klm},n_b,\Delta_m,\ell\boldsymbol{e}_1)}\cdot\tCF{b}{k}{l}{m}{3}{4})_{D\{E''\}}&=g_{DE''_1}\bar{\bar{\eta}}_{4E''_2}\cdots\bar{\bar{\eta}}_{4E''_\ell},
}
with the corresponding transformation matrix $R_\ell$.  Although it is not necessary here, the latter can be easily computed from the three-point correlation functions.

Adapting the steps leading to the conformal blocks for scalar-scalar-scalar-scalar four-point correlation functions, while being careful with the explicit symmetrizations appearing in the tensor structures as in the scalar-scalar-scalar-$\boldsymbol{e}_2$ four-point correlation functions, the conformal blocks in the mixed basis are given by
\eqn{\mathscr{G}_{(1,1]}^{\ell\boldsymbol{e}_1}=\lambda_{(\ell+1)\boldsymbol{e}_1}\frac{(-1)^\ell\ell!}{(d/2)_\ell}\left[\left(C_\ell^{d/2}(X)\right)_{s_1^{(1,1]}}-\left(C_{\ell-1}^{d/2}(X)\right)_{s_2^{(1,1]}}\right],}[EqCBle1in0e10e1TS11]
with the conformal substitutions
\eqna{
s_1^{(1,1]}:\alpha_2^{s_2}\alpha_3^{s_3}\alpha_4^{s_4}x_3^{r_3}x_4^{r_4}&\to G_{(0,1,3,1,-1)BD}^{ij|m|kl},\\
s_2^{(1,1]}:\alpha_2^{s_2}\alpha_3^{s_3}\alpha_4^{s_4}x_3^{r_3}x_4^{r_4}&\to\bar{\eta}_{4B}G_{(1,1,2,0,-2)D}^{ij|m|kl}-G_{(1,1,4,0,-2)BD}^{ij|m|kl}-\bar{\eta}_{3B}G_{(1,1,2,2,0)D}^{ij|m|kl}+G_{(1,1,4,2,0)BD}^{ij|m|kl},
}
as well as
\eqna{
\mathscr{G}_{(1,2]}^{\ell\boldsymbol{e}_1}&=\lambda_{(\ell+1)\boldsymbol{e}_1}\frac{(-1)^{\ell+1}(\ell-1)!}{(d/2)_\ell}\left[\left(C_{\ell-1}^{d/2}(X)\right)_{s_1^{(1,2]}}-\frac{d}{2}\left(C_{\ell-1}^{d/2+1}(X)\right)_{s_2^{(1,2]}}\right.\\
&\phantom{=}\qquad\left.+\frac{d}{2}\left(C_{\ell-2}^{d/2+1}(X)\right)_{s_3^{(1,2]}}-\frac{d}{2}\left(C_{\ell-3}^{d/2+1}(X)\right)_{s_4^{(1,2]}}\right],
}[EqCBle1in0e10e1TS12]
with the conformal substitutions
\begingroup\makeatletter\def\f@size{10}\check@mathfonts\def\maketag@@@#1{\hbox{\m@th\large\normalfont#1}}%
\eqna{
s_1^{(1,2]}:\alpha_2^{s_2}\alpha_3^{s_3}\alpha_4^{s_4}x_3^{r_3}x_4^{r_4}&\to g_{BD}G_{(1,1,0,0,0)}^{ij|m|kl}-\bar{\eta}_{1D}G_{(1,1,2,0,0)B}^{ij|m|kl}-\bar{\eta}_{3B}G_{(1,1,2,2,0)D}^{ij|m|kl}+G_{(1,1,4,2,0)BD}^{ij|m|kl},\\
s_2^{(1,2]}:\alpha_2^{s_2}\alpha_3^{s_3}\alpha_4^{s_4}x_3^{r_3}x_4^{r_4}&\to\bar{\eta}_{2D}G_{(-1,1,0,0,0)B}^{ij|m|kl}-\bar{\eta}_{1D}G_{(1,1,2,0,0)B}^{ij|m|kl}-x_3^{-1}G_{(-1,1,2,2,0)BD}^{ij|m|kl}+G_{(1,1,4,2,0)BD}^{ij|m|kl},\\
s_3^{(1,2]}:\alpha_2^{s_2}\alpha_3^{s_3}\alpha_4^{s_4}x_3^{r_3}x_4^{r_4}&\to\bar{\eta}_{2D}\left[\bar{\eta}_{4B}G_{(0,1,-1,-1,-1)}^{ij|m|kl}-G_{(0,1,1,-1,-1)B}^{ij|m|kl}-\bar{\eta}_{3B}G_{(0,1,-1,1,1)}^{ij|m|kl}+G_{(0,1,1,1,1)B}^{ij|m|kl}\right]\\
&\phantom{\to}\qquad-\bar{\eta}_{1D}\left[\bar{\eta}_{4B}G_{(2,1,1,-1,-1)}^{ij|m|kl}-G_{(2,1,3,-1,-1)B}^{ij|m|kl}-\bar{\eta}_{3B}G_{(2,1,1,1,1)}^{ij|m|kl}+G_{(2,1,3,1,1)B}^{ij|m|kl}\right]\\
&\phantom{\to}\qquad-x_3^{-1}\left[\bar{\eta}_{4B}G_{(0,1,1,1,-1)D}^{ij|m|kl}-G_{(0,1,3,1,-1)BD}^{ij|m|kl}-\bar{\eta}_{3B}G_{(0,1,1,3,1)D}^{ij|m|kl}+G_{(0,1,3,3,1)BD}^{ij|m|kl}\right]\\
&\phantom{\to}\qquad+\bar{\eta}_{4B}G_{(2,1,3,1,-1)D}^{ij|m|kl}-G_{(2,1,5,1,-1)BD}^{ij|m|kl}-\bar{\eta}_{3B}G_{(2,1,3,3,1)D}^{ij|m|kl}+G_{(2,1,5,3,1)BD}^{ij|m|kl}\\
&\phantom{\to}\qquad+G_{(0,1,3,3,1)BD}^{ij|m|kl},\\
s_4^{(1,2]}:\alpha_2^{s_2}\alpha_3^{s_3}\alpha_4^{s_4}x_3^{r_3}x_4^{r_4}&\to\bar{\eta}_{4B}G_{(1,1,2,0,-2)D}^{ij|m|kl}-G_{(1,1,4,0,-2)BD}^{ij|m|kl}-\bar{\eta}_{3B}G_{(1,1,2,2,0)D}^{ij|m|kl}+G_{(1,1,4,2,0)BD}^{ij|m|kl},
}
\endgroup
and
\eqn{\mathscr{G}_{(2,1]}^{\ell\boldsymbol{e}_1}=\lambda_{\ell\boldsymbol{e}_1}\frac{(-1)^{\ell-1}(\ell-1)!}{(d/2)_{\ell-1}}\left[\left(C_{\ell-1}^{d/2}(X)\right)_{s_1^{(2,1]}}-\left(C_{\ell-2}^{d/2}(X)\right)_{s_2^{(2,1]}}\right],}[EqCBle1in0e10e1TS21]
with the conformal substitutions
\eqna{
s_1^{(2,1]}:\alpha_2^{s_2}\alpha_3^{s_3}\alpha_4^{s_4}x_3^{r_3}x_4^{r_4}&\to\bar{\eta}_{4B}G_{(-1,-1,2,0,-2)D}^{ij|m|kl}-G_{(-1,-1,4,0,-2)BD}^{ij|m|kl}\\
&\phantom{\to}\qquad-\bar{\eta}_{3B}G_{(-1,-1,2,2,0)D}^{ij|m|kl}+G_{(-1,-1,4,2,0)BD}^{ij|m|kl},\\
s_2^{(2,1]}:\alpha_2^{s_2}\alpha_3^{s_3}\alpha_4^{s_4}x_3^{r_3}x_4^{r_4}&\to G_{(-2,-1,3,1,-1)BD}^{ij|m|kl},
}
and finally
\eqna{
\mathscr{G}_{(2,2]}^{\ell\boldsymbol{e}_1}&=\lambda_{\ell\boldsymbol{e}_1}\frac{(-1)^{\ell-1}(\ell-1)!}{\ell(d/2+1)_{\ell-1}}\left[\left(C_{\ell-1}^{d/2}(X)\right)_{s_1^{(2,2]}}+\frac{d}{2}\left(C_{\ell-2}^{d/2+1}(X)\right)_{s_2^{(2,2]}}\right.\\
&\phantom{=}\qquad\left.-\frac{d}{2}\left(C_{\ell-3}^{d/2+1}(X)\right)_{s_3^{(2,2]}}-\left(C_{\ell-2}^{d/2}(X)-\frac{d}{2}C_{\ell-4}^{d/2+1}(X)\right)_{s_4^{(2,2]}}\right],
}[EqCBle1in0e10e1TS22]
with the conformal substitutions
\begingroup\makeatletter\def\f@size{10}\check@mathfonts\def\maketag@@@#1{\hbox{\m@th\large\normalfont#1}}%
\eqna{
s_1^{(2,2]}:\alpha_2^{s_2}\alpha_3^{s_3}\alpha_4^{s_4}x_3^{r_3}x_4^{r_4}&\to g_{BD}G_{(-1,-1,0,0,0)}^{ij|m|kl}-\bar{\eta}_{1D}G_{(-1,-1,2,0,0)B}^{ij|m|kl}-\bar{\eta}_{3B}G_{(-1,-1,2,2,0)D}^{ij|m|kl}+G_{(-1,-1,4,2,0)BD}^{ij|m|kl},\\
s_2^{(2,2]}:\alpha_2^{s_2}\alpha_3^{s_3}\alpha_4^{s_4}x_3^{r_3}x_4^{r_4}&\to\bar{\eta}_{2D}\left[\bar{\eta}_{4B}G_{(-2,-1,-1,-1,-1)}^{ij|m|kl}-G_{(-2,-1,1,-1,-1)B}^{ij|m|kl}-\bar{\eta}_{3B}G_{(-2,-1,-1,1,1)}^{ij|m|kl}+G_{(-2,-1,1,1,1)B}^{ij|m|kl}\right]\\
&\phantom{\to}\qquad-\bar{\eta}_{1D}\left[\bar{\eta}_{4B}G_{(0,-1,1,-1,-1)}^{ij|m|kl}-G_{(0,-1,3,-1,-1)B}^{ij|m|kl}-\bar{\eta}_{3B}G_{(0,-1,1,1,1)}^{ij|m|kl}+G_{(0,-1,3,1,1)B}^{ij|m|kl}\right]\\
&\phantom{\to}\qquad-x_3^{-1}\left[\bar{\eta}_{4B}G_{(-2,-1,1,1,-1)D}^{ij|m|kl}-G_{(-2,-1,3,1,-1)BD}^{ij|m|kl}-\bar{\eta}_{3B}G_{(-2,-1,1,3,1)D}^{ij|m|kl}+G_{(-2,-1,3,3,1)BD}^{ij|m|kl}\right]\\
&\phantom{\to}\qquad+\bar{\eta}_{4B}G_{(0,-1,3,1,-1)D}^{ij|m|kl}-G_{(0,-1,5,1,-1)BD}^{ij|m|kl}-\bar{\eta}_{3B}G_{(0,-1,3,3,1)D}^{ij|m|kl}+G_{(0,-1,5,3,1)BD}^{ij|m|kl},\\
s_3^{(2,2]}:\alpha_2^{s_2}\alpha_3^{s_3}\alpha_4^{s_4}x_3^{r_3}x_4^{r_4}&\to\bar{\eta}_{2D}G_{(-3,-1,0,0,0)B}^{ij|m|kl}-\bar{\eta}_{1D}G_{(-1,-1,2,0,0)B}^{ij|m|kl}-x_3^{-1}G_{(-3,-1,2,2,0)BD}^{ij|m|kl}+\textcolor[rgb]{1,0,0}{G_{(-1,-1,4,2,0)BD}^{ij|m|kl}}\\
&\phantom{\to}\qquad-\left[\bar{\eta}_{4B}G_{(-1,-1,2,0,-2)D}^{ij|m|kl}-G_{(-1,-1,4,0,-2)BD}^{ij|m|kl}-\bar{\eta}_{3B}G_{(-1,-1,2,2,0)D}^{ij|m|kl}+\textcolor[rgb]{1,0,0}{G_{(-1,-1,4,2,0)BD}^{ij|m|kl}}\right],\\
s_4^{(2,2]}:\alpha_2^{s_2}\alpha_3^{s_3}\alpha_4^{s_4}x_3^{r_3}x_4^{r_4}&\to G_{(-2,-1,3,1,-1)BD}^{ij|m|kl}.
}
\endgroup
Clearly, only \eqref{EqCBle1in0e10e1TS11} exists for $\ell=0$ and matches with \eqref{EqCB0in0e10e1} once the proper rescaling necessary to convert from the purely OPE basis of the latter to the mixed basis of the former is done.  Moreover, for $\ell=1$, all the conformal blocks \eqref{EqCBle1in0e10e1TS11}, \eqref{EqCBle1in0e10e1TS12}, \eqref{EqCBle1in0e10e1TS21} and \eqref{EqCBle1in0e10e1TS22} match the conformal blocks \eqref{EqCBe1in0e10e1I} and \eqref{EqCBe1in0e10e1II} obtained from the pre-conformal blocks rotated to the mixed basis.  Finally, as for all previous four-point correlation functions, the conformal blocks are easily displayed as Gegenbauer polynomials in terms of the variable $X$ \eqref{EqX}.  They can be expanded explicitly as in \eqref{EqCBScalar}, and recurrence relations can be found as in \eqref{EqRRScalar}.

Before proceeding, it is interesting to note the similarities between the conformal blocks \eqref{EqCBle1in0e10e1TS11} and \eqref{EqCBle1in0e10e1TS21} and their respective conformal substitutions.  As can be seen above, other similarities occur, mostly due to their common origin, mainly the OPE.  Moreover, in $s_3^{(2,2]}$ the two terms in red cancel each other.  They were kept to exhibit the similarities.


\subsubsection{\texorpdfstring{$\ell\boldsymbol{e}_1+\boldsymbol{e}_2$}{le1+e2} Exchange in Scalar-Vector-Scalar-Vector}

In the case of $\ell\boldsymbol{e}_1+\boldsymbol{e}_2$ exchange in scalar-vector-scalar-vector four-point correlation functions, there is only one tensor structure per OPE.  As before, the easiest way to obtain the conformal blocks is to work in the mixed basis.  The OPE and three-point tensor structures are simply
\eqn{
\begin{gathered}
(\tOPE{1}{i}{j}{m}{1}{2})_B^{\phantom{B}E_1\cdots E_{\ell+2}F_1\cdots F_{\ell+1}}=\lambda_{\ell\boldsymbol{e}_1+\boldsymbol{e}_2}((g)^{\ell+2}\hat{\mathcal{P}}_{12}^{\ell\boldsymbol{e}_1+\boldsymbol{e}_2}g)^{E_1\cdots E_{\ell+2}F_{\ell+1}\cdots F_1}{}_B,\\
\lambda_{\ell\boldsymbol{e}_1+\boldsymbol{e}_2}R_\ell(\bar{\bar{J}}_{34;2}^{(d,h_{klm},\ell+1,\Delta_m,\ell\boldsymbol{e}_1+\boldsymbol{e}_2)}\cdot\tCF{1}{k}{l}{m}{3}{4})_{D\{E''\}}=g_{DE''_1}\bar{\bar{\eta}}_{4E''_2}\cdots\bar{\bar{\eta}}_{4E''_{\ell+2}},
\end{gathered}
}
where on the right-hand side the indices $B$ and $F_1$ are matched to $E_1$ and $E_2$, respectively, and $R_\ell$ is the transformation matrix which is just a multiplicative factor introduced for proper normalization of the three-point correlation functions.  It is understood that $E_1$ and $E_2$ (respectively $B$ and $F_1$) are the $\boldsymbol{e}_2$ indices of the $\ell\boldsymbol{e}_1+\boldsymbol{e}_2$, hence they are antisymmetrized as in \eqref{EqPle1pe2}.

Following the arguments presented above, it is easy to obtain
\begingroup\makeatletter\def\f@size{10}\check@mathfonts\def\maketag@@@#1{\hbox{\m@th\large\normalfont#1}}%
\eqna{
\mathscr{G}_{(1,1]}^{\ell\boldsymbol{e}_1+\boldsymbol{e}_2}&=\lambda_{\ell\boldsymbol{e}_1+\boldsymbol{e}_2}\frac{(-1)^{\ell+1}2\ell!}{(\ell+2)(d/2)_\ell}\\
&\phantom{=}\qquad\times\left[\left(XC_\ell^{d/2}(X)-\frac{2\ell-2+3d/2}{\ell-2+d}C_{\ell-1}^{d/2}(X)+\frac{d}{2}X^2C_{\ell-1}^{d/2+1}(X)-dXC_{\ell-2}^{d/2+1}(X)+\frac{d}{2}C_{\ell-3}^{d/2+1}(X)\right)_{s_1}\right.\\
&\phantom{=}\qquad-\frac{1}{2}\left(C_\ell^{d/2}(X)+\frac{d}{2}XC_{\ell-1}^{d/2+1}(X)-\frac{d(d/2-2)}{\ell-2+d}C_{\ell-2}^{d/2+1}(X)\right)_{s_2}\\
&\phantom{=}\qquad+\frac{1}{2}\left(\frac{2\ell-2+3d/2}{\ell-2+d}C_{\ell-1}^{d/2}(X)+\frac{d(\ell+d/2)}{\ell-2+d}XC_{\ell-2}^{d/2+1}(X)-\frac{d}{2}C_{\ell-3}^{d/2+1}(X)\right)_{s_3}\\
&\phantom{=}\left.\qquad-\frac{1}{2}\left(\frac{2\ell-2+3d/2}{\ell-2+d}XC_{\ell-1}^{d/2}(X)+\frac{d(\ell+d/2)}{\ell-2+d}X^2C_{\ell-2}^{d/2+1}(X)-\frac{d}{2}XC_{\ell-3}^{d/2+1}(X)\right)_{s_4}\right],
}[EqCBle1pe2in0e10e1TS11]
\endgroup
with the conformal substitutions
\begingroup\makeatletter\def\f@size{10}\check@mathfonts\def\maketag@@@#1{\hbox{\m@th\large\normalfont#1}}%
\eqna{
s_1:\alpha_2^{s_2}\alpha_3^{s_3}\alpha_4^{s_4}x_3^{r_3}x_4^{r_4}&\to g_{BD}G_{(1,1,0,0,0)}^{ij|m|kl}-\bar{\eta}_{1D}G_{(1,1,2,0,0)B}^{ij|m|kl}-\bar{\eta}_{3B}G_{(1,1,2,2,0)D}^{ij|m|kl}+G_{(1,1,4,2,0)BD}^{ij|m|kl}\\
s_2:\alpha_2^{s_2}\alpha_3^{s_3}\alpha_4^{s_4}x_3^{r_3}x_4^{r_4}&\to\bar{\eta}_{2D}\left[\bar{\eta}_{4B}G_{(0,1,-1,-1,-1)}^{ij|m|kl}-G_{(0,1,1,-1,-1)B}^{ij|m|kl}-\bar{\eta}_{3B}G_{(0,1,-1,1,1)}^{ij|m|kl}+G_{(0,1,1,1,1)B}^{ij|m|kl}\right]\\
&\phantom{\to}\qquad-\bar{\eta}_{1D}\left[\bar{\eta}_{4B}G_{(2,1,1,-1,-1)}^{ij|m|kl}-G_{(2,1,3,-1,-1)B}^{ij|m|kl}-\bar{\eta}_{3B}G_{(2,1,1,1,1)}^{ij|m|kl}+G_{(2,1,3,1,1)B}^{ij|m|kl}\right]\\
&\phantom{\to}\qquad-x_3^{-1}\left[\bar{\eta}_{4B}G_{(0,1,1,1,-1)D}^{ij|m|kl}-G_{(0,1,3,1,-1)BD}^{ij|m|kl}-\bar{\eta}_{3B}G_{(0,1,1,3,1)D}^{ij|m|kl}+G_{(0,1,3,3,1)BD}^{ij|m|kl}\right]\\
&\phantom{\to}\qquad+\bar{\eta}_{4B}G_{(2,1,3,1,-1)D}^{ij|m|kl}-G_{(2,1,5,1,-1)BD}^{ij|m|kl}-\bar{\eta}_{3B}G_{(2,1,3,3,1)D}^{ij|m|kl}+G_{(2,1,5,3,1)BD}^{ij|m|kl},\\
s_3:\alpha_2^{s_2}\alpha_3^{s_3}\alpha_4^{s_4}x_3^{r_3}x_4^{r_4}&\to\bar{\eta}_{4B}G_{(1,1,2,0,-2)D}^{ij|m|kl}-G_{(1,1,4,0,-2)BD}^{ij|m|kl}-\bar{\eta}_{3B}G_{(1,1,2,2,0)D}^{ij|m|kl}+G_{(1,1,4,2,0)BD}^{ij|m|kl}\\
&\phantom{\to}\qquad+\bar{\eta}_{2D}G_{(-1,1,0,0,0)B}^{ij|m|kl}-\bar{\eta}_{1D}G_{(1,1,2,0,0)B}^{ij|m|kl}-x_3^{-1}G_{(-1,1,2,2,0)BD}^{ij|m|kl}+G_{(1,1,4,2,0)BD}^{ij|m|kl},\\
s_4:\alpha_2^{s_2}\alpha_3^{s_3}\alpha_4^{s_4}x_3^{r_3}x_4^{r_4}&\to G_{(0,1,3,1,-1)BD}^{ij|m|kl}.
}
\endgroup
The observation that the coefficients in the projection operator for $\ell\boldsymbol{e}_1+\boldsymbol{e}_2$ are related to those in the projection operator for $\ell\boldsymbol{e}_1$ and the fact that the latter lead to Gegenbauer polynomials explain why all conformal blocks can be displayed as appropriate conformal substitutions of Gegenbauer polynomials in the variable $X$.


\subsubsection{Symmetric-Traceless Exchange in Scalar-Scalar-Vector-Vector}

For completeness, in our final example, we determine the conformal blocks for scalar-scalar-vector-vector four-point correlation functions, which would then empower us to fully implement the bootstrap for correlation functions of two scalars and two vectors.

In the mixed basis, the necessary inputs are the tensor structures for symmetric-traceless exchange, which are
\eqn{
\begin{gathered}
(\tOPE{1}{i}{j}{m}{1}{2})^{E_1\cdots E_\ell F_1\cdots F_\ell}=\lambda_{\ell\boldsymbol{e}_1}(g)^\ell\hat{\mathcal{P}}_{12}^{\ell\boldsymbol{e}_1},\\
\lambda_{\ell\boldsymbol{e}_1}(R_\ell)_1^{\phantom{b}b}(\bar{\bar{J}}_{34;2}^{(d,h_{klm},n_b,\Delta_m,\ell\boldsymbol{e}_1)}\cdot\tCF{b}{k}{l}{m}{3}{4})_{CD\{E''\}}=\bar{\bar{\eta}}_{2C}\bar{\bar{\eta}}_{2D}\bar{\bar{\eta}}_{4E''_1}\cdots\bar{\bar{\eta}}_{4E''_\ell},\\
\lambda_{\ell\boldsymbol{e}_1}(R_\ell)_2^{\phantom{b}b}(\bar{\bar{J}}_{34;2}^{(d,h_{klm},n_b,\Delta_m,\ell\boldsymbol{e}_1)}\cdot\tCF{b}{k}{l}{m}{3}{4})_{CD\{E''\}}=g_{CD}\bar{\bar{\eta}}_{4E''_1}\cdots\bar{\bar{\eta}}_{4E''_\ell},\\
\lambda_{\ell\boldsymbol{e}_1}(R_\ell)_3^{\phantom{b}b}(\bar{\bar{J}}_{34;2}^{(d,h_{klm},n_b,\Delta_m,\ell\boldsymbol{e}_1)}\cdot\tCF{b}{k}{l}{m}{3}{4})_{CD\{E''\}}=g_{CE''_1}\bar{\bar{\eta}}_{2D}\bar{\bar{\eta}}_{4E''_2}\cdots\bar{\bar{\eta}}_{4E''_\ell},\\
\lambda_{\ell\boldsymbol{e}_1}(R_\ell)_4^{\phantom{b}b}(\bar{\bar{J}}_{34;2}^{(d,h_{klm},n_b,\Delta_m,\ell\boldsymbol{e}_1)}\cdot\tCF{b}{k}{l}{m}{3}{4})_{CD\{E''\}}=g_{DE''_1}\bar{\bar{\eta}}_{2C}\bar{\bar{\eta}}_{4E''_2}\cdots\bar{\bar{\eta}}_{4E''_\ell},\\
\lambda_{\ell\boldsymbol{e}_1}(R_\ell)_5^{\phantom{b}b}(\bar{\bar{J}}_{34;2}^{(d,h_{klm},n_b,\Delta_m,\ell\boldsymbol{e}_1)}\cdot\tCF{b}{k}{l}{m}{3}{4})_{CD\{E''\}}=g_{CE''_1}g_{DE''_2}\bar{\bar{\eta}}_{4E''_3}\cdots\bar{\bar{\eta}}_{4E''_\ell}.
\end{gathered}
}
Once again, the indices were suppressed on the right-hand side of the natural OPE tensor structure, and the transformation matrix $R_\ell$ leads to the natural three-point tensor structures.

The conformal blocks are thus
\eqn{\mathscr{G}_{(1,1]}^{\ell\boldsymbol{e}_1}=\lambda_{\ell\boldsymbol{e}_1}\frac{(-1)^\ell\ell!}{(d/2-1)_\ell}\left(C_\ell^{d/2-1}(X)\right)_{s^{(1,1]}},}[EqCBle1in00e1e1TS11]
with the conformal substitution
\eqn{s^{(1,1]}:\alpha_2^{s_2}\alpha_3^{s_3}\alpha_4^{s_4}x_3^{r_3}x_4^{r_4}\to G_{(0,0,4,2,-2)CD}^{ij|m|kl},}
followed by
\eqn{\mathscr{G}_{(1,2]}^{\ell\boldsymbol{e}_1}=\lambda_{\ell\boldsymbol{e}_1}\frac{(-1)^\ell\ell!}{(d/2-1)_\ell}\left(C_\ell^{d/2-1}(X)\right)_{s^{(1,2]}},}[EqCBle1in00e1e1TS12]
with the conformal substitution
\eqn{s^{(1,2]}:\alpha_2^{s_2}\alpha_3^{s_3}\alpha_4^{s_4}x_3^{r_3}x_4^{r_4}\to g_{CD}G_{(0,0,0,0,0)}^{ij|m|kl},}
as well as
\eqn{\mathscr{G}_{(1,3]}^{\ell\boldsymbol{e}_1}=\lambda_{\ell\boldsymbol{e}_1}\frac{(-1)^\ell(\ell-1)!}{(d/2)_{\ell-1}}\left[\left(C_{\ell-1}^{d/2}(X)\right)_{s_1^{(1,3]}}-\left(C_{\ell-2}^{d/2}(X)\right)_{s_2^{(1,3]}}\right],}[EqCBle1in00e1e1TS13]
with the conformal substitutions
\eqna{
s_1^{(1,3]}:\alpha_2^{s_2}\alpha_3^{s_3}\alpha_4^{s_4}x_3^{r_3}x_4^{r_4}&\to\bar{\eta}_{2C}G_{(-1,0,1,1,-1)D}^{ij|m|kl}-\bar{\eta}_{1C}G_{(1,0,3,1,-1)D}^{ij|m|kl}\\
&\phantom{\to}\qquad-x_3^{-1}G_{(-1,0,3,3,-1)CD}^{ij|m|kl}+G_{(1,0,5,3,-1)CD}^{ij|m|kl},\\
s_2^{(1,3]}:\alpha_2^{s_2}\alpha_3^{s_3}\alpha_4^{s_4}x_3^{r_3}x_4^{r_4}&\to g_{CD}G_{(0,0,4,2,-2)}^{ij|m|kl},
}
and
\eqn{\mathscr{G}_{(1,4]}^{\ell\boldsymbol{e}_1}=\lambda_{\ell\boldsymbol{e}_1}\frac{(-1)^\ell(\ell-1)!}{(d/2)_{\ell-1}}\left[\left(C_{\ell-1}^{d/2}(X)\right)_{s_1^{(1,4]}}-\left(C_{\ell-2}^{d/2}(X)\right)_{s_2^{(1,4]}}\right],}[EqCBle1in00e1e1TS14]
with the conformal substitutions
\eqna{
s_1^{(1,4]}:\alpha_2^{s_2}\alpha_3^{s_3}\alpha_4^{s_4}x_3^{r_3}x_4^{r_4}&\to\bar{\eta}_{2D}G_{(-1,0,1,1,-1)C}^{ij|m|kl}-\bar{\eta}_{1D}G_{(1,0,3,1,-1)C}^{ij|m|kl}\\
&\phantom{\to}\qquad-x_3^{-1}G_{(-1,0,3,3,-1)CD}^{ij|m|kl}+G_{(1,0,5,3,-1)CD}^{ij|m|kl},\\
s_2^{(1,4]}:\alpha_2^{s_2}\alpha_3^{s_3}\alpha_4^{s_4}x_3^{r_3}x_4^{r_4}&\to g_{CD}G_{(0,0,4,2,-2)}^{ij|m|kl},
}
and finally
\eqna{
\mathscr{G}_{(1,5]}^{\ell\boldsymbol{e}_1}&=\lambda_{\ell\boldsymbol{e}_1}\frac{(-1)^\ell(\ell-2)!}{(d/2)_{\ell-1}}\left[\left(C_{\ell-2}^{d/2}(X)\right)_{s_1^{(1,5]}}+\frac{d}{2}\left(C_{\ell-2}^{d/2+1}(X)\right)_{s_2^{(1,5]}}\right.\\
&\phantom{=}\left.\qquad-\frac{d}{2}\left(C_{\ell-3}^{d/2+1}(X)\right)_{s_3^{(1,5]}}+\frac{d}{2}\left(C_{\ell-4}^{d/2+1}(X)\right)_{s_4^{(1,5]}}\right],
}[EqCBle1in00e1e1TS15]
with the conformal substitutions
\begingroup\makeatletter\def\f@size{10}\check@mathfonts\def\maketag@@@#1{\hbox{\m@th\large\normalfont#1}}%
\eqna{
s_1^{(1,5]}:\alpha_2^{s_2}\alpha_3^{s_3}\alpha_4^{s_4}x_3^{r_3}x_4^{r_4}&\to g_{CD}G_{(0,0,0,0,0)}^{ij|m|kl},\\
s_2^{(1,5]}:\alpha_2^{s_2}\alpha_3^{s_3}\alpha_4^{s_4}x_3^{r_3}x_4^{r_4}&\to\bar{\eta}_{2C}\left[\bar{\eta}_{2D}G_{(-2,0,-2,0,0)}^{ij|m|kl}-\bar{\eta}_{1D}G_{(0,0,0,0,0)}^{ij|m|kl}-x_3^{-1}G_{(-2,0,0,2,0)D}^{ij|m|kl}+G_{(0,0,2,2,0)D}^{ij|m|kl}\right]\\
&\phantom{\to}\qquad-\bar{\eta}_{1C}\left[\bar{\eta}_{2D}G_{(0,0,0,0,0)}^{ij|m|kl}-\bar{\eta}_{1D}G_{(2,0,2,0,0)}^{ij|m|kl}-x_3^{-1}G_{(0,0,2,2,0)D}^{ij|m|kl}+G_{(2,0,4,2,0)D}^{ij|m|kl}\right]\\
&\phantom{\to}\qquad-x_3^{-1}\left[\bar{\eta}_{2D}G_{(-2,0,0,2,0)C}^{ij|m|kl}-\bar{\eta}_{1D}G_{(0,0,2,2,0)C}^{ij|m|kl}-x_3^{-1}G_{(-2,0,2,4,0)CD}^{ij|m|kl}+G_{(0,0,4,4,0)CD}^{ij|m|kl}\right]\\
&\phantom{\to}\qquad+\bar{\eta}_{2D}G_{(0,0,2,2,0)C}^{ij|m|kl}-\bar{\eta}_{1D}G_{(2,0,4,2,0)C}^{ij|m|kl}-x_3^{-1}G_{(0,0,4,4,0)CD}^{ij|m|kl}+G_{(2,0,6,4,0)CD}^{ij|m|kl},\\
s_3^{(1,5]}:\alpha_2^{s_2}\alpha_3^{s_3}\alpha_4^{s_4}x_3^{r_3}x_4^{r_4}&\to\bar{\eta}_{2D}G_{(-1,0,1,1,-1)C}^{ij|m|kl}-\bar{\eta}_{1D}G_{(1,0,3,1,-1)C}^{ij|m|kl}-x_3^{-1}G_{(-1,0,3,3,-1)CD}^{ij|m|kl}+G_{(1,0,5,3,-1)CD}^{ij|m|kl}\\
&\phantom{\to}\qquad+\{C\leftrightarrow D\},\\
s_4^{(1,5]}:\alpha_2^{s_2}\alpha_3^{s_3}\alpha_4^{s_4}x_3^{r_3}x_4^{r_4}&\to G_{(0,0,4,2,-2)CD}^{ij|m|kl}.
}
\endgroup
In this example, there are two conformal blocks for $\ell=0$, four conformal blocks for $\ell=1$, and five conformal blocks for $\ell>1$, as expected from the tensor product decomposition.


\subsubsection{Conformal Blocks as Linear Combinations of Gegenbauer Polynomials with Substitutions}

All of the examples above led to expressions for conformal blocks given by linear combinations of Gegenbauer polynomials with appropriate conformal substitutions.  On the one hand, noting the identical simplifications that occur in the procedure leading to the conformal blocks, we anticipate that there are generic Feynman-like rules for the corresponding conformal substitutions that can be deduced from the previous examples, starting from the mixed basis.

On the other hand, the presence of Gegenbauer polynomials in terms of the variable $X$ might at first seem intriguing.  The origin of the variable $X$ is clear, as it is directly obtained from the $\A$-metric contractions.  We can also argue that the Gegenbauer polynomials appear for any tower of conformal blocks with exchanged quasi-primary operators in $\boldsymbol{N}+\ell\boldsymbol{e}_1$.  Indeed, starting from the mixed basis, the three-point correlation function does not have any special features.  Then, the three-point function is multiplied by hatted projection operators at different embedding space coordinates.  In \cite{Fortin:2019pep}, it was proved that the hatted projection operators merged into one hatted projection operator constructed from the two $\A$-metrics.  Subsequently, the result is transformed into the conformal block with the help of the conformal substitution \eqref{EqJbSub4} and contractions with the tensor structure.  At this point, the implicit hatted projection operator can be extracted from the tensor structure, as described in \cite{Fortin:2019pep}, moving all the nontrivial $\ell$-dependence to the hatted projection operators.  Now, from the tensor product decomposition, we know that the hatted projection operator for $\boldsymbol{N}+\ell\boldsymbol{e}_1$ can be obtained from the tensor product of $\boldsymbol{N}$ and $\ell\boldsymbol{e}_1$.  In that product, one must subtract the smaller irreducible representations.  The trace ones are easily discarded, while the non-trace ones can be removed by simply demanding that the resulting projection operator satisfy the proper symmetries.  Hence, the hatted projection operator for $\boldsymbol{N}+\ell\boldsymbol{e}_1$ is built from the fixed projection operator for $\boldsymbol{N}$ and the projection operator for $\ell\boldsymbol{e}_1$.  The latter carries the $\ell$-dependence through its coefficients, see \eqref{EqPle1}.  The coefficients, which re-sum into simple Gegenbauer polynomials, ultimately lead to linear combinations of Gegenbauer polynomials after the steps necessary to determine the conformal substitutions are completed.  Hence, in a fixed four-point correlation function, conformal blocks for a tower of quasi-primary operators in irreducible representations $\boldsymbol{N}+\ell\boldsymbol{e}_1$ are expressed as linear combinations of Gegenbauer polynomials with proper conformal substitutions, in agreement with the examples above.  Moreover, the conformal substitutions replace the variable $X$ by $\bar{I}$-functions, which are tensorial generalizations of the Exton $G$-function, without derivatives.


\section{Discussion and Conclusion}\label{SecConc}

We have shown how to obtain conformal blocks using the method described in \cite{Fortin:2019fvx,Fortin:2019dnq}.  Given the agreement with several results in the literature obtained using other methods, and our earlier calculations of two- and three-point functions in \cite{Fortin:2019xyr,Fortin:2019pep}, it is clear that the approach is sound.  Using the OPE in embedding space, one can indeed systematically build up $M$-point functions from $(M-1)$-point functions and so on and obtain explicit expressions for $M$-point functions.  As we have already stressed, the method is universal and not limited to any particular Lorentz representation or spacetime dimension. 

The general procedure for obtaining conformal blocks is described in Section \ref{SecFour}, and it involves starting with the hatted projection operators and then performing substitutions.  The two required substitutions involve first the three-point tensorial function and then the four-point tensorial function on the outcome of the first substitution.  Carrying out the conformal substitutions is straightforward but can become tedious for four-point correlation functions of quasi-primary operators in large irreducible representations.  Obtaining the hatted projection operators is perhaps less straightforward, but can also be done systematically by starting with small representations and then working up to larger ones.  It is likely that the procedure for computing conformal blocks could be automated and handled by a computer program.

The intermediate expressions for the blocks involving the Gegenbauer polynomials, which lead to the actual conformal blocks through the $s$-substitutions, are certainly intriguing.  Based on the examples we have worked out, we argued that this feature is general and applicable to other conformal blocks.  Moreover, it should be possible to codify the procedure for obtaining the appropriate conformal substitutions as a set of Feynman-like rules.  This will be addressed in a future publication.

While many of the conformal blocks derived here were already known, it was useful to rederive such results using the new method.  Looking ahead to what can be accomplished with this method soon, we consider tackling several specific conformal blocks.  One of the most fundamental objects for exploring CFTs is the four-point function involving four energy-momentum tensors.  Because of the sheer number of blocks involved in a four-point function of energy-momentum tensors, this will be the subject of a separate publication.

Looking further ahead, we hope that our approach will be useful for both the numerical and analytic bootstrap.  Despite an enormous amount of progress, it is apparent that CFTs have a rich and complicated structure that has not been fully explored yet.


\ack{
The authors would like to thank Wenjie Ma for useful discussions.  The work of JFF and VP is supported by NSERC and FRQNT.  JFF would like to thank the organizers and participants of the Bootstrap 2019 conference, held at Perimeter Institute, for their hospitality.  WS would like to thank the KITP Santa Barbara for its hospitality, his work was supported in part by the National Science Foundation under Grant No.~NSF PHY-1748958.
}


\bibliography{FourPtFcts}

\begin{thebibliography}{100}
\ifx\href\asklfhas\newcommand{\href}[2]{#2}\fi
\ifx\arxivref\asklfhas\newcommand{\arxivref}[2]{\href{http://arxiv.org/abs/#1}{#2}}\fi
\ifx\doiref\asklfhas\newcommand{\doiref}[2]{\href{http://dx.doi.org/#1}{#2}}\fi
\parskip 0pt
\normalsize

\bibitem{Fortin:2019fvx}
J.-F. Fortin \& W.~Skiba,
\textit{``{A recipe for conformal blocks}''},
\normalsize{\texttt{\arxivref{1905.00036}{arXiv:1905.00036}}}\ignorespaces
\bibitem{Fortin:2019dnq}
J.-F. Fortin \& W.~Skiba,
\textit{``{New Methods for Conformal Correlation Functions}''},
\normalsize{\texttt{\arxivref{1905.00434}{arXiv:1905.00434}}}\ignorespaces
\bibitem{Ferrara:1973yt}
S.~Ferrara, A.~F. Grillo \& R.~Gatto,
\textit{``{Tensor representations of conformal algebra and conformally
  covariant operator product expansion}''},
\doiref{10.1016/0003-4916(73)90446-6}{Annals~Phys. \textbf{76}, 161
  (1973)\ignorespaces}\ignorespaces
\bibitem{Polyakov:1974gs}
A.~M. Polyakov,
\textit{``{Nonhamiltonian approach to conformal quantum field theory}''},
Zh.~Eksp.~Teor.~Fiz. \textbf{66}, 23 (1974)\ignorespaces\ignorespaces,
[Sov. Phys. JETP39,9(1974)]\ignorespaces
\bibitem{Rattazzi:2008pe}
R.~Rattazzi, V.~S. Rychkov, E.~Tonni \& A.~Vichi,
\textit{``{Bounding scalar operator dimensions in 4D CFT}''},
\doiref{10.1088/1126-6708/2008/12/031}{JHEP \textbf{0812}, 031
  (2008)\ignorespaces}\ignorespaces,
\normalsize{\texttt{\arxivref{0807.0004}{arXiv:0807.0004}}}\ignorespaces
\bibitem{Rychkov:2009ij}
V.~S. Rychkov \& A.~Vichi,
\textit{``{Universal Constraints on Conformal Operator Dimensions}''},
\doiref{10.1103/PhysRevD.80.045006}{Phys.~Rev. \textbf{D80}, 045006
  (2009)\ignorespaces}\ignorespaces,
\normalsize{\texttt{\arxivref{0905.2211}{arXiv:0905.2211}}}\ignorespaces
\bibitem{Caracciolo:2009bx}
F.~Caracciolo \& V.~S. Rychkov,
\textit{``{Rigorous Limits on the Interaction Strength in Quantum Field
  Theory}''},
\doiref{10.1103/PhysRevD.81.085037}{Phys.~Rev. \textbf{D81}, 085037
  (2010)\ignorespaces}\ignorespaces,
\normalsize{\texttt{\arxivref{0912.2726}{arXiv:0912.2726}}}\ignorespaces
\bibitem{Poland:2010wg}
D.~Poland \& D.~Simmons-Duffin,
\textit{``{Bounds on 4D Conformal and Superconformal Field Theories}''},
\doiref{10.1007/JHEP05(2011)017}{JHEP \textbf{1105}, 017
  (2011)\ignorespaces}\ignorespaces,
\normalsize{\texttt{\arxivref{1009.2087}{arXiv:1009.2087}}}\ignorespaces
\bibitem{Rattazzi:2010gj}
R.~Rattazzi, S.~Rychkov \& A.~Vichi,
\textit{``{Central Charge Bounds in 4D Conformal Field Theory}''},
\doiref{10.1103/PhysRevD.83.046011}{Phys.~Rev. \textbf{D83}, 046011
  (2011)\ignorespaces}\ignorespaces,
\normalsize{\texttt{\arxivref{1009.2725}{arXiv:1009.2725}}}\ignorespaces
\bibitem{Poland:2011ey}
D.~Poland, D.~Simmons-Duffin \& A.~Vichi,
\textit{``{Carving Out the Space of 4D CFTs}''},
\doiref{10.1007/JHEP05(2012)110}{JHEP \textbf{1205}, 110
  (2012)\ignorespaces}\ignorespaces,
\normalsize{\texttt{\arxivref{1109.5176}{arXiv:1109.5176}}}\ignorespaces
\bibitem{Rychkov:2011et}
S.~Rychkov,
\textit{``{Conformal Bootstrap in Three Dimensions?}''},
\normalsize{\texttt{\arxivref{1111.2115}{arXiv:1111.2115}}}\ignorespaces
\bibitem{ElShowk:2012ht}
S.~El-Showk, M.~F. Paulos, D.~Poland, S.~Rychkov, D.~Simmons-Duffin \&
  A.~Vichi,
\textit{``{Solving the 3D Ising Model with the Conformal Bootstrap}''},
\doiref{10.1103/PhysRevD.86.025022}{Phys.~Rev. \textbf{D86}, 025022
  (2012)\ignorespaces}\ignorespaces,
\normalsize{\texttt{\arxivref{1203.6064}{arXiv:1203.6064}}}\ignorespaces
\bibitem{Liendo:2012hy}
P.~Liendo, L.~Rastelli \& B.~C. van~Rees,
\textit{``{The Bootstrap Program for Boundary CFT$_d$}''},
\doiref{10.1007/JHEP07(2013)113}{JHEP \textbf{1307}, 113
  (2013)\ignorespaces}\ignorespaces,
\normalsize{\texttt{\arxivref{1210.4258}{arXiv:1210.4258}}}\ignorespaces
\bibitem{ElShowk:2012hu}
S.~El-Showk \& M.~F. Paulos,
\textit{``{Bootstrapping Conformal Field Theories with the Extremal Functional
  Method}''},
\doiref{10.1103/PhysRevLett.111.241601}{Phys.~Rev.~Lett. \textbf{111}, 241601
  (2013)\ignorespaces}\ignorespaces,
\normalsize{\texttt{\arxivref{1211.2810}{arXiv:1211.2810}}}\ignorespaces
\bibitem{Gliozzi:2013ysa}
F.~Gliozzi,
\textit{``{More constraining conformal bootstrap}''},
\doiref{10.1103/PhysRevLett.111.161602}{Phys.~Rev.~Lett. \textbf{111}, 161602
  (2013)\ignorespaces}\ignorespaces,
\normalsize{\texttt{\arxivref{1307.3111}{arXiv:1307.3111}}}\ignorespaces
\bibitem{Alday:2013opa}
L.~F. Alday \& A.~Bissi,
\textit{``{The superconformal bootstrap for structure constants}''},
\doiref{10.1007/JHEP09(2014)144}{JHEP \textbf{1409}, 144
  (2014)\ignorespaces}\ignorespaces,
\normalsize{\texttt{\arxivref{1310.3757}{arXiv:1310.3757}}}\ignorespaces
\bibitem{Gaiotto:2013nva}
D.~Gaiotto, D.~Mazac \& M.~F. Paulos,
\textit{``{Bootstrapping the 3d Ising twist defect}''},
\doiref{10.1007/JHEP03(2014)100}{JHEP \textbf{1403}, 100
  (2014)\ignorespaces}\ignorespaces,
\normalsize{\texttt{\arxivref{1310.5078}{arXiv:1310.5078}}}\ignorespaces
\bibitem{El-Showk:2014dwa}
S.~El-Showk, M.~F. Paulos, D.~Poland, S.~Rychkov, D.~Simmons-Duffin \&
  A.~Vichi,
\textit{``{Solving the 3d Ising Model with the Conformal Bootstrap II.
  c-Minimization and Precise Critical Exponents}''},
\doiref{10.1007/s10955-014-1042-7}{J.~Stat.~Phys. \textbf{157}, 869
  (2014)\ignorespaces}\ignorespaces,
\normalsize{\texttt{\arxivref{1403.4545}{arXiv:1403.4545}}}\ignorespaces
\bibitem{Chester:2014fya}
S.~M. Chester, J.~Lee, S.~S. Pufu \& R.~Yacoby,
\textit{``{The $ \mathcal{N}=8 $ superconformal bootstrap in three
  dimensions}''},
\doiref{10.1007/JHEP09(2014)143}{JHEP \textbf{1409}, 143
  (2014)\ignorespaces}\ignorespaces,
\normalsize{\texttt{\arxivref{1406.4814}{arXiv:1406.4814}}}\ignorespaces
\bibitem{Kos:2014bka}
F.~Kos, D.~Poland \& D.~Simmons-Duffin,
\textit{``{Bootstrapping Mixed Correlators in the 3D Ising Model}''},
\doiref{10.1007/JHEP11(2014)109}{JHEP \textbf{1411}, 109
  (2014)\ignorespaces}\ignorespaces,
\normalsize{\texttt{\arxivref{1406.4858}{arXiv:1406.4858}}}\ignorespaces
\bibitem{Caracciolo:2014cxa}
F.~Caracciolo, A.~Castedo~Echeverri, B.~von~Harling \& M.~Serone,
\textit{``{Bounds on OPE Coefficients in 4D Conformal Field Theories}''},
\doiref{10.1007/JHEP10(2014)020}{JHEP \textbf{1410}, 020
  (2014)\ignorespaces}\ignorespaces,
\normalsize{\texttt{\arxivref{1406.7845}{arXiv:1406.7845}}}\ignorespaces
\bibitem{Paulos:2014vya}
M.~F. Paulos,
\textit{``{JuliBootS: a hands-on guide to the conformal bootstrap}''},
\normalsize{\texttt{\arxivref{1412.4127}{arXiv:1412.4127}}}\ignorespaces
\bibitem{Beem:2014zpa}
C.~Beem, M.~Lemos, P.~Liendo, L.~Rastelli \& B.~C. van~Rees,
\textit{``{The $ \mathcal{N}=2 $ superconformal bootstrap}''},
\doiref{10.1007/JHEP03(2016)183}{JHEP \textbf{1603}, 183
  (2016)\ignorespaces}\ignorespaces,
\normalsize{\texttt{\arxivref{1412.7541}{arXiv:1412.7541}}}\ignorespaces
\bibitem{Simmons-Duffin:2015qma}
D.~Simmons-Duffin,
\textit{``{A Semidefinite Program Solver for the Conformal Bootstrap}''},
\doiref{10.1007/JHEP06(2015)174}{JHEP \textbf{1506}, 174
  (2015)\ignorespaces}\ignorespaces,
\normalsize{\texttt{\arxivref{1502.02033}{arXiv:1502.02033}}}\ignorespaces
\bibitem{Bobev:2015jxa}
N.~Bobev, S.~El-Showk, D.~Mazac \& M.~F. Paulos,
\textit{``{Bootstrapping SCFTs with Four Supercharges}''},
\doiref{10.1007/JHEP08(2015)142}{JHEP \textbf{1508}, 142
  (2015)\ignorespaces}\ignorespaces,
\normalsize{\texttt{\arxivref{1503.02081}{arXiv:1503.02081}}}\ignorespaces
\bibitem{Beem:2015aoa}
C.~Beem, M.~Lemos, L.~Rastelli \& B.~C. van~Rees,
\textit{``{The (2, 0) superconformal bootstrap}''},
\doiref{10.1103/PhysRevD.93.025016}{Phys.~Rev. \textbf{D93}, 025016
  (2016)\ignorespaces}\ignorespaces,
\normalsize{\texttt{\arxivref{1507.05637}{arXiv:1507.05637}}}\ignorespaces
\bibitem{Iliesiu:2015qra}
L.~Iliesiu, F.~Kos, D.~Poland, S.~S. Pufu, D.~Simmons-Duffin \& R.~Yacoby,
\textit{``{Bootstrapping 3D Fermions}''},
\doiref{10.1007/JHEP03(2016)120}{JHEP \textbf{1603}, 120
  (2016)\ignorespaces}\ignorespaces,
\normalsize{\texttt{\arxivref{1508.00012}{arXiv:1508.00012}}}\ignorespaces
\bibitem{Poland:2015mta}
D.~Poland \& A.~Stergiou,
\textit{``{Exploring the Minimal 4D $\mathcal{N}=1$ SCFT}''},
\doiref{10.1007/JHEP12(2015)121}{JHEP \textbf{1512}, 121
  (2015)\ignorespaces}\ignorespaces,
\normalsize{\texttt{\arxivref{1509.06368}{arXiv:1509.06368}}}\ignorespaces
\bibitem{Lemos:2015awa}
M.~Lemos \& P.~Liendo,
\textit{``{Bootstrapping $ \mathcal{N}=2 $ chiral correlators}''},
\doiref{10.1007/JHEP01(2016)025}{JHEP \textbf{1601}, 025
  (2016)\ignorespaces}\ignorespaces,
\normalsize{\texttt{\arxivref{1510.03866}{arXiv:1510.03866}}}\ignorespaces
\bibitem{Lin:2015wcg}
Y.-H. Lin, S.-H. Shao, D.~Simmons-Duffin, Y.~Wang \& X.~Yin,
\textit{``{$ \mathcal{N} $ = 4 superconformal bootstrap of the K3 CFT}''},
\doiref{10.1007/JHEP05(2017)126}{JHEP \textbf{1705}, 126
  (2017)\ignorespaces}\ignorespaces,
\normalsize{\texttt{\arxivref{1511.04065}{arXiv:1511.04065}}}\ignorespaces
\bibitem{Chester:2016wrc}
S.~M. Chester \& S.~S. Pufu,
\textit{``{Towards bootstrapping QED$_{3}$}''},
\doiref{10.1007/JHEP08(2016)019}{JHEP \textbf{1608}, 019
  (2016)\ignorespaces}\ignorespaces,
\normalsize{\texttt{\arxivref{1601.03476}{arXiv:1601.03476}}}\ignorespaces
\bibitem{Behan:2016dtz}
C.~Behan,
\textit{``{PyCFTBoot: A flexible interface for the conformal bootstrap}''},
\doiref{10.4208/cicp.OA-2016-0107}{Commun.~Comput.~Phys. \textbf{22}, 1
  (2017)\ignorespaces}\ignorespaces,
\normalsize{\texttt{\arxivref{1602.02810}{arXiv:1602.02810}}}\ignorespaces
\bibitem{El-Showk:2016mxr}
S.~El-Showk \& M.~F. Paulos,
\textit{``{Extremal bootstrapping: go with the flow}''},
\doiref{10.1007/JHEP03(2018)148}{JHEP \textbf{1803}, 148
  (2018)\ignorespaces}\ignorespaces,
\normalsize{\texttt{\arxivref{1605.08087}{arXiv:1605.08087}}}\ignorespaces
\bibitem{Lin:2016gcl}
Y.-H. Lin, S.-H. Shao, Y.~Wang \& X.~Yin,
\textit{``{(2, 2) superconformal bootstrap in two dimensions}''},
\doiref{10.1007/JHEP05(2017)112}{JHEP \textbf{1705}, 112
  (2017)\ignorespaces}\ignorespaces,
\normalsize{\texttt{\arxivref{1610.05371}{arXiv:1610.05371}}}\ignorespaces
\bibitem{Lemos:2016xke}
M.~Lemos, P.~Liendo, C.~Meneghelli \& V.~Mitev,
\textit{``{Bootstrapping $\mathcal{N}=3$ superconformal theories}''},
\doiref{10.1007/JHEP04(2017)032}{JHEP \textbf{1704}, 032
  (2017)\ignorespaces}\ignorespaces,
\normalsize{\texttt{\arxivref{1612.01536}{arXiv:1612.01536}}}\ignorespaces
\bibitem{Beem:2016wfs}
C.~Beem, L.~Rastelli \& B.~C. van~Rees,
\textit{``{More ${\mathcal N}=4$ superconformal bootstrap}''},
\doiref{10.1103/PhysRevD.96.046014}{Phys.~Rev. \textbf{D96}, 046014
  (2017)\ignorespaces}\ignorespaces,
\normalsize{\texttt{\arxivref{1612.02363}{arXiv:1612.02363}}}\ignorespaces
\bibitem{Li:2017ddj}
D.~Li, D.~Meltzer \& A.~Stergiou,
\textit{``{Bootstrapping mixed correlators in 4D $ \mathcal{N} $ = 1 SCFTs}''},
\doiref{10.1007/JHEP07(2017)029}{JHEP \textbf{1707}, 029
  (2017)\ignorespaces}\ignorespaces,
\normalsize{\texttt{\arxivref{1702.00404}{arXiv:1702.00404}}}\ignorespaces
\bibitem{Collier:2017shs}
S.~Collier, P.~Kravchuk, Y.-H. Lin \& X.~Yin,
\textit{``{Bootstrapping the Spectral Function: On the Uniqueness of Liouville
  and the Universality of BTZ}''},
\doiref{10.1007/JHEP09(2018)150}{JHEP \textbf{1809}, 150
  (2018)\ignorespaces}\ignorespaces,
\normalsize{\texttt{\arxivref{1702.00423}{arXiv:1702.00423}}}\ignorespaces
\bibitem{Cornagliotto:2017dup}
M.~Cornagliotto, M.~Lemos \& V.~Schomerus,
\textit{``{Long Multiplet Bootstrap}''},
\doiref{10.1007/JHEP10(2017)119}{JHEP \textbf{1710}, 119
  (2017)\ignorespaces}\ignorespaces,
\normalsize{\texttt{\arxivref{1702.05101}{arXiv:1702.05101}}}\ignorespaces
\bibitem{Rychkov:2017tpc}
J.~Qiao \& S.~Rychkov,
\textit{``{Cut-touching linear functionals in the conformal bootstrap}''},
\doiref{10.1007/JHEP06(2017)076}{JHEP \textbf{1706}, 076
  (2017)\ignorespaces}\ignorespaces,
\normalsize{\texttt{\arxivref{1705.01357}{arXiv:1705.01357}}}\ignorespaces
\bibitem{Nakayama:2017vdd}
Y.~Nakayama,
\textit{``{Bootstrap experiments on higher dimensional CFTs}''},
\doiref{10.1142/S0217751X18500367}{Int.~J.~Mod.~Phys. \textbf{A33}, 1850036
  (2018)\ignorespaces}\ignorespaces,
\normalsize{\texttt{\arxivref{1705.02744}{arXiv:1705.02744}}}\ignorespaces
\bibitem{Chang:2017xmr}
C.-M. Chang \& Y.-H. Lin,
\textit{``{Carving Out the End of the World or (Superconformal Bootstrap in Six
  Dimensions)}''},
\doiref{10.1007/JHEP08(2017)128}{JHEP \textbf{1708}, 128
  (2017)\ignorespaces}\ignorespaces,
\normalsize{\texttt{\arxivref{1705.05392}{arXiv:1705.05392}}}\ignorespaces
\bibitem{Dymarsky:2017yzx}
A.~Dymarsky, F.~Kos, P.~Kravchuk, D.~Poland \& D.~Simmons-Duffin,
\textit{``{The 3d Stress-Tensor Bootstrap}''},
\doiref{10.1007/JHEP02(2018)164}{JHEP \textbf{1802}, 164
  (2018)\ignorespaces}\ignorespaces,
\normalsize{\texttt{\arxivref{1708.05718}{arXiv:1708.05718}}}\ignorespaces,
[,343(2017)]\ignorespaces
\bibitem{Karateev:2019pvw}
D.~Karateev, P.~Kravchuk, M.~Serone \& A.~Vichi,
\textit{``{Fermion Conformal Bootstrap in 4d}''},
\normalsize{\texttt{\arxivref{1902.05969}{arXiv:1902.05969}}}\ignorespaces
\bibitem{Cornalba:2007fs}
L.~Cornalba,
\textit{``{Eikonal methods in AdS/CFT: Regge theory and multi-reggeon
  exchange}''},
\normalsize{\texttt{\arxivref{0710.5480}{arXiv:0710.5480}}}\ignorespaces
\bibitem{Cornalba:2009ax}
L.~Cornalba, M.~S. Costa \& J.~Penedones,
\textit{``{Deep Inelastic Scattering in Conformal QCD}''},
\doiref{10.1007/JHEP03(2010)133}{JHEP \textbf{1003}, 133
  (2010)\ignorespaces}\ignorespaces,
\normalsize{\texttt{\arxivref{0911.0043}{arXiv:0911.0043}}}\ignorespaces
\bibitem{Pappadopulo:2012jk}
D.~Pappadopulo, S.~Rychkov, J.~Espin \& R.~Rattazzi,
\textit{``{OPE Convergence in Conformal Field Theory}''},
\doiref{10.1103/PhysRevD.86.105043}{Phys.~Rev. \textbf{D86}, 105043
  (2012)\ignorespaces}\ignorespaces,
\normalsize{\texttt{\arxivref{1208.6449}{arXiv:1208.6449}}}\ignorespaces
\bibitem{Costa:2012cb}
M.~S. Costa, V.~Goncalves \& J.~Penedones,
\textit{``{Conformal Regge theory}''},
\doiref{10.1007/JHEP12(2012)091}{JHEP \textbf{1212}, 091
  (2012)\ignorespaces}\ignorespaces,
\normalsize{\texttt{\arxivref{1209.4355}{arXiv:1209.4355}}}\ignorespaces
\bibitem{Hogervorst:2013sma}
M.~Hogervorst \& S.~Rychkov,
\textit{``{Radial Coordinates for Conformal Blocks}''},
\doiref{10.1103/PhysRevD.87.106004}{Phys.~Rev. \textbf{D87}, 106004
  (2013)\ignorespaces}\ignorespaces,
\normalsize{\texttt{\arxivref{1303.1111}{arXiv:1303.1111}}}\ignorespaces
\bibitem{Hartman:2015lfa}
T.~Hartman, S.~Jain \& S.~Kundu,
\textit{``{Causality Constraints in Conformal Field Theory}''},
\doiref{10.1007/JHEP05(2016)099}{JHEP \textbf{1605}, 099
  (2016)\ignorespaces}\ignorespaces,
\normalsize{\texttt{\arxivref{1509.00014}{arXiv:1509.00014}}}\ignorespaces
\bibitem{Kim:2015oca}
H.~Kim, P.~Kravchuk \& H.~Ooguri,
\textit{``{Reflections on Conformal Spectra}''},
\doiref{10.1007/JHEP04(2016)184}{JHEP \textbf{1604}, 184
  (2016)\ignorespaces}\ignorespaces,
\normalsize{\texttt{\arxivref{1510.08772}{arXiv:1510.08772}}}\ignorespaces,
[,322(2015)]\ignorespaces
\bibitem{Li:2015itl}
D.~Li, D.~Meltzer \& D.~Poland,
\textit{``{Conformal Collider Physics from the Lightcone Bootstrap}''},
\doiref{10.1007/JHEP02(2016)143}{JHEP \textbf{1602}, 143
  (2016)\ignorespaces}\ignorespaces,
\normalsize{\texttt{\arxivref{1511.08025}{arXiv:1511.08025}}}\ignorespaces
\bibitem{Hartman:2016dxc}
T.~Hartman, S.~Jain \& S.~Kundu,
\textit{``{A New Spin on Causality Constraints}''},
\doiref{10.1007/JHEP10(2016)141}{JHEP \textbf{1610}, 141
  (2016)\ignorespaces}\ignorespaces,
\normalsize{\texttt{\arxivref{1601.07904}{arXiv:1601.07904}}}\ignorespaces
\bibitem{Hofman:2016awc}
D.~M. Hofman, D.~Li, D.~Meltzer, D.~Poland \& F.~Rejon-Barrera,
\textit{``{A Proof of the Conformal Collider Bounds}''},
\doiref{10.1007/JHEP06(2016)111}{JHEP \textbf{1606}, 111
  (2016)\ignorespaces}\ignorespaces,
\normalsize{\texttt{\arxivref{1603.03771}{arXiv:1603.03771}}}\ignorespaces
\bibitem{Hartman:2016lgu}
T.~Hartman, S.~Kundu \& A.~Tajdini,
\textit{``{Averaged Null Energy Condition from Causality}''},
\doiref{10.1007/JHEP07(2017)066}{JHEP \textbf{1707}, 066
  (2017)\ignorespaces}\ignorespaces,
\normalsize{\texttt{\arxivref{1610.05308}{arXiv:1610.05308}}}\ignorespaces
\bibitem{Afkhami-Jeddi:2016ntf}
N.~Afkhami-Jeddi, T.~Hartman, S.~Kundu \& A.~Tajdini,
\textit{``{Einstein gravity 3-point functions from conformal field theory}''},
\doiref{10.1007/JHEP12(2017)049}{JHEP \textbf{1712}, 049
  (2017)\ignorespaces}\ignorespaces,
\normalsize{\texttt{\arxivref{1610.09378}{arXiv:1610.09378}}}\ignorespaces
\bibitem{Gadde:2017sjg}
A.~Gadde,
\textit{``{In search of conformal theories}''},
\normalsize{\texttt{\arxivref{1702.07362}{arXiv:1702.07362}}}\ignorespaces
\bibitem{Hogervorst:2017sfd}
M.~Hogervorst \& B.~C. van~Rees,
\textit{``{Crossing symmetry in alpha space}''},
\doiref{10.1007/JHEP11(2017)193}{JHEP \textbf{1711}, 193
  (2017)\ignorespaces}\ignorespaces,
\normalsize{\texttt{\arxivref{1702.08471}{arXiv:1702.08471}}}\ignorespaces
\bibitem{Caron-Huot:2017vep}
S.~Caron-Huot,
\textit{``{Analyticity in Spin in Conformal Theories}''},
\doiref{10.1007/JHEP09(2017)078}{JHEP \textbf{1709}, 078
  (2017)\ignorespaces}\ignorespaces,
\normalsize{\texttt{\arxivref{1703.00278}{arXiv:1703.00278}}}\ignorespaces
\bibitem{Hogervorst:2017kbj}
M.~Hogervorst,
\textit{``{Crossing Kernels for Boundary and Crosscap CFTs}''},
\normalsize{\texttt{\arxivref{1703.08159}{arXiv:1703.08159}}}\ignorespaces
\bibitem{Kulaxizi:2017ixa}
M.~Kulaxizi, A.~Parnachev \& A.~Zhiboedov,
\textit{``{Bulk Phase Shift, CFT Regge Limit and Einstein Gravity}''},
\doiref{10.1007/JHEP06(2018)121}{JHEP \textbf{1806}, 121
  (2018)\ignorespaces}\ignorespaces,
\normalsize{\texttt{\arxivref{1705.02934}{arXiv:1705.02934}}}\ignorespaces
\bibitem{Li:2017lmh}
D.~Li, D.~Meltzer \& D.~Poland,
\textit{``{Conformal Bootstrap in the Regge Limit}''},
\doiref{10.1007/JHEP12(2017)013}{JHEP \textbf{1712}, 013
  (2017)\ignorespaces}\ignorespaces,
\normalsize{\texttt{\arxivref{1705.03453}{arXiv:1705.03453}}}\ignorespaces
\bibitem{Cuomo:2017wme}
G.~F. Cuomo, D.~Karateev \& P.~Kravchuk,
\textit{``{General Bootstrap Equations in 4D CFTs}''},
\doiref{10.1007/JHEP01(2018)130}{JHEP \textbf{1801}, 130
  (2018)\ignorespaces}\ignorespaces,
\normalsize{\texttt{\arxivref{1705.05401}{arXiv:1705.05401}}}\ignorespaces,
[,57(2017)]\ignorespaces
\bibitem{Dey:2017oim}
P.~Dey \& A.~Kaviraj,
\textit{``{Towards a Bootstrap approach to higher orders of epsilon
  expansion}''},
\doiref{10.1007/JHEP02(2018)153}{JHEP \textbf{1802}, 153
  (2018)\ignorespaces}\ignorespaces,
\normalsize{\texttt{\arxivref{1711.01173}{arXiv:1711.01173}}}\ignorespaces
\bibitem{Simmons-Duffin:2017nub}
D.~Simmons-Duffin, D.~Stanford \& E.~Witten,
\textit{``{A spacetime derivation of the Lorentzian OPE inversion formula}''},
\doiref{10.1007/JHEP07(2018)085}{JHEP \textbf{1807}, 085
  (2018)\ignorespaces}\ignorespaces,
\normalsize{\texttt{\arxivref{1711.03816}{arXiv:1711.03816}}}\ignorespaces
\bibitem{Elkhidir:2017iov}
E.~Elkhidir \& D.~Karateev,
\textit{``{Scalar-Fermion Analytic Bootstrap in 4D}''},
\normalsize{\texttt{\arxivref{1712.01554}{arXiv:1712.01554}}}\ignorespaces
\bibitem{Kravchuk:2018htv}
P.~Kravchuk \& D.~Simmons-Duffin,
\textit{``{Light-ray operators in conformal field theory}''},
\doiref{10.1007/JHEP11(2018)102}{JHEP \textbf{1811}, 102
  (2018)\ignorespaces}\ignorespaces,
\normalsize{\texttt{\arxivref{1805.00098}{arXiv:1805.00098}}}\ignorespaces,
[,236(2018)]\ignorespaces
\bibitem{Karateev:2018oml}
D.~Karateev, P.~Kravchuk \& D.~Simmons-Duffin,
\textit{``{Harmonic Analysis and Mean Field Theory}''},
\normalsize{\texttt{\arxivref{1809.05111}{arXiv:1809.05111}}}\ignorespaces
\bibitem{Liendo:2019jpu}
P.~Liendo, Y.~Linke \& V.~Schomerus,
\textit{``{A Lorentzian inversion formula for defect CFT}''},
\normalsize{\texttt{\arxivref{1903.05222}{arXiv:1903.05222}}}\ignorespaces
\bibitem{Albayrak:2019gnz}
S.~Albayrak, D.~Meltzer \& D.~Poland,
\textit{``{More Analytic Bootstrap: Nonperturbative Effects and Fermions}''},
\normalsize{\texttt{\arxivref{1904.00032}{arXiv:1904.00032}}}\ignorespaces
\bibitem{Li:2019twz}
Z.~Li,
\textit{``{Bootstrapping Veneziano Amplitude of Vasiliev Theory and $3D$
  Bosonization}''},
\normalsize{\texttt{\arxivref{1906.05834}{arXiv:1906.05834}}}\ignorespaces
\bibitem{Rattazzi:2010yc}
R.~Rattazzi, S.~Rychkov \& A.~Vichi,
\textit{``{Bounds in 4D Conformal Field Theories with Global Symmetry}''},
\doiref{10.1088/1751-8113/44/3/035402}{J.~Phys. \textbf{A44}, 035402
  (2011)\ignorespaces}\ignorespaces,
\normalsize{\texttt{\arxivref{1009.5985}{arXiv:1009.5985}}}\ignorespaces
\bibitem{Vichi:2011ux}
A.~Vichi,
\textit{``{Improved bounds for CFT's with global symmetries}''},
\doiref{10.1007/JHEP01(2012)162}{JHEP \textbf{1201}, 162
  (2012)\ignorespaces}\ignorespaces,
\normalsize{\texttt{\arxivref{1106.4037}{arXiv:1106.4037}}}\ignorespaces
\bibitem{Kos:2013tga}
F.~Kos, D.~Poland \& D.~Simmons-Duffin,
\textit{``{Bootstrapping the $O(N)$ vector models}''},
\doiref{10.1007/JHEP06(2014)091}{JHEP \textbf{1406}, 091
  (2014)\ignorespaces}\ignorespaces,
\normalsize{\texttt{\arxivref{1307.6856}{arXiv:1307.6856}}}\ignorespaces
\bibitem{Berkooz:2014yda}
M.~Berkooz, R.~Yacoby \& A.~Zait,
\textit{``{Bounds on $\mathcal{N} = 1$ superconformal theories with global
  symmetries}''},
\doiref{10.1007/JHEP01(2015)132, 10.1007/JHEP08(2014)008}{JHEP \textbf{1408},
  008 (2014)\ignorespaces}\ignorespaces,
\normalsize{\texttt{\arxivref{1402.6068}{arXiv:1402.6068}}}\ignorespaces,
[Erratum: JHEP01,132(2015)]\ignorespaces
\bibitem{Nakayama:2014lva}
Y.~Nakayama \& T.~Ohtsuki,
\textit{``{Approaching the conformal window of $O(n)\times O(m)$ symmetric
  Landau-Ginzburg models using the conformal bootstrap}''},
\doiref{10.1103/PhysRevD.89.126009}{Phys.~Rev. \textbf{D89}, 126009
  (2014)\ignorespaces}\ignorespaces,
\normalsize{\texttt{\arxivref{1404.0489}{arXiv:1404.0489}}}\ignorespaces
\bibitem{Nakayama:2014yia}
Y.~Nakayama \& T.~Ohtsuki,
\textit{``{Five dimensional $O(N)$-symmetric CFTs from conformal bootstrap}''},
\doiref{10.1016/j.physletb.2014.05.058}{Phys.~Lett. \textbf{B734}, 193
  (2014)\ignorespaces}\ignorespaces,
\normalsize{\texttt{\arxivref{1404.5201}{arXiv:1404.5201}}}\ignorespaces
\bibitem{Nakayama:2014sba}
Y.~Nakayama \& T.~Ohtsuki,
\textit{``{Bootstrapping phase transitions in QCD and frustrated spin
  systems}''},
\doiref{10.1103/PhysRevD.91.021901}{Phys.~Rev. \textbf{D91}, 021901
  (2015)\ignorespaces}\ignorespaces,
\normalsize{\texttt{\arxivref{1407.6195}{arXiv:1407.6195}}}\ignorespaces
\bibitem{Bae:2014hia}
J.-B. Bae \& S.-J. Rey,
\textit{``{Conformal Bootstrap Approach to O(N) Fixed Points in Five
  Dimensions}''},
\normalsize{\texttt{\arxivref{1412.6549}{arXiv:1412.6549}}}\ignorespaces
\bibitem{Chester:2014gqa}
S.~M. Chester, S.~S. Pufu \& R.~Yacoby,
\textit{``{Bootstrapping $O(N)$ vector models in 4 $< d <$ 6}''},
\doiref{10.1103/PhysRevD.91.086014}{Phys.~Rev. \textbf{D91}, 086014
  (2015)\ignorespaces}\ignorespaces,
\normalsize{\texttt{\arxivref{1412.7746}{arXiv:1412.7746}}}\ignorespaces
\bibitem{Kos:2015mba}
F.~Kos, D.~Poland, D.~Simmons-Duffin \& A.~Vichi,
\textit{``{Bootstrapping the O(N) Archipelago}''},
\doiref{10.1007/JHEP11(2015)106}{JHEP \textbf{1511}, 106
  (2015)\ignorespaces}\ignorespaces,
\normalsize{\texttt{\arxivref{1504.07997}{arXiv:1504.07997}}}\ignorespaces
\bibitem{Chester:2015qca}
S.~M. Chester, S.~Giombi, L.~V. Iliesiu, I.~R. Klebanov, S.~S. Pufu \&
  R.~Yacoby,
\textit{``{Accidental Symmetries and the Conformal Bootstrap}''},
\doiref{10.1007/JHEP01(2016)110}{JHEP \textbf{1601}, 110
  (2016)\ignorespaces}\ignorespaces,
\normalsize{\texttt{\arxivref{1507.04424}{arXiv:1507.04424}}}\ignorespaces
\bibitem{Chester:2015lej}
S.~M. Chester, L.~V. Iliesiu, S.~S. Pufu \& R.~Yacoby,
\textit{``{Bootstrapping $O(N)$ Vector Models with Four Supercharges in $3 \leq
  d \leq4$}''},
\doiref{10.1007/JHEP05(2016)103}{JHEP \textbf{1605}, 103
  (2016)\ignorespaces}\ignorespaces,
\normalsize{\texttt{\arxivref{1511.07552}{arXiv:1511.07552}}}\ignorespaces
\bibitem{Dey:2016zbg}
P.~Dey, A.~Kaviraj \& K.~Sen,
\textit{``{More on analytic bootstrap for O(N) models}''},
\doiref{10.1007/JHEP06(2016)136}{JHEP \textbf{1606}, 136
  (2016)\ignorespaces}\ignorespaces,
\normalsize{\texttt{\arxivref{1602.04928}{arXiv:1602.04928}}}\ignorespaces
\bibitem{Nakayama:2016knq}
Y.~Nakayama,
\textit{``{Bootstrap bound for conformal multi-flavor QCD on lattice}''},
\doiref{10.1007/JHEP07(2016)038}{JHEP \textbf{1607}, 038
  (2016)\ignorespaces}\ignorespaces,
\normalsize{\texttt{\arxivref{1605.04052}{arXiv:1605.04052}}}\ignorespaces
\bibitem{Li:2016wdp}
Z.~Li \& N.~Su,
\textit{``{Bootstrapping Mixed Correlators in the Five Dimensional Critical
  O(N) Models}''},
\doiref{10.1007/JHEP04(2017)098}{JHEP \textbf{1704}, 098
  (2017)\ignorespaces}\ignorespaces,
\normalsize{\texttt{\arxivref{1607.07077}{arXiv:1607.07077}}}\ignorespaces
\bibitem{Pang:2016xno}
Y.~Pang, J.~Rong \& N.~Su,
\textit{``{$\phi^{3}$ theory with F$_{4}$ flavor symmetry in 6 − 2$\epsilon$
  dimensions: 3-loop renormalization and conformal bootstrap}''},
\doiref{10.1007/JHEP12(2016)057}{JHEP \textbf{1612}, 057
  (2016)\ignorespaces}\ignorespaces,
\normalsize{\texttt{\arxivref{1609.03007}{arXiv:1609.03007}}}\ignorespaces
\bibitem{Dymarsky:2017xzb}
A.~Dymarsky, J.~Penedones, E.~Trevisani \& A.~Vichi,
\textit{``{Charting the space of 3D CFTs with a continuous global symmetry}''},
\normalsize{\texttt{\arxivref{1705.04278}{arXiv:1705.04278}}}\ignorespaces
\bibitem{Stergiou:2018gjj}
A.~Stergiou,
\textit{``{Bootstrapping hypercubic and hypertetrahedral theories in three
  dimensions}''},
\doiref{10.1007/JHEP05(2018)035}{JHEP \textbf{1805}, 035
  (2018)\ignorespaces}\ignorespaces,
\normalsize{\texttt{\arxivref{1801.07127}{arXiv:1801.07127}}}\ignorespaces
\bibitem{Kousvos:2018rhl}
S.~R. Kousvos \& A.~Stergiou,
\textit{``{Bootstrapping Mixed Correlators in Three-Dimensional Cubic
  Theories}''},
\normalsize{\texttt{\arxivref{1810.10015}{arXiv:1810.10015}}}\ignorespaces
\bibitem{Stergiou:2019dcv}
A.~Stergiou,
\textit{``{Bootstrapping MN and Tetragonal CFTs in Three Dimensions}''},
\normalsize{\texttt{\arxivref{1904.00017}{arXiv:1904.00017}}}\ignorespaces
\bibitem{Alday:2015eya}
L.~F. Alday, A.~Bissi \& T.~Lukowski,
\textit{``{Large spin systematics in CFT}''},
\doiref{10.1007/JHEP11(2015)101}{JHEP \textbf{1511}, 101
  (2015)\ignorespaces}\ignorespaces,
\normalsize{\texttt{\arxivref{1502.07707}{arXiv:1502.07707}}}\ignorespaces
\bibitem{Alday:2015ota}
L.~F. Alday \& A.~Zhiboedov,
\textit{``{Conformal Bootstrap With Slightly Broken Higher Spin Symmetry}''},
\doiref{10.1007/JHEP06(2016)091}{JHEP \textbf{1606}, 091
  (2016)\ignorespaces}\ignorespaces,
\normalsize{\texttt{\arxivref{1506.04659}{arXiv:1506.04659}}}\ignorespaces
\bibitem{Alday:2015ewa}
L.~F. Alday \& A.~Zhiboedov,
\textit{``{An Algebraic Approach to the Analytic Bootstrap}''},
\doiref{10.1007/JHEP04(2017)157}{JHEP \textbf{1704}, 157
  (2017)\ignorespaces}\ignorespaces,
\normalsize{\texttt{\arxivref{1510.08091}{arXiv:1510.08091}}}\ignorespaces
\bibitem{Alday:2016mxe}
L.~F. Alday \& A.~Bissi,
\textit{``{Crossing symmetry and Higher spin towers}''},
\doiref{10.1007/JHEP12(2017)118}{JHEP \textbf{1712}, 118
  (2017)\ignorespaces}\ignorespaces,
\normalsize{\texttt{\arxivref{1603.05150}{arXiv:1603.05150}}}\ignorespaces
\bibitem{Alday:2016njk}
L.~F. Alday,
\textit{``{Large Spin Perturbation Theory for Conformal Field Theories}''},
\doiref{10.1103/PhysRevLett.119.111601}{Phys.~Rev.~Lett. \textbf{119}, 111601
  (2017)\ignorespaces}\ignorespaces,
\normalsize{\texttt{\arxivref{1611.01500}{arXiv:1611.01500}}}\ignorespaces
\bibitem{Alday:2016jfr}
L.~F. Alday,
\textit{``{Solving CFTs with Weakly Broken Higher Spin Symmetry}''},
\doiref{10.1007/JHEP10(2017)161}{JHEP \textbf{1710}, 161
  (2017)\ignorespaces}\ignorespaces,
\normalsize{\texttt{\arxivref{1612.00696}{arXiv:1612.00696}}}\ignorespaces
\bibitem{Rychkov:2016iqz}
S.~Rychkov,
\textit{``{EPFL Lectures on Conformal Field Theory in D>= 3 Dimensions}''},
\normalsize{\texttt{\arxivref{1601.05000}{arXiv:1601.05000}}}\ignorespaces
\bibitem{Simmons-Duffin:2016gjk}
D.~Simmons-Duffin,
\textit{``{The Conformal Bootstrap}''},
\normalsize{\texttt{\arxivref{1602.07982}{arXiv:1602.07982}}}\ignorespaces,
in \textit{``{Proceedings, Theoretical Advanced Study Institute in Elementary
  Particle Physics: New Frontiers in Fields and Strings (TASI 2015): Boulder,
  CO, USA, June 1-26, 2015}''},
1-74\ignorespaces
\bibitem{Poland:2018epd}
D.~Poland, S.~Rychkov \& A.~Vichi,
\textit{``{The Conformal Bootstrap: Theory, Numerical Techniques, and
  Applications}''},
\doiref{10.1103/RevModPhys.91.015002}{Rev.~Mod.~Phys. \textbf{91}, 15002
  (2019)\ignorespaces}\ignorespaces,
\normalsize{\texttt{\arxivref{1805.04405}{arXiv:1805.04405}}}\ignorespaces,
[Rev. Mod. Phys.91,015002(2019)]\ignorespaces
\bibitem{Chester:2019wfx}
S.~M. Chester,
\textit{``{Weizmann Lectures on the Numerical Conformal Bootstrap}''},
\normalsize{\texttt{\arxivref{1907.05147}{arXiv:1907.05147}}}\ignorespaces
\bibitem{Dolan:2000ut}
F.~A. Dolan \& H.~Osborn,
\textit{``{Conformal four point functions and the operator product
  expansion}''},
\doiref{10.1016/S0550-3213(01)00013-X}{Nucl.~Phys. \textbf{B599}, 459
  (2001)\ignorespaces}\ignorespaces,
\normalsize{\texttt{\arxivref{hep-th/0011040}{hep-th/0011040}}}\ignorespaces
\bibitem{Dolan:2003hv}
F.~A. Dolan \& H.~Osborn,
\textit{``{Conformal partial waves and the operator product expansion}''},
\doiref{10.1016/j.nuclphysb.2003.11.016}{Nucl.~Phys. \textbf{B678}, 491
  (2004)\ignorespaces}\ignorespaces,
\normalsize{\texttt{\arxivref{hep-th/0309180}{hep-th/0309180}}}\ignorespaces
\bibitem{Ferrara:1973vz}
S.~Ferrara, A.~F. Grillo, G.~Parisi \& R.~Gatto,
\textit{``{Covariant expansion of the conformal four-point function}''},
\doiref{10.1016/0550-3213(72)90587-1, 10.1016/0550-3213(73)90467-7}{Nucl.~Phys.
  \textbf{B49}, 77 (1972)\ignorespaces}\ignorespaces,
[Erratum: Nucl. Phys.B53,643(1973)]\ignorespaces
\bibitem{Ferrara:1974nf}
S.~Ferrara, A.~F. Grillo, R.~Gatto \& G.~Parisi,
\textit{``{Analyticity properties and asymptotic expansions of conformal
  covariant green's functions}''},
\doiref{10.1007/BF02813413}{Nuovo~Cim. \textbf{A19}, 667
  (1974)\ignorespaces}\ignorespaces
\bibitem{Dobrev:1977qv}
V.~K. Dobrev, G.~Mack, V.~B. Petkova, S.~G. Petrova \& I.~T. Todorov,
\textit{``{Harmonic Analysis on the n-Dimensional Lorentz Group and Its
  Application to Conformal Quantum Field Theory}''},
\doiref{10.1007/BFb0009678}{Lect.~Notes~Phys. \textbf{63}, 1
  (1977)\ignorespaces}\ignorespaces
\bibitem{Exton_1995}
H.~Exton,
\textit{``On the system of partial differential equations associated with
  Appell's function F4''},
\doiref{10.1088/0305-4470/28/3/017}{Journal~of~Physics~A:~Mathematical~and~General
  \textbf{28}, 631 (1995)\ignorespaces}\ignorespaces
\bibitem{Giombi:2011rz}
S.~Giombi, S.~Prakash \& X.~Yin,
\textit{``{A Note on CFT Correlators in Three Dimensions}''},
\doiref{10.1007/JHEP07(2013)105}{JHEP \textbf{1307}, 105
  (2013)\ignorespaces}\ignorespaces,
\normalsize{\texttt{\arxivref{1104.4317}{arXiv:1104.4317}}}\ignorespaces
\bibitem{Costa:2011mg}
M.~S. Costa, J.~Penedones, D.~Poland \& S.~Rychkov,
\textit{``{Spinning Conformal Correlators}''},
\doiref{10.1007/JHEP11(2011)071}{JHEP \textbf{1111}, 071
  (2011)\ignorespaces}\ignorespaces,
\normalsize{\texttt{\arxivref{1107.3554}{arXiv:1107.3554}}}\ignorespaces
\bibitem{Dolan:2011dv}
F.~A. Dolan \& H.~Osborn,
\textit{``{Conformal Partial Waves: Further Mathematical Results}''},
\normalsize{\texttt{\arxivref{1108.6194}{arXiv:1108.6194}}}\ignorespaces
\bibitem{Costa:2011dw}
M.~S. Costa, J.~Penedones, D.~Poland \& S.~Rychkov,
\textit{``{Spinning Conformal Blocks}''},
\doiref{10.1007/JHEP11(2011)154}{JHEP \textbf{1111}, 154
  (2011)\ignorespaces}\ignorespaces,
\normalsize{\texttt{\arxivref{1109.6321}{arXiv:1109.6321}}}\ignorespaces
\bibitem{SimmonsDuffin:2012uy}
D.~Simmons-Duffin,
\textit{``{Projectors, Shadows, and Conformal Blocks}''},
\doiref{10.1007/JHEP04(2014)146}{JHEP \textbf{1404}, 146
  (2014)\ignorespaces}\ignorespaces,
\normalsize{\texttt{\arxivref{1204.3894}{arXiv:1204.3894}}}\ignorespaces
\bibitem{Costa:2014rya}
M.~S. Costa \& T.~Hansen,
\textit{``{Conformal correlators of mixed-symmetry tensors}''},
\doiref{10.1007/JHEP02(2015)151}{JHEP \textbf{1502}, 151
  (2015)\ignorespaces}\ignorespaces,
\normalsize{\texttt{\arxivref{1411.7351}{arXiv:1411.7351}}}\ignorespaces
\bibitem{Elkhidir:2014woa}
E.~Elkhidir, D.~Karateev \& M.~Serone,
\textit{``{General Three-Point Functions in 4D CFT}''},
\doiref{10.1007/JHEP01(2015)133}{JHEP \textbf{1501}, 133
  (2015)\ignorespaces}\ignorespaces,
\normalsize{\texttt{\arxivref{1412.1796}{arXiv:1412.1796}}}\ignorespaces
\bibitem{Echeverri:2015rwa}
A.~Castedo~Echeverri, E.~Elkhidir, D.~Karateev \& M.~Serone,
\textit{``{Deconstructing Conformal Blocks in 4D CFT}''},
\doiref{10.1007/JHEP08(2015)101}{JHEP \textbf{1508}, 101
  (2015)\ignorespaces}\ignorespaces,
\normalsize{\texttt{\arxivref{1505.03750}{arXiv:1505.03750}}}\ignorespaces
\bibitem{Hijano:2015zsa}
E.~Hijano, P.~Kraus, E.~Perlmutter \& R.~Snively,
\textit{``{Witten Diagrams Revisited: The AdS Geometry of Conformal Blocks}''},
\doiref{10.1007/JHEP01(2016)146}{JHEP \textbf{1601}, 146
  (2016)\ignorespaces}\ignorespaces,
\normalsize{\texttt{\arxivref{1508.00501}{arXiv:1508.00501}}}\ignorespaces
\bibitem{Rejon-Barrera:2015bpa}
F.~Rejon-Barrera \& D.~Robbins,
\textit{``{Scalar-Vector Bootstrap}''},
\doiref{10.1007/JHEP01(2016)139}{JHEP \textbf{1601}, 139
  (2016)\ignorespaces}\ignorespaces,
\normalsize{\texttt{\arxivref{1508.02676}{arXiv:1508.02676}}}\ignorespaces
\bibitem{Penedones:2015aga}
J.~Penedones, E.~Trevisani \& M.~Yamazaki,
\textit{``{Recursion Relations for Conformal Blocks}''},
\doiref{10.1007/JHEP09(2016)070}{JHEP \textbf{1609}, 070
  (2016)\ignorespaces}\ignorespaces,
\normalsize{\texttt{\arxivref{1509.00428}{arXiv:1509.00428}}}\ignorespaces
\bibitem{Iliesiu:2015akf}
L.~Iliesiu, F.~Kos, D.~Poland, S.~S. Pufu, D.~Simmons-Duffin \& R.~Yacoby,
\textit{``{Fermion-Scalar Conformal Blocks}''},
\doiref{10.1007/JHEP04(2016)074}{JHEP \textbf{1604}, 074
  (2016)\ignorespaces}\ignorespaces,
\normalsize{\texttt{\arxivref{1511.01497}{arXiv:1511.01497}}}\ignorespaces
\bibitem{Echeverri:2016dun}
A.~Castedo~Echeverri, E.~Elkhidir, D.~Karateev \& M.~Serone,
\textit{``{Seed Conformal Blocks in 4D CFT}''},
\doiref{10.1007/JHEP02(2016)183}{JHEP \textbf{1602}, 183
  (2016)\ignorespaces}\ignorespaces,
\normalsize{\texttt{\arxivref{1601.05325}{arXiv:1601.05325}}}\ignorespaces
\bibitem{Isachenkov:2016gim}
M.~Isachenkov \& V.~Schomerus,
\textit{``{Superintegrability of $d$-dimensional Conformal Blocks}''},
\doiref{10.1103/PhysRevLett.117.071602}{Phys.~Rev.~Lett. \textbf{117}, 071602
  (2016)\ignorespaces}\ignorespaces,
\normalsize{\texttt{\arxivref{1602.01858}{arXiv:1602.01858}}}\ignorespaces
\bibitem{Costa:2016hju}
M.~S. Costa, T.~Hansen, J.~Penedones \& E.~Trevisani,
\textit{``{Projectors and seed conformal blocks for traceless mixed-symmetry
  tensors}''},
\doiref{10.1007/JHEP07(2016)018}{JHEP \textbf{1607}, 018
  (2016)\ignorespaces}\ignorespaces,
\normalsize{\texttt{\arxivref{1603.05551}{arXiv:1603.05551}}}\ignorespaces
\bibitem{Costa:2016xah}
M.~S. Costa, T.~Hansen, J.~Penedones \& E.~Trevisani,
\textit{``{Radial expansion for spinning conformal blocks}''},
\doiref{10.1007/JHEP07(2016)057}{JHEP \textbf{1607}, 057
  (2016)\ignorespaces}\ignorespaces,
\normalsize{\texttt{\arxivref{1603.05552}{arXiv:1603.05552}}}\ignorespaces
\bibitem{Chen:2016bxc}
H.-Y. Chen \& J.~D. Qualls,
\textit{``{Quantum Integrable Systems from Conformal Blocks}''},
\doiref{10.1103/PhysRevD.95.106011}{Phys.~Rev. \textbf{D95}, 106011
  (2017)\ignorespaces}\ignorespaces,
\normalsize{\texttt{\arxivref{1605.05105}{arXiv:1605.05105}}}\ignorespaces
\bibitem{Nishida:2016vds}
M.~Nishida \& K.~Tamaoka,
\textit{``{Geodesic Witten diagrams with an external spinning field}''},
\doiref{10.1093/ptep/ptx055}{PTEP \textbf{2017}, 053B06
  (2017)\ignorespaces}\ignorespaces,
\normalsize{\texttt{\arxivref{1609.04563}{arXiv:1609.04563}}}\ignorespaces
\bibitem{Cordova:2016emh}
C.~Cordova, T.~T. Dumitrescu \& K.~Intriligator,
\textit{``{Multiplets of Superconformal Symmetry in Diverse Dimensions}''},
\doiref{10.1007/JHEP03(2019)163}{JHEP \textbf{1903}, 163
  (2019)\ignorespaces}\ignorespaces,
\normalsize{\texttt{\arxivref{1612.00809}{arXiv:1612.00809}}}\ignorespaces
\bibitem{Schomerus:2016epl}
V.~Schomerus, E.~Sobko \& M.~Isachenkov,
\textit{``{Harmony of Spinning Conformal Blocks}''},
\doiref{10.1007/JHEP03(2017)085}{JHEP \textbf{1703}, 085
  (2017)\ignorespaces}\ignorespaces,
\normalsize{\texttt{\arxivref{1612.02479}{arXiv:1612.02479}}}\ignorespaces
\bibitem{Kravchuk:2016qvl}
P.~Kravchuk \& D.~Simmons-Duffin,
\textit{``{Counting Conformal Correlators}''},
\doiref{10.1007/JHEP02(2018)096}{JHEP \textbf{1802}, 096
  (2018)\ignorespaces}\ignorespaces,
\normalsize{\texttt{\arxivref{1612.08987}{arXiv:1612.08987}}}\ignorespaces
\bibitem{Gliozzi:2017hni}
F.~Gliozzi, A.~L. Guerrieri, A.~C. Petkou \& C.~Wen,
\textit{``{The analytic structure of conformal blocks and the generalized
  Wilson-Fisher fixed points}''},
\doiref{10.1007/JHEP04(2017)056}{JHEP \textbf{1704}, 056
  (2017)\ignorespaces}\ignorespaces,
\normalsize{\texttt{\arxivref{1702.03938}{arXiv:1702.03938}}}\ignorespaces
\bibitem{Castro:2017hpx}
A.~Castro, E.~Llabrés \& F.~Rejon-Barrera,
\textit{``{Geodesic Diagrams, Gravitational Interactions \& OPE Structures}''},
\doiref{10.1007/JHEP06(2017)099}{JHEP \textbf{1706}, 099
  (2017)\ignorespaces}\ignorespaces,
\normalsize{\texttt{\arxivref{1702.06128}{arXiv:1702.06128}}}\ignorespaces
\bibitem{Dyer:2017zef}
E.~Dyer, D.~Z. Freedman \& J.~Sully,
\textit{``{Spinning Geodesic Witten Diagrams}''},
\doiref{10.1007/JHEP11(2017)060}{JHEP \textbf{1711}, 060
  (2017)\ignorespaces}\ignorespaces,
\normalsize{\texttt{\arxivref{1702.06139}{arXiv:1702.06139}}}\ignorespaces
\bibitem{Sleight:2017fpc}
C.~Sleight \& M.~Taronna,
\textit{``{Spinning Witten Diagrams}''},
\doiref{10.1007/JHEP06(2017)100}{JHEP \textbf{1706}, 100
  (2017)\ignorespaces}\ignorespaces,
\normalsize{\texttt{\arxivref{1702.08619}{arXiv:1702.08619}}}\ignorespaces
\bibitem{Chen:2017yia}
H.-Y. Chen, E.-J. Kuo \& H.~Kyono,
\textit{``{Anatomy of Geodesic Witten Diagrams}''},
\doiref{10.1007/JHEP05(2017)070}{JHEP \textbf{1705}, 070
  (2017)\ignorespaces}\ignorespaces,
\normalsize{\texttt{\arxivref{1702.08818}{arXiv:1702.08818}}}\ignorespaces
\bibitem{Pasterski:2017kqt}
S.~Pasterski \& S.-H. Shao,
\textit{``{Conformal basis for flat space amplitudes}''},
\doiref{10.1103/PhysRevD.96.065022}{Phys.~Rev. \textbf{D96}, 065022
  (2017)\ignorespaces}\ignorespaces,
\normalsize{\texttt{\arxivref{1705.01027}{arXiv:1705.01027}}}\ignorespaces
\bibitem{Cardoso:2017qmj}
V.~Cardoso, T.~Houri \& M.~Kimura,
\textit{``{Mass Ladder Operators from Spacetime Conformal Symmetry}''},
\doiref{10.1103/PhysRevD.96.024044}{Phys.~Rev. \textbf{D96}, 024044
  (2017)\ignorespaces}\ignorespaces,
\normalsize{\texttt{\arxivref{1706.07339}{arXiv:1706.07339}}}\ignorespaces
\bibitem{Karateev:2017jgd}
D.~Karateev, P.~Kravchuk \& D.~Simmons-Duffin,
\textit{``{Weight Shifting Operators and Conformal Blocks}''},
\doiref{10.1007/JHEP02(2018)081}{JHEP \textbf{1802}, 081
  (2018)\ignorespaces}\ignorespaces,
\normalsize{\texttt{\arxivref{1706.07813}{arXiv:1706.07813}}}\ignorespaces,
[,91(2017)]\ignorespaces
\bibitem{Kravchuk:2017dzd}
P.~Kravchuk,
\textit{``{Casimir recursion relations for general conformal blocks}''},
\doiref{10.1007/JHEP02(2018)011}{JHEP \textbf{1802}, 011
  (2018)\ignorespaces}\ignorespaces,
\normalsize{\texttt{\arxivref{1709.05347}{arXiv:1709.05347}}}\ignorespaces,
[,164(2017)]\ignorespaces
\bibitem{Dey:2017fab}
P.~Dey, K.~Ghosh \& A.~Sinha,
\textit{``{Simplifying large spin bootstrap in Mellin space}''},
\doiref{10.1007/JHEP01(2018)152}{JHEP \textbf{1801}, 152
  (2018)\ignorespaces}\ignorespaces,
\normalsize{\texttt{\arxivref{1709.06110}{arXiv:1709.06110}}}\ignorespaces
\bibitem{Hollands:2017chb}
S.~Hollands,
\textit{``{Action principle for OPE}''},
\doiref{10.1016/j.nuclphysb.2017.11.013}{Nucl.~Phys. \textbf{B926}, 614
  (2018)\ignorespaces}\ignorespaces,
\normalsize{\texttt{\arxivref{1710.05601}{arXiv:1710.05601}}}\ignorespaces
\bibitem{Schomerus:2017eny}
V.~Schomerus \& E.~Sobko,
\textit{``{From Spinning Conformal Blocks to Matrix Calogero-Sutherland
  Models}''},
\doiref{10.1007/JHEP04(2018)052}{JHEP \textbf{1804}, 052
  (2018)\ignorespaces}\ignorespaces,
\normalsize{\texttt{\arxivref{1711.02022}{arXiv:1711.02022}}}\ignorespaces
\bibitem{Isachenkov:2017qgn}
M.~Isachenkov \& V.~Schomerus,
\textit{``{Integrability of conformal blocks. Part I. Calogero-Sutherland
  scattering theory}''},
\doiref{10.1007/JHEP07(2018)180}{JHEP \textbf{1807}, 180
  (2018)\ignorespaces}\ignorespaces,
\normalsize{\texttt{\arxivref{1711.06609}{arXiv:1711.06609}}}\ignorespaces
\bibitem{Faller:2017hyt}
J.~Faller, S.~Sarkar \& M.~Verma,
\textit{``{Mellin Amplitudes for Fermionic Conformal Correlators}''},
\doiref{10.1007/JHEP03(2018)106}{JHEP \textbf{1803}, 106
  (2018)\ignorespaces}\ignorespaces,
\normalsize{\texttt{\arxivref{1711.07929}{arXiv:1711.07929}}}\ignorespaces
\bibitem{Rong:2017cow}
J.~Rong \& N.~Su,
\textit{``{Scalar CFTs and Their Large N Limits}''},
\doiref{10.1007/JHEP09(2018)103}{JHEP \textbf{1809}, 103
  (2018)\ignorespaces}\ignorespaces,
\normalsize{\texttt{\arxivref{1712.00985}{arXiv:1712.00985}}}\ignorespaces
\bibitem{Chen:2017xdz}
H.-Y. Chen, E.-J. Kuo \& H.~Kyono,
\textit{``{Towards Spinning Mellin Amplitudes}''},
\doiref{10.1016/j.nuclphysb.2018.04.019}{Nucl.~Phys. \textbf{B931}, 291
  (2018)\ignorespaces}\ignorespaces,
\normalsize{\texttt{\arxivref{1712.07991}{arXiv:1712.07991}}}\ignorespaces
\bibitem{Sleight:2018epi}
C.~Sleight \& M.~Taronna,
\textit{``{Spinning Mellin Bootstrap: Conformal Partial Waves, Crossing Kernels
  and Applications}''},
\doiref{10.1002/prop.201800038}{Fortsch.~Phys. \textbf{66}, 1800038
  (2018)\ignorespaces}\ignorespaces,
\normalsize{\texttt{\arxivref{1804.09334}{arXiv:1804.09334}}}\ignorespaces
\bibitem{Costa:2018mcg}
M.~S. Costa \& T.~Hansen,
\textit{``{AdS Weight Shifting Operators}''},
\doiref{10.1007/JHEP09(2018)040}{JHEP \textbf{1809}, 040
  (2018)\ignorespaces}\ignorespaces,
\normalsize{\texttt{\arxivref{1805.01492}{arXiv:1805.01492}}}\ignorespaces
\bibitem{Kobayashi:2018okw}
N.~Kobayashi \& T.~Nishioka,
\textit{``{Spinning conformal defects}''},
\doiref{10.1007/JHEP09(2018)134}{JHEP \textbf{1809}, 134
  (2018)\ignorespaces}\ignorespaces,
\normalsize{\texttt{\arxivref{1805.05967}{arXiv:1805.05967}}}\ignorespaces
\bibitem{Bhatta:2018gjb}
A.~Bhatta, P.~Raman \& N.~V. Suryanarayana,
\textit{``{Scalar Blocks as Gravitational Wilson Networks}''},
\doiref{10.1007/JHEP12(2018)125}{JHEP \textbf{1812}, 125
  (2018)\ignorespaces}\ignorespaces,
\normalsize{\texttt{\arxivref{1806.05475}{arXiv:1806.05475}}}\ignorespaces
\bibitem{Lauria:2018klo}
E.~Lauria, M.~Meineri \& E.~Trevisani,
\textit{``{Spinning operators and defects in conformal field theory}''},
\normalsize{\texttt{\arxivref{1807.02522}{arXiv:1807.02522}}}\ignorespaces
\bibitem{Liu:2018jhs}
J.~Liu, E.~Perlmutter, V.~Rosenhaus \& D.~Simmons-Duffin,
\textit{``{$d$-dimensional SYK, AdS Loops, and $6j$ Symbols}''},
\doiref{10.1007/JHEP03(2019)052}{JHEP \textbf{1903}, 052
  (2019)\ignorespaces}\ignorespaces,
\normalsize{\texttt{\arxivref{1808.00612}{arXiv:1808.00612}}}\ignorespaces
\bibitem{Gromov:2018hut}
N.~Gromov, V.~Kazakov \& G.~Korchemsky,
\textit{``{Exact Correlation Functions in Conformal Fishnet Theory}''},
\normalsize{\texttt{\arxivref{1808.02688}{arXiv:1808.02688}}}\ignorespaces
\bibitem{Rosenhaus:2018zqn}
V.~Rosenhaus,
\textit{``{Multipoint Conformal Blocks in the Comb Channel}''},
\doiref{10.1007/JHEP02(2019)142}{JHEP \textbf{1902}, 142
  (2019)\ignorespaces}\ignorespaces,
\normalsize{\texttt{\arxivref{1810.03244}{arXiv:1810.03244}}}\ignorespaces
\bibitem{Zhou:2018sfz}
X.~Zhou,
\textit{``{Recursion Relations in Witten Diagrams and Conformal Partial
  Waves}''},
\normalsize{\texttt{\arxivref{1812.01006}{arXiv:1812.01006}}}\ignorespaces
\bibitem{Kazakov:2018gcy}
V.~Kazakov, E.~Olivucci \& M.~Preti,
\textit{``{Generalized fishnets and exact four-point correlators in chiral
  CFT$_{4}$}''},
\doiref{10.1007/JHEP06(2019)078}{JHEP \textbf{1906}, 078
  (2019)\ignorespaces}\ignorespaces,
\normalsize{\texttt{\arxivref{1901.00011}{arXiv:1901.00011}}}\ignorespaces
\bibitem{Li:2019dix}
W.~Li,
\textit{``{Closed-form expression for cross-channel conformal blocks near the
  lightcone}''},
\normalsize{\texttt{\arxivref{1906.00707}{arXiv:1906.00707}}}\ignorespaces
\bibitem{Goncalves:2019znr}
V.~Gonçalves, R.~Pereira \& X.~Zhou,
\textit{``{$20'$ Five-Point Function from $AdS_5\times S^5$ Supergravity}''},
\normalsize{\texttt{\arxivref{1906.05305}{arXiv:1906.05305}}}\ignorespaces
\bibitem{Jepsen:2019svc}
C.~B. Jepsen \& S.~Parikh,
\textit{``{Propagator identities, holographic conformal blocks, and
  higher-point AdS diagrams}''},
\normalsize{\texttt{\arxivref{1906.08405}{arXiv:1906.08405}}}\ignorespaces
\bibitem{Dirac:1936fq}
P.~A.~M. Dirac,
\textit{``{Wave equations in conformal space}''},
\doiref{10.2307/1968455}{Annals~Math. \textbf{37}, 429
  (1936)\ignorespaces}\ignorespaces
\bibitem{Mack:1969rr}
G.~Mack \& A.~Salam,
\textit{``{Finite component field representations of the conformal group}''},
\doiref{10.1016/0003-4916(69)90278-4}{Annals~Phys. \textbf{53}, 174
  (1969)\ignorespaces}\ignorespaces
\bibitem{Weinberg:2010fx}
S.~Weinberg,
\textit{``{Six-dimensional Methods for Four-dimensional Conformal Field
  Theories}''},
\doiref{10.1103/PhysRevD.82.045031}{Phys.Rev. \textbf{D82}, 045031
  (2010)\ignorespaces}\ignorespaces,
\normalsize{\texttt{\arxivref{1006.3480}{arXiv:1006.3480}}}\ignorespaces
\bibitem{Weinberg:2012mz}
S.~Weinberg,
\textit{``{Six-dimensional Methods for Four-dimensional Conformal Field
  Theories II: Irreducible Fields}''},
\doiref{10.1103/PhysRevD.86.085013}{Phys.Rev. \textbf{D86}, 085013
  (2012)\ignorespaces}\ignorespaces,
\normalsize{\texttt{\arxivref{1209.4659}{arXiv:1209.4659}}}\ignorespaces
\bibitem{Ferrara:1971vh}
S.~Ferrara, A.~F. Grillo \& R.~Gatto,
\textit{``{Manifestly conformal covariant operator-product expansion}''},
\doiref{10.1007/BF02770435}{Lett.~Nuovo~Cim. \textbf{2S2}, 1363
  (1971)\ignorespaces}\ignorespaces,
[Lett. Nuovo Cim.2,1363(1971)]\ignorespaces
\bibitem{Ferrara:1971zy}
S.~Ferrara, R.~Gatto \& A.~F. Grillo,
\textit{``{Conformal invariance on the light cone and canonical dimensions}''},
\doiref{10.1016/0550-3213(71)90333-6}{Nucl.~Phys. \textbf{B34}, 349
  (1971)\ignorespaces}\ignorespaces
\bibitem{Ferrara:1972cq}
S.~Ferrara, A.~F. Grillo \& R.~Gatto,
\textit{``{Manifestly conformal-covariant expansion on the light cone}''},
\doiref{10.1103/PhysRevD.5.3102}{Phys.~Rev. \textbf{D5}, 3102
  (1972)\ignorespaces}\ignorespaces
\bibitem{Ferrara:1973eg}
S.~Ferrara, P.~Gatto \& A.~F. Grilla,
\textit{``{Conformal algebra in spacetime and operator product expansion}''},
\doiref{10.1007/BFb0111104}{Springer~Tracts~Mod.~Phys. \textbf{67}, 1
  (1973)\ignorespaces}\ignorespaces
\bibitem{Dobrev:1975ru}
V.~K. Dobrev, V.~B. Petkova, S.~G. Petrova \& I.~T. Todorov,
\textit{``{Dynamical Derivation of Vacuum Operator Product Expansion in
  Euclidean Conformal Quantum Field Theory}''},
\doiref{10.1103/PhysRevD.13.887}{Phys.~Rev. \textbf{D13}, 887
  (1976)\ignorespaces}\ignorespaces
\bibitem{Mack:1976pa}
G.~Mack,
\textit{``{Convergence of Operator Product Expansions on the Vacuum in
  Conformal Invariant Quantum Field Theory}''},
\doiref{10.1007/BF01609130}{Commun.~Math.~Phys. \textbf{53}, 155
  (1977)\ignorespaces}\ignorespaces
\bibitem{Fortin:2016lmf}
J.-F. Fortin \& W.~Skiba,
\textit{``{Conformal Bootstrap in Embedding Space}''},
\doiref{10.1103/PhysRevD.93.105047}{Phys.~Rev. \textbf{D93}, 105047
  (2016)\ignorespaces}\ignorespaces,
\normalsize{\texttt{\arxivref{1602.05794}{arXiv:1602.05794}}}\ignorespaces
\bibitem{Fortin:2016dlj}
J.-F. Fortin \& W.~Skiba,
\textit{``{Conformal Differential Operator in Embedding Space and its
  Applications}''},
\normalsize{\texttt{\arxivref{1612.08672}{arXiv:1612.08672}}}\ignorespaces
\bibitem{Comeau:2019xco}
V.~Comeau, J.-F. Fortin \& W.~Skiba,
\textit{``{Further Results on a Function Relevant for Conformal Blocks}''},
\normalsize{\texttt{\arxivref{1902.08598}{arXiv:1902.08598}}}\ignorespaces
\bibitem{Fortin:2019xyr}
J.-F. Fortin, V.~Prilepina \& W.~Skiba,
\textit{``{Conformal Two-Point Correlation Functions from the Operator Product
  Expansion}''},
\normalsize{\texttt{\arxivref{1906.12349}{arXiv:1906.12349}}}\ignorespaces
\bibitem{Fortin:2019pep}
J.-F. Fortin, V.~Prilepina \& W.~Skiba,
\textit{``{Conformal Three-Point Correlation Functions from the Operator
  Product Expansion}''},
\normalsize{\texttt{\arxivref{1907.08599}{arXiv:1907.08599}}}\ignorespaces
\end{thebibliography}

\end{document}